\documentclass[twocolumn,10pt]{article}

\usepackage{bm}
\usepackage{comment}
\usepackage{soul}
\setstcolor{red}
\usepackage{booktabs} 

\usepackage[a4paper, left=1.5cm, right=1.5cm, top=2cm, bottom=2cm]{geometry}

\usepackage[dvips]{graphicx}
\usepackage{epsfig}
\usepackage{pstricks}
\usepackage{amsmath,amssymb,amsfonts}
\usepackage{color}
\usepackage{enumitem}
\usepackage{titletoc}
\usepackage{hyperref}

\newcommand{\tr}{\text{Tr}}
\renewcommand{\dag}{\dagger}
\newcommand{\Ito}{It\={o}}

\newcommand{\q}[1]{|#1\rangle}
\newcommand{\qd}[1]{\langle #1|}

\newtheorem{prop}{Proposition}
\newtheorem{cor}{Corollary}[prop]

\title{Confinement to deterministic manifolds and low-dimensional solution formulas for continuously measured quantum systems}

\author{Alain Sarlette
\thanks{Laboratoire de Physique de l'Ecole normale supérieure, ENS-PSL, CNRS, Inria, Centre Automatique et Systèmes (CAS), Mines Paris, Université PSL, Sorbonne Université, Université Paris Cité, Paris, France ; Department of Electronics and Information Systems, Ghent University, Belgium} \and
Cyril Elourard
\thanks{Universit\'e de Lorraine, CNRS, LPCT, F-54000 Nancy, France}
\and
Pierre Rouchon
\thanks{Laboratoire de Physique de l'Ecole normale supérieure, ENS-PSL, CNRS, Inria, Centre Automatique et Systèmes (CAS), Mines Paris, Université PSL, Sorbonne Université, Université Paris Cité, Paris, France}}

\begin{document}

\maketitle

\begin{abstract}
Quantum systems subjected to a continuous weak measurement process evolve according to stochastic differential equations (SDE). Depending on the outcomes of these stochastic measurements, the quantum state may diffuse in various directions across the state space. This note points out that in many scenarios relevant to quantum engineering, this diffusion is effectively constrained to a low-dimensional space. Specifically, the quantum state remains confined to a low-dimensional, nonlinear manifold --- often time-dependent, yet independent of the specific measurement outcomes. This note derives the corresponding low-dimensional formulations for expressing the stochastically evolving state in several prototypical cases: quantum non-demolition measurements in arbitrary dimensions; quadrature measurements of a harmonic oscillator (linear quantum system); and measurements of subsystems within multipartite quantum systems. Additionally, it introduces an algebraic criterion to determine whether such low-dimensional manifolds exist or persist when additional dynamics are present.
\end{abstract}


\section{Introduction}

The collapse of the wavefunction under projective measurements is among the most distinctive features of quantum mechanics \cite{WCphil0,WCphil1,WCphil2}. For more general measurement processes, the update of the wavefunction $\q{\psi_t}$, or equivalently of the density operator $\rho_t$, conditioned on measurement outcomes, follows the standard framework of conditional dynamics or quantum filtering \cite{Haroche,QF0}. In the continuous-time limit, the evolution equations governing this conditional state are known by various names, including the Belavkin filter, quantum trajectories, and the stochastic master equation; see, for instance, \cite{QF1,QF2}. The corresponding \emph{quantum trajectory} solution, denoted $\rho_t(y)$, describes the density operator that governs the statistics of all observables at time $t$, conditioned on an initial state $\rho_0$, on the known system dynamics (including Hamiltonian evolution, decoherence channels, and measurement channels), and on the measurement record $y$ accumulated over the time interval $[0,t)$. This formulation constitutes the most widely used quantum filter for real-time estimation and control operation, and it is the central object of study in this note. Alternative filtering approaches also exist. Notably, the past quantum state filter \cite{PQS0,PQS1,PQS2} provides a conditioned estimate of $\rho_t$ based on measurement outcomes collected over the entire interval $[0,T]$, with $t \in [0,T]$. This approach is particularly useful in post-processing contexts, such as characterization of an experimental setting, retrospective estimation when an ancillary system has interacted with a system of interest in the past, or for estimating the initial state (e.g., when $t=0$).

In the continuous-time weak measurement setting considered in this note, the evolution of the quantum state $\rho_t$ is governed by a nonlinear stochastic master equation, typically expressed as a stochastic differential equation (SDE) \cite{Qsde1,Qsde2,gardiner2004quantum,barchielli2009quantum,QF3}). The stochastic component of this equation reflects the randomness of the measurement outomces. \emph{A posteriori} --- that is, given a specific measurement record $y$ over the time interval $[0,t]$ --- this equation can be integrated explicitly to yield a conditional state $\rho_t(y)$. In contrast, when considered \emph{a priori}, the equation describes a probability distribution over possible future trajectories of the system. Taking this distribution into account entails significant computational complexity, particularly in tasks such as numerical simulation, system and control design, or system identification. This stands in contrast to the (Gorini–Kossakowski–Sudarshan-) Lindblad master equation \cite{Ldb1,Ldb2,Ldb3}, which is deterministic and linear, and describes the ensemble-averaged evolution of the system. The Lindblad equation accurately predicts the statistics of future measurement outcomes, only in the \emph{absence of conditioning} on previous measurements. However, whenever feedback control is applied based on measurement outcomes \cite{QFB0,QFB1,QFB2,QFB3,cardona2020exponential,amini2024exponential}, or when quantities such as purity or multi-tume correlations --- nonlinear functions of $\rho_t$ --- are of interest \cite{NC,tilloy2018exact}, a more detailed characterization of the full distribution over quantum trajectories becomes essential.

In principle, a system following an SDE can diffuse in all directions of its state space, regardless of the number of underlying stochastic processes. It can therefore reach states arbitrarily far from the ensemble-averaged dynamics, and its observables can exhibit extensive (co-)variance over the repetitions of the experiment. To make trajectory prediction tractable in this context, efficient model reduction techniques are necessary. For example, Refs.~\cite{ZFfilter,Gfilter} aim to significantly reduce the dimension of the quantum filter by expressing the state at each time as a superposition over a small, carefully selected set of basis states. The selection of relevant basis states relies on a good \emph{a priori} estimate of the likely region of state space the system will explore. Such model reduction strategies may also play a key role in the efficient estimation of ``likely trajectories'' \cite{JordanMLT0} and their practical applications in high-dimensional quantum systems.\\

This note highlights that, in several quantum SDE models of particular interest in quantum engineering, such \emph{a priori} model reduction can in fact be performed \emph{exactly}. More precisely, we show that for these systems, the evolution of the state remains confined to a low-dimensional, nonlinear manifold that evolves deterministically in time, independently of the measurement record. Only the motion \emph{along} the manifold depends on the specific measurement outcomes. Our attention was first drawn to this property by the experimental observations reported in \cite{PhCManifolds}, where the distribution of conditioned qubit states was seen to remain constrained to a ``parachute-like'' surface converging deterministically to the ground state. A complete theoretical characterization of this behavior in the qubit case is given in \cite{SRqubit}. The objective of the present work is to extend this analysis to higher-dimensional systems. 

\begin{figure*}
1D manifold \hspace{33mm} $|$ \hspace{12mm} 2D manifold \hspace{35mm} $|$ \hspace{10mm} diffusion\\
\includegraphics[trim=0.3cm 0cm 0cm 0cm, clip=true,width=2.7cm]{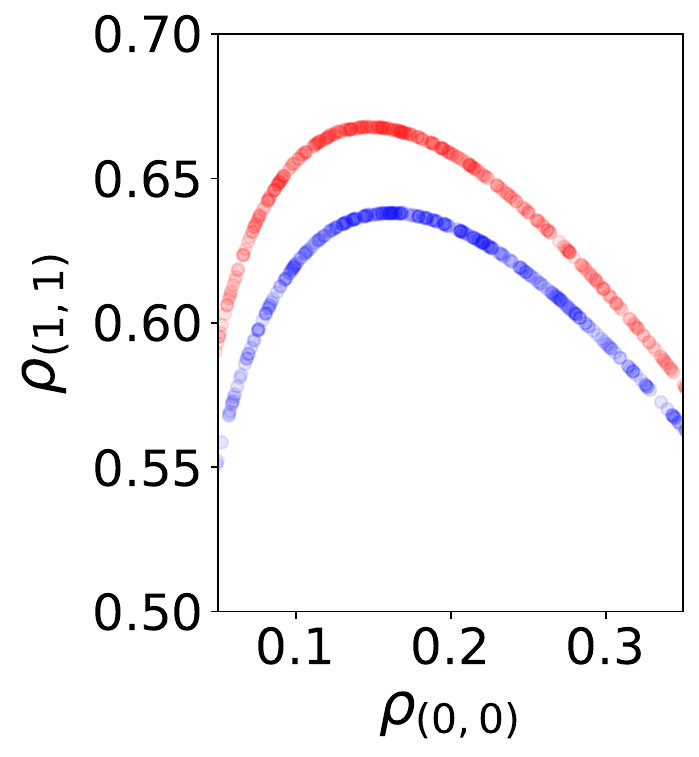}~~\includegraphics[trim=2.5cm 1cm 3.2cm 2cm, clip=true,width=3.6cm]{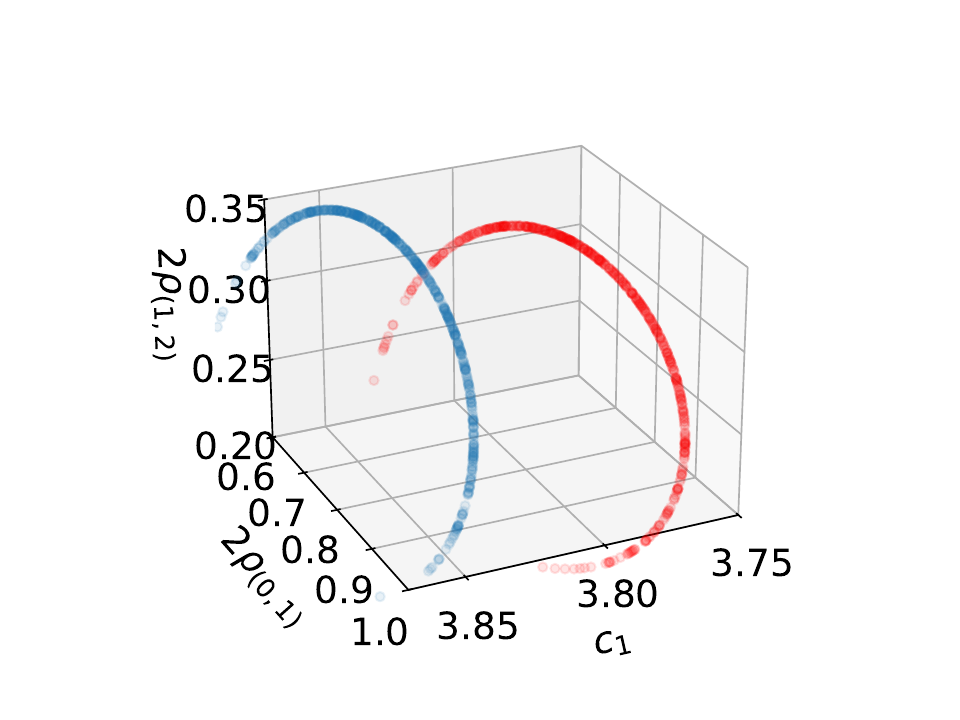}~~~~
\includegraphics[trim=0.3cm 0cm 0cm 0cm, clip=true,width=2.7cm]{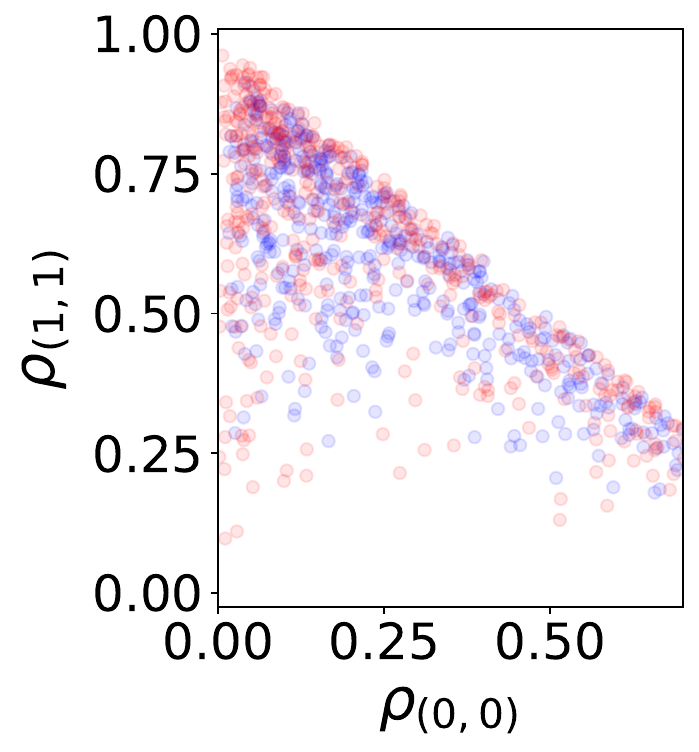}~
\includegraphics[trim=2.5cm 1cm 3.5cm 2cm, clip=true,width=3.6cm]{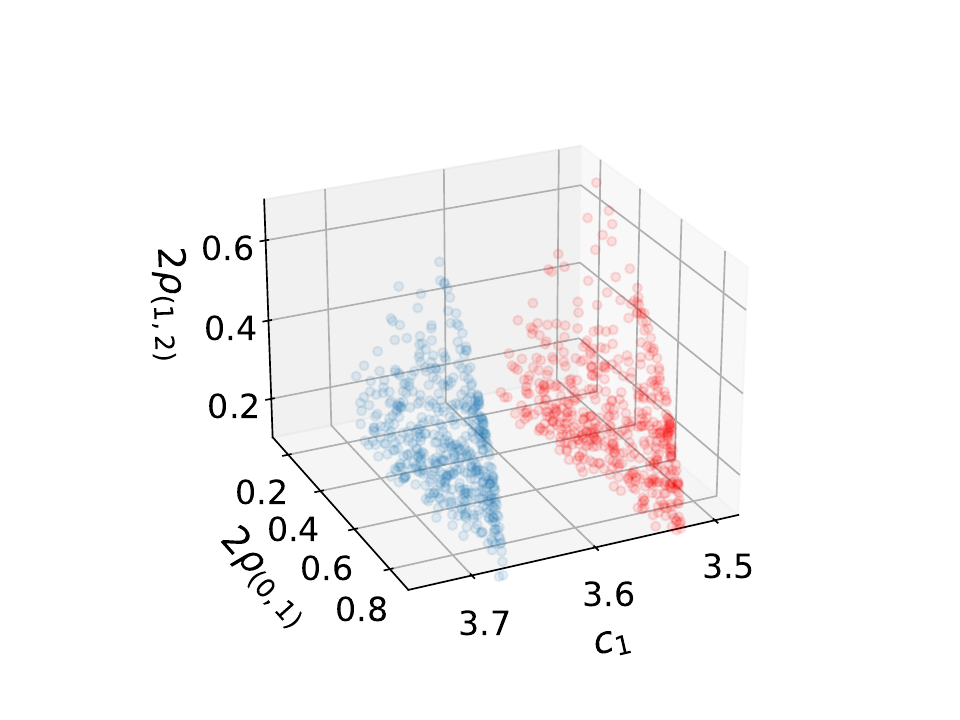}~~~~\includegraphics[trim=1.5cm 0.5cm 3.2cm 2cm, clip=true,width=3.6cm]{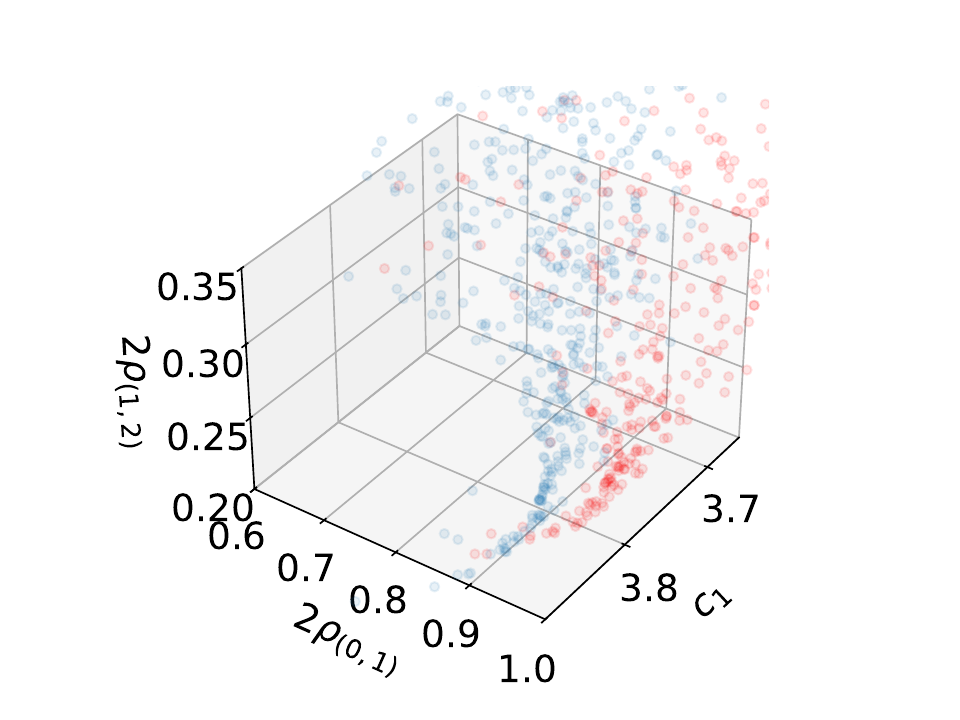}  \caption{\emph{Simulation results for a qutrit undergoing QND population measurement.} Each panel shows 500 sample states at two time points: $t=0.2$ (blue) and $t=0.3$ (red), all starting from the same initial state $\q{\psi_0} = \sqrt{.3} \q{0} + \sqrt{.55}\q{1}+\sqrt{.15}\q{2}$. (Time is in units of the QND measurement rate.) Panels 1 and 3 display the populations in levels $\q{0}$ and $\q{1}$. Panels 2, 4 and 5 show the coherences $\qd{0} \rho \q{1}+\qd{1}\rho\q{0}$, $\qd{1} \rho \q{2}+\qd{2}\rho\q{1}$, and $c_1 := \frac{(\qd{0} \rho \q{1}+\qd{1}\rho\q{0})^2}{\qd{0}\rho\q{0}\qd{1}\rho\q{1}}{}$. The two leftmost panels correspond to a single QND measurement operator with no other dynamics. The samples remain confined to a time-dependent curve. The two next panels correspond to two commuting QND measurement operators. Here, the populations can evolve independently, but the ensemble still remains confined to a time-dependent, two-dimensional surface. The rightmost panel corresponds to a single QND measurement operator, combined with a Rabi Hamiltonian coupling levels $\q{0}$ and $\q{1}$. The samples now appear to diffuse across all three represented dimensions. See the main text for further details.}\label{fig:ill:1}
\end{figure*}

\begin{figure*}
2D manifold \hspace{25mm} $|$ \hspace{20mm} diffusion \hspace{60mm} $\phantom{.}$\\[3mm]
\includegraphics[trim=1.5cm 0.5cm 1cm 2cm, clip=true,width=5.6cm]{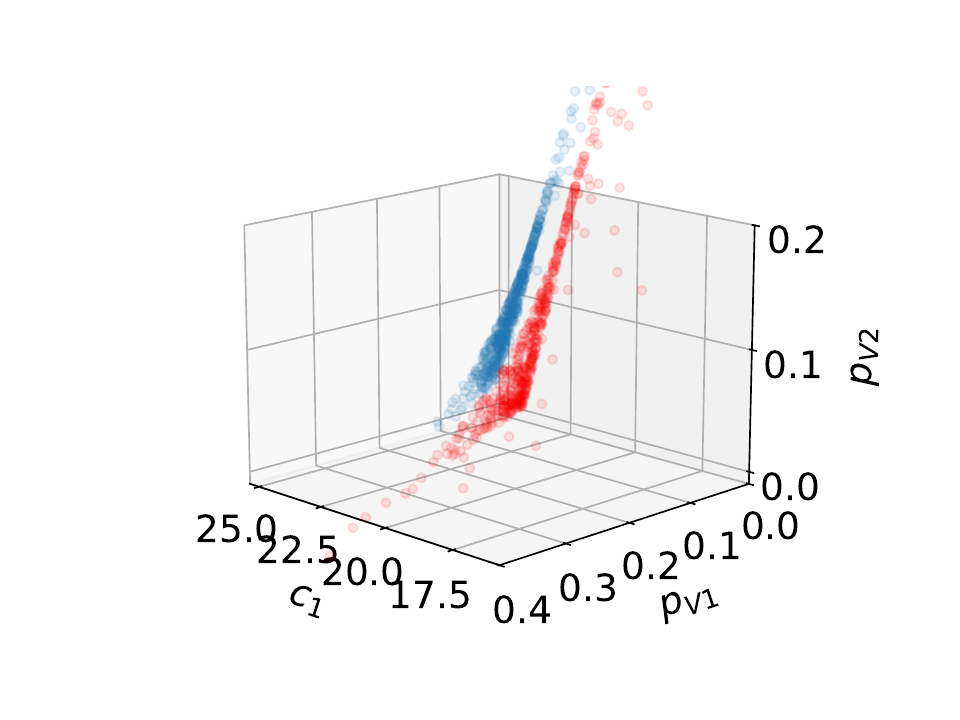}~~~\includegraphics[trim=1.5cm 0.5cm 1cm 2cm, clip=true,width=5.6cm]{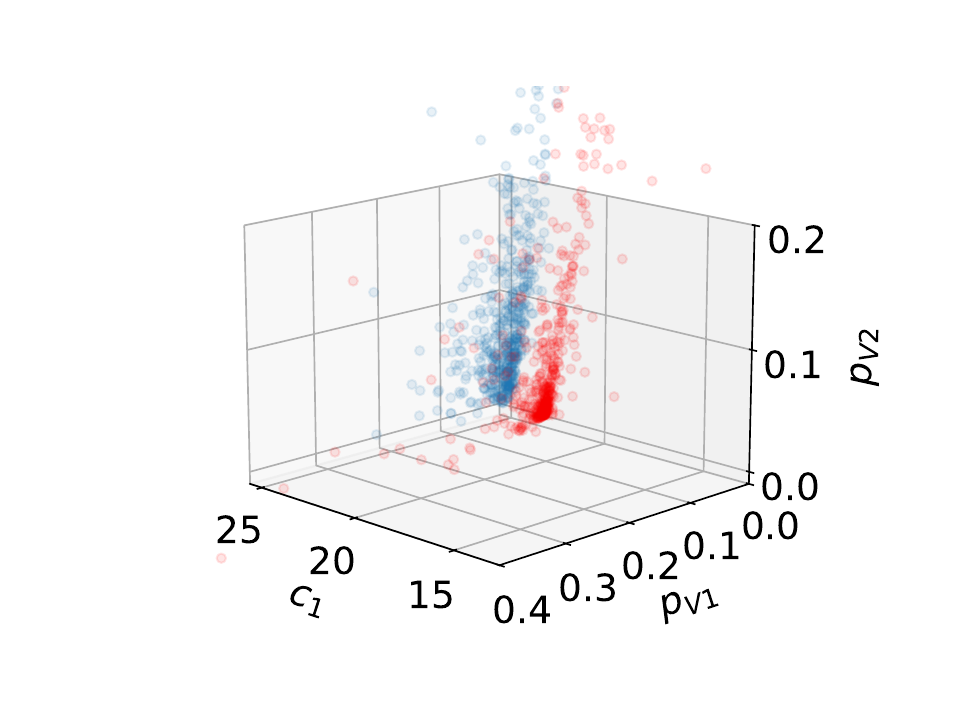}~~~\includegraphics[trim=1.5cm 0.5cm 1cm 2cm, clip=true,width=5.6cm]{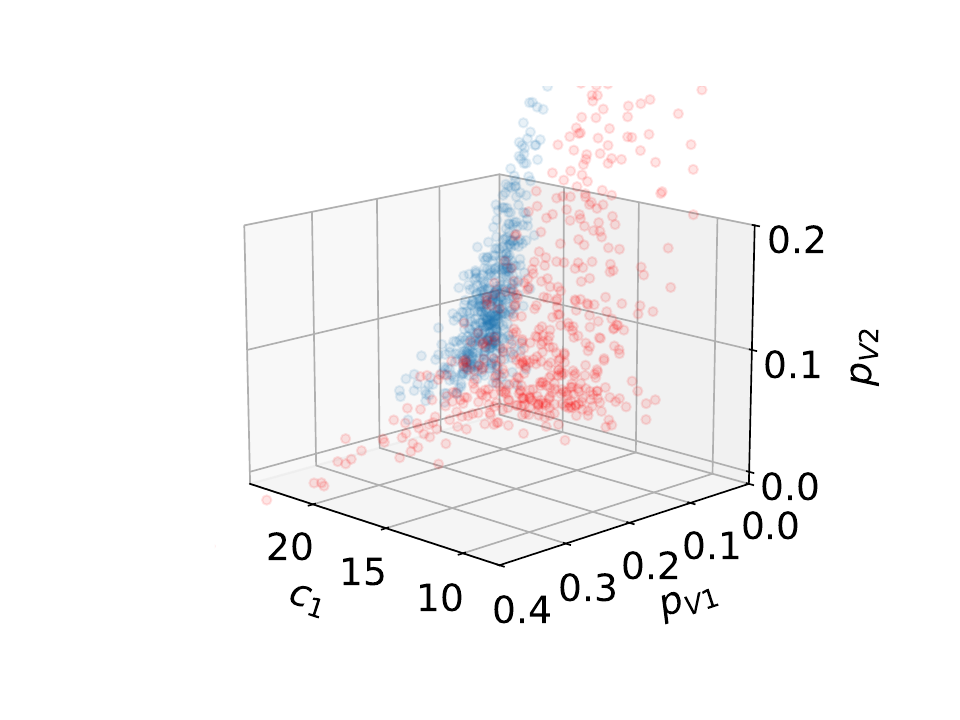}     \caption{\emph{Simulation results for a 3-qubit repetition code.} Each panel shows 500 sample states at two time points: $t=.025$ (blue) and $t=.1$ (red), all starting from the same initial state $\q{\psi_0}$. (Time is in units of the QND measurement rate.) This initial state satisfies $|\langle 111 | \psi_0\rangle|^2 = 0.3$ and $|\langle 000 | \psi_0\rangle|^2 = 0.5$ and is otherwise arbitrary. Each sample is plotted using the following three coordinates: $p_{V1}=$ population in the subspace where qubit 1 is flipped with respect to the two others; $p_{V2}=$ population where qubit 2 is flipped with respect to the two others; $c_1 =$ a linear combination of the population in the subspace with all qubits equal in the canonical basis, and
its relative coherence with the $V_1$ subspace; this particular combination is just chosen to aid 3D visualization. Panel 1 considers two QND syndrome measurements (efficiency $\eta=0.8$), comparing qubits $(1,3)$ and $(2,3)$ respectively. The state at each time remains confined to a two-dimensional surface. Panel 2 adds a third QND syndrome measurement, comparing qubits $(1,2)$. This appears to break the surface confinement. Panel 3 returns to just two syndrome measurements but includes a Lindbladian term modeling spurious bit-flips at rate $\gamma=0.3$. Although this added dynamics is deterministic --- unlike for Panel 2 --- it increases the spread of trajectories, which now diffuse independently in all three displayed directions, similarly to Panel 2.}\label{fig:ill:2}
\end{figure*}

\begin{figure*}
1D manifold \hspace{25mm} $|$ \hspace{15mm} 2D manifold \hspace{25mm} $|$ \hspace{20mm} diffusion \\[3mm]
\includegraphics[trim=2.8cm 0.7cm 1cm 2cm, clip=true,width=6cm]{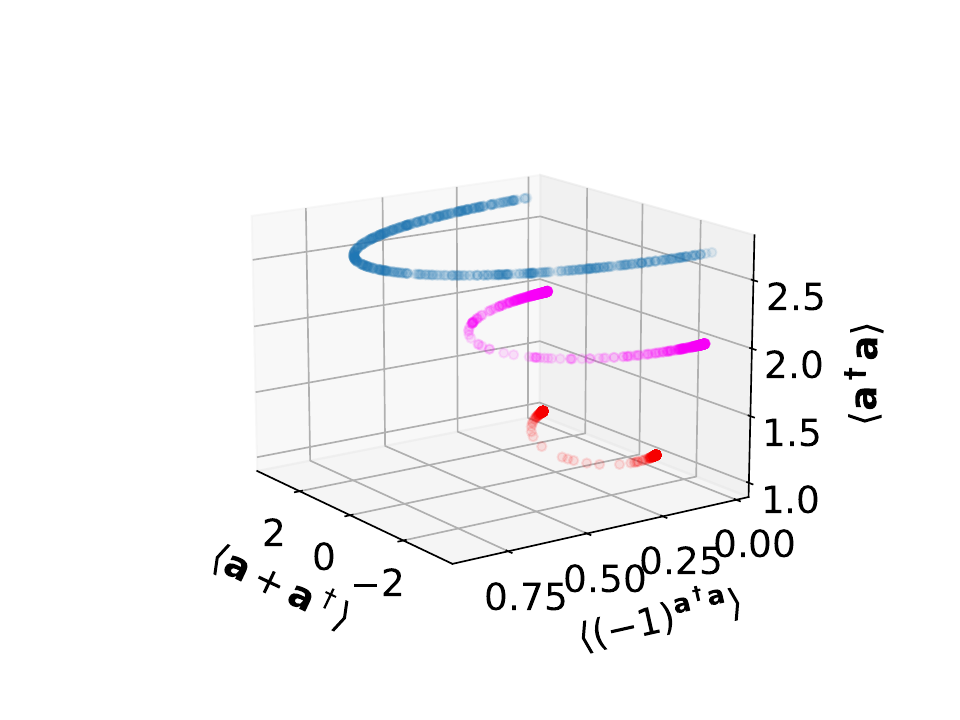}~~~\includegraphics[trim=2.2cm 0.3cm 1.6cm 2cm, clip=true,width=6cm]{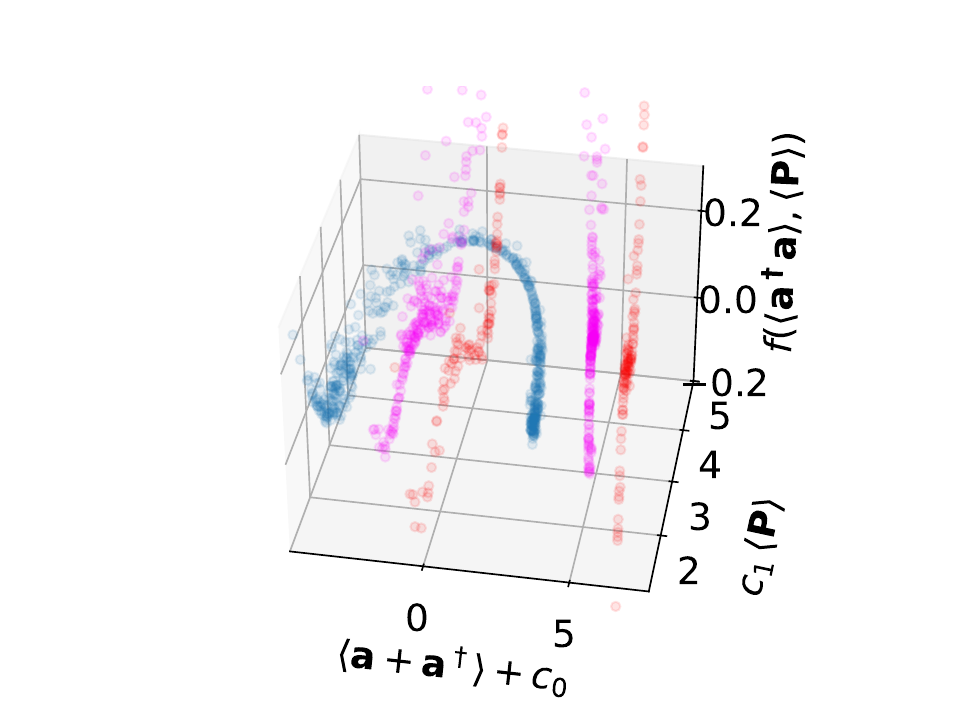}~~~\includegraphics[trim=1.8cm 0.7cm 1.8cm 2cm, clip=true,width=6cm]{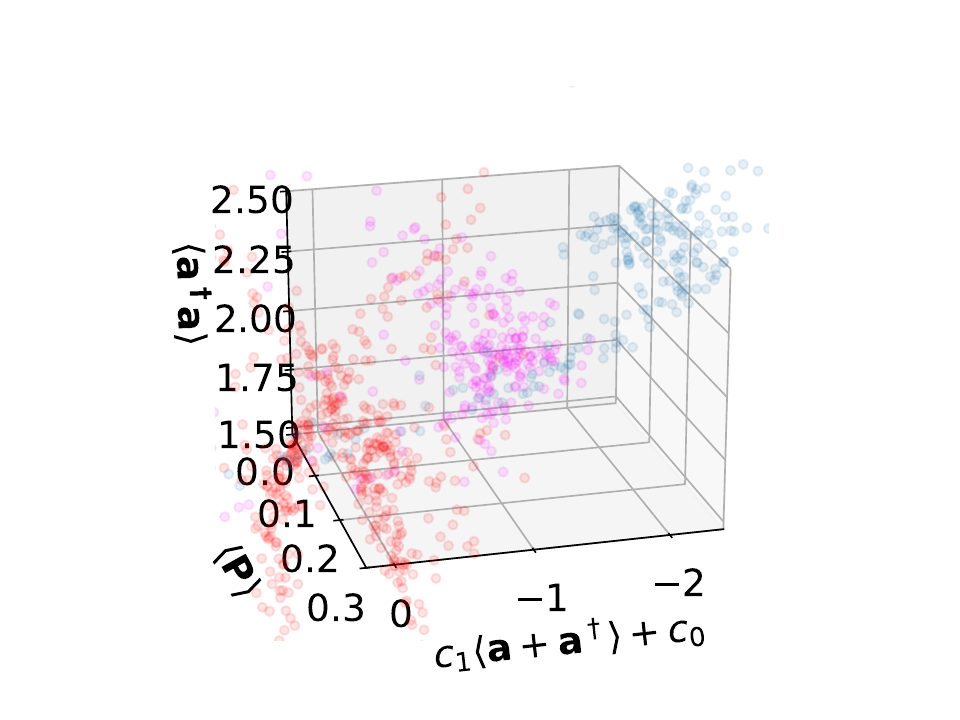}   \caption{\emph{Simulation results for homodyne fluorescence measurement on a cavity} (see Section \ref{sec:aandia} and Appendix \ref{ssec:app:aandia} for details). Each panel shows 500 sample states at three time points, increasing from blue to purple to red. All trajectories start in the cat state $\q{\psi_0} \propto \tfrac{\q{\alpha}+\q{-\alpha}}{\sqrt{2}}$. Panel 1 considers homodyne detection of fluorescence, monitoring the annihilation operator $L={\bf a}$ with measurement efficiency $\eta=0.8$, and a coherent drive $H \propto ({\bf a} + {\bf a}^\dagger)$. States are plotted using the position operator ${\bf a} + {\bf a}^\dagger$, the parity operator ${\bf \sf P} = (-1)^{{\bf a}^\dagger {\bf a}}$, and the number operator ${\bf a}^\dagger {\bf a}$. At each fixed time, the data lies on a curve. As time progresses, the samples show decay in both energy and position, as well as localization along the position axis, reflecting the influence of the measurement on the observed quadrature. Panel 2 adds thermal relaxation ($n_{th}=2.3$). The sample states now lie on time-dependent surfaces. (Note that the scatterplot is left-right symmetric. On the left side of this view, the points appear to extend in at least 2 dimensions, while on the right, their alignment on a surface becomes apparent. To aid 3D visualization, we show time-dependent combinations and shifts of the base coordinates from Panel 1.) Panel 3 retains ${\bf a}$ as the only dissipation and measurement channel, but adds a Kerr Hamiltonian $H \propto ({\bf a} {\bf a}^\dagger)^2$. The samples now appear to spread across all three coordinate directions, indicating higher-dimensional diffusion.}\label{fig:ill:3}
\end{figure*}

Figures \ref{fig:ill:1} through \ref{fig:ill:3} illustrate the simulated distribution of conditioned state components for three representative cases, respectively: a qutrit undergoing quantum non-demolition (QND) population measurement, a three-qubit repetition code subject to continuous syndrome measurement, and a harmonic oscillator under continuous fluorescence measurement. Depending on the specifics of each model, the resulting distributions are observed to remain confined to a one-dimensional curve, a surface, or to diffuse across all represented directions of the state space. Which case arises depends on the interplay between the measurement process and the system's deterministic dynamics. We will explain how a well-known algebraic criterion from control theory \cite{AstromMurray} can be applied to efficiently determine whether such manifolds exist and to compute their dimension.

Moreover, systems that exhibit such confinement are likely to admit relatively simple, explicit expressions for the quantum state as a function of time and of the measurement record. We derive such explicit solutions for a set of physically relevant models in quantum engineering, namely: 
\begin{itemize}[noitemsep,topsep=0pt]
\item quantum non-demolition (QND) measurements of commuting observables;
\item certain so-called linear quantum systems on the Hilbert space of a harmonic oscillator;
\item multi-partite systems, particularly bipartite settings in which a measured subsystem is coupled via Hamiltonian interaction to an unmeasured subsystem.
\end{itemize}
From a system-theoretic perspective, this corresponds to identifying a minimal representation --- significantly lower in dimension than the full density operator $\rho(t)$ --- that nonetheless exactly reproduces the output signal statistics for arbitrary initial conditions. Conversely, in models where the quantum state diffuses substantially across the state space as a function of the measurement history, it is unlikely that such compact and explicit representations of the state evolution can be obtained. In this context, the algebraic criterion serves as a valuable tool for identifying cases that are analytically more tractable.

The remainder of this note is organized as follows. Section \ref{sec:background} relates the present work to prior literature. Section \ref{sec:theory} introduces the mathematical framework, including the quantum filter SDE and algebraic criterion for investigating confining manifolds. The qutrit case illustrated on Figure \ref{fig:ill:1} is used as a running example. Section \ref{sec:applications} summarizes the results obtained by applying this framework to the three classes of systems described above. Detailed derivations and explicit formulas for each case are provided in Appendices A, B, and C, respectively.

\subsection{Related Work and Concepts}\label{sec:background} 

Following the observations in \cite{PhCManifolds,SRqubit}, the existence of confining manifolds has also been demonstrated in the \emph{backward} (smoothing) equation for a qubit's past quantum state filter \cite{MMmanifolds}.

A prominent setting where low-dimensional dynamics naturally arise is that of \emph{linear quantum systems} \cite{LQS,GaussPreserved}, where Gaussian initial states remain Gaussian under measurement-conditioned evolution. In this case, the state is fully characterized by its mean and covariance matrix, which evolve within a low-dimensional space. In Section \ref{sec:HO}, we extend these compact descriptions to arbitrary initial states.

Recent work on model reduction for open quantum systems includes algebraic methods \cite{Grigoletto2024,grigoletto2024exact} and spectral techniques \cite{essig2021multiplexed}, which simplify the model with invariant linear subspaces associated to specific initial conditions and observables of interest. In contrast, the present work aims to provide reduced models valid for \emph{any} initial condition and \emph{all} observables.

A more radical form of \emph{a priori} model reduction considers only the \emph{most likely trajectory} of a quantum SDE \cite{JordanMLT0}. While this yields a single pure-state evolution, it ignores the actual measurement record --- like Lindblad evolution --- and thus generally fails to capture the full dynamics. Nevertheless, it can be useful in contexts like simplified feedback design. In such context, a confining manifold --- when available --- can serve either as a more accurate but still low-cost alternative to the most likely trajectory, or as a restricted domain for more efficient search of this trajectory.

Our framework can be viewed as a nonlinear generalization of decoherence-free subspaces (DFS) \cite{LidarChuang1998}. Specifically, the confining manifolds induce a foliation of the state space. For a manifold of dimension $M$ and any fixed initial state, the combined deterministic drift and measurement-dependent diffusion drive the state evolution along at most $M+1$ directions. The remaining directions correspond to invariant quantities. However, in contrast to DFS, these invariants are often nonlinear functions of the state $\rho$. Investigating whether they can be exploited for quantum computation or error correction is left for future work.

A key tool in our analysis is the algebraic \emph{reachability criterion} from control theory \cite{ControlOnLiegGroups,AstromMurray}. Beyond technical details discussed in Section \ref{ssec:criterion}, our use of this criterion differs from standard control applications in two main ways. First, we seek a compact state description valid \emph{everywhere} in state space, so we conservatively bound where the state can go ``at most'', retaining possibly oversized models in regions where the criterion becomes singular (e.g.,~at pure states). In contrast, control design must ensure trajectory \emph{feasibility}, thus bounding where the system can go ``at least'', often requiring special treatment of singularities. Second, we care about the directions of diffusion induced by measurement backaction. Whether this is accompanied by deterministic, irreversible drift is irrelevant from the perspective of model reduction, while it poses significant challenges in control theory.

One might expect that confinement improves state estimation by already reducing uncertainty \emph{a priori}. In fact, the opposite may hold: Lindblad evolution features no diffusion but also no information gain, while increasing the number of measurement channels typically induces greater information extraction \emph{a posteriori}, but also stronger diffusion, reflecting stronger measurement backaction --- see, e.g., the syndrome measurements on Fig.~\ref{fig:ill:2}). When measurement-induced diffusion is confined to a manifold, the resulting state estimate becomes dependent on the initial guess. In this light, confinement may rather offer advantages for \emph{parameter estimation} or \emph{calibration}, where reducing the accessible state space can help isolate dynamical parameters. That being said, it remains an open question whether quantum systems should be explicitly engineered to exhibit such confinement. In this work, we take the system model as given and ask whether it admits an exact low-dimensional description of its quantum trajectories. From this standpoint, in an experimental context, characterizing confining manifolds can help determine which integrals of the measurement signals need to be accurately tracked to retain maximal information about the quantum state.


\section{Capturing deterministic manifolds: general method}\label{sec:theory}

Weak continuous-time measurement processes can be physically pictured as monitoring an electromagnetic wave after its interaction with the quantum system of interest. If this wave is probed using photon counters, the resulting dynamics are governed by stochastic equations with discrete jumps (i.e.,~ Poisson processes) \cite{PoissMeas1,PoissMeas2}. This work, however, focuses on the continuous monitoring of wave amplitudes, as formalized in \cite{wiseman1993quantum,Qsde1,Qsde2,gardiner2004quantum,barchielli2009quantum,QF3}, and realized experimentally in \cite{murch2013observing,hatridge2013quantum,PhCManifolds}, among others. Each measurement channel is associated with an operator $L_k$ acting on the system's Hilbert space $\mathcal{H}$. The evolution of the density operator $\rho_t$, conditioned on the measurement record up to time $t$, follows a stochastic differential equation (SDE) in \Ito~sense  \cite{QF2,QF3}:
\begin{eqnarray}
\label{eq:SDE} d\rho_t &=& -i\, [H, \rho_t]\, dt \\
\nonumber && + \sum_{k=1}^{\bar k} \left(L_k \rho_t L_k^\dag - \tfrac{1}{2} (L_k^\dag L_k \rho_t + \tfrac{1}{2} \rho_t L_k^\dag L_k) \right) dt \\
\nonumber		 && \!\!\!\!\!\!\!\!\!\!\!\!+ \sum_{k=1}^{\bar k} \left(L_k \rho_t + \rho_t L_k^\dag -\tr(L_k \rho_t + \rho_t L_k^\dag)\rho_t \right) \sqrt{\eta_k} \, dw_{t,k} \\
\nonumber		 &=:& -i\, [H, \rho_t]\, dt + \sum_{k=1}^{\bar k} F_{L_k}(\rho_t)\, dt \\
\nonumber && +  \sum_{k=1}^{\bar k} G_{L_k}(\rho_t) \, \sqrt{\eta_k} \, dw_{t,k} \;\; ; \\[3mm]
\nonumber dy_{t,k}	&=& \sqrt{\eta_k}\, \tr(L_k \rho_t + \rho_t L_k^\dag)\, dt + dw_{t,k} \; ,\\ \nonumber && \quad \text{ for } k=1,2,...,{\bar k}. 
\end{eqnarray}
Here, $H$ is the system Hamiltonian, $\eta_k \in [0,1]$ is the measurement efficiency of channel $k$, and $\bar{k}$ is the total number of monitored channels. Measurement signal $dy_{t,k}$ describes the weak continuous-time equivalent of probabilistic detection results. The terms $dw_{t,k}$ denote independent Wiener increments, which are Gaussian random processes each of mean $0$ and variance $dt$, capturing the intrinsically probabilistic character of quantum measurements. The evolution $d\rho_t$ represents both measurement backaction, as a continuous analog of wavefunction collapse, and unitary dynamics when $H \neq 0$. Note that the measurement operators $L_k$ need not be Hermitian nor mutually commuting. In this continuous-time limit, their effects just add up in the evolution $d\rho_t$. Unmonitored channels, i.e.~with $\eta_k=0$, correspond to decoherence.

Different stochastic measurement outcomes $dy_{t,k}$ are thus in one-to-one correspondence with different realizations of the processes $dw_{t,k}$ in \eqref{eq:SDE}.
The analog of renormalization at wavefunction collapse makes \eqref{eq:SDE} nonlinear in $\rho_t$. One could alternatively consider unnormalized states to obtain a linear equation, but this offers only modest simplifications and is not pursued here.

In the qubit setting ($\mathcal{H} \simeq \mathbb{C}^2$), Ref.~\cite{SRqubit} showed that under certain conditions that are relevant in practice \cite{PhCManifolds}, the solution $\rho_t$ of Eq.~\eqref{eq:SDE} does not explore the full Bloch sphere but rigorously remains confined to a surface or curve, regardless of the measurement record. Such confinement significantly facilitates trajectory computation and enables closed-form expressions for both the conditional state and its statistical distribution over time. The present note extends this analysis to higher-dimensional quantum systems.
\vspace{2mm}

\noindent \textbf{Running example (1):} Figure \ref{fig:ill:1} illustrates simulations of several qutrit measurement scenarios, according to Eq.~\eqref{eq:SDE}, all with measurement operators diagonal in the orthonormal basis $\{ \q{0},\q{1},\q{2} \}$. The first two panels correspond to $H=0$ and a single measurement operator $L_1 = \lambda_0 \q{0}\qd{0}+\lambda_1 \q{1}\qd{1}+\lambda_2 \q{2}\qd{2}$, with $\lambda_0=0$, $\lambda_1=1$, $\lambda_2=1.8$ and measusrement efficiency $\eta_1 = 0.8$ (arbitrary values). On the next two panels, we have added a second measurement operator $L_2=\lambda'_0\q{0}\qd{0}+\lambda'_1\q{1}\qd{1}) + \lambda'_2 \q{2}\qd{2}$, where $\lambda'_0=0$, $\lambda'_1=1$, $\lambda'_2=0.2$ and $\eta_2 = 0.8$. For the last panel, we have taken the single measurement channel $L_1$, now associated with coherent evolution governed by $H=\Omega\, (\q{0}\qd{1}+\q{1}\qd{0})$, with arbitrary Rabi frequency $\Omega=1.35$.

This example involves $L_1$ and $L_2$ Hermitian and commuting, enabling visualization of low-dimensional behavior. However, neither Hermiticity nor commutativity of the $L_k$ is required in the general theory. For Hermitian $L_k$, the measurement signal $dy_{t,k}$ is centered around $2\sqrt{\eta_k}$ times the expectation value of the observable $\langle L_k \rangle$. In the absence of additional dynamics, $dy_{t,k}$ gradually converges towards an eigenvalue of $L_k$, randomly selected according to the Born rule applied to the initial state $\rho_0$; simultaneously, the state $\rho_t$ converges towards the corresponding eigenprojector.  \hfill  \textbf{R.e.(1)}~$\square$


\subsection{Algebraic criterion}\label{ssec:criterion}

The existence of confining manifolds is governed by the interplay between measurement-induced stochastic backaction and the system's deterministic dynamics. In general, even a single Wiener process can result in full-dimensional diffusion if the stochastic and deterministic contributions do not commute. A classical example is a vertical wheel rolling on a plane: when its orientation changes stochastically (one angular variable), applying a deterministic rolling velocity causes the position to spread randomly in the plane. After any given time $t>0$, the system can be spread arbitrarily over all degrees of freedom (position and orientation) despite being driven by a single noise source. Such reachability properties (and more refined versions) are well- characterized in control theory, when replacing the noise process by a control input \cite{AstromMurray}. In contrast, model reduction seeks scenarios where some degrees of freedom determine the others, indicating the state is confined to a lower-dimensional manifold.

Faced with equation \eqref{eq:SDE}, it is therefore natural to ask where the different realizations of the noise processes $dw_{t,k}$ can drive the system, interpreting them as control signals and applying methods from control theory. As recalled in the appendix of \cite{SRqubit}, this viewpoint is correct provided the SDE is rewritten in Stratonovich form. Indeed, unlike in \Ito~calculus, where coordinate transformations introduce second-order correction terms, Stratonovich SDE's behave like ordinary differential equations under coordinate changes. The transformation from \Ito~to Stratonovich involves a deterministic correction term $d_k(x)$ for each Wiener process:
\Ito~
$$dx_t = f(x_t) \, dt + \sum_{k=1}^{\bar{k}} \, g_k(x)\, dw_{t,k}$$
transforms into Stratonovich 
$$dx_t = \left(f(x_t) + \sum_{k=1}^{\bar{k}} d_k(x_t)\right) \, dt + \sum_{k=1}^{\bar{k}} \, g_k(x)\, \circ\, dw_{t,k} \; ,$$ where 
$$d_k(x) = - \frac{1}{2} \sum_{\ell=1}^N \, \frac{\partial g_k}{\partial (x)_\ell} (g_k)_\ell(x) \; ,$$ and $(v)_\ell$ denotes the $\ell$the component of vector $v$. The symbol $\circ$ just indicates that the SDE is interpreted in the Stratonovich sense. 
For \eqref{eq:SDE}, translating this to the quantum setting gives the correction terms $D_{L_k}(\rho)$ corresponding to the vector fields $G_{L_k}$:
\begin{align*}
D_{L_k}(\rho) = -\frac{\eta_k}{2}\Big(& L_k G_{L_k}(\rho) + G_{L_k}(\rho) L_k^\dagger \\
& - \tr\big(L_k G_{L_k}(\rho) + G_{L_k}(\rho) L_k^\dagger \big) \; \rho \\
& - \tr\big(L_k \rho + \rho L_k^\dagger\big)\; G_{L_k}(\rho) \Big) \; .
\end{align*}
A general result from \cite{SVtheorem} then yields the following criterion:

\begin{prop}\label{prop:1}
Consider the SDE \eqref{eq:SDE} with a given initial state $\rho_0$. At any time $t$, the distribution of $\rho_t$ --- over all measurement realizations --- is supported on a manifold $\mathcal{M}(t)$. The dimension $M$ of this manifold is given by the dimension of the smallest real Lie algebra $\mathfrak{G}_F$ generated by the vector fields $\{ G_{L_k} \}_{k=1}^{\bar k}$ and closed under repeated Lie brackets with the vector field $-i\, [H, \cdot ] + \sum_{k=1}^{\bar k} \; F_{L_k} + D_{L_k}$.
\end{prop}
We recall that the Lie bracket, which generates the Lie algebra, is given by
$$[f_1,f_2] = \sum_\ell \, \tfrac{\partial f_1}{\partial x_\ell} (f_2)_\ell\; - \; \tfrac{\partial f_2}{\partial x_\ell} (f_1)_\ell \; $$
for two arbitrary vector fields $f_1(x)$ and $f_2(x)$. Some of its relevant properties in the present context are listed below.
The deterministic part of the dynamics thus contributes to the dimension of $\mathcal{M}(t)$ indirectly, through its non-commutativity with the stochastic vector fields. Proposition \ref{prop:1} enables rapid identification of scenarios in which the dynamics remain confined to a lower-dimensional manifold, which can guide the search for simplified explicit expressions of $\rho_t$. It plays a central role in organizing the examples discussed  in Section \ref{sec:applications}.
\vspace{2mm}

\noindent \textbf{Running example (2):} For the first qutrit scenario in Figure \ref{fig:ill:1}, the dimension $M$ of the confinement manifold is generated by the vector field $G_{L_1}$ and its repeated commutators with $F_{L_1}+ D_{L_1}$. Note that these are both nonlinear vector fields, so their Lie bracket involves more than matrix commutators. We do not explicitly compute these terms here; instead, the next section outlines general properties that help simplify such computations. As observed in Fig.~\ref{fig:ill:1}, the two vector fields appear to commute, implying that even at long times, diffusion remains restricted to a one-dimensional curve tangent to $G_{L_1}$.

In contrast, on the last panel of Fig.~\ref{fig:ill:1}, adding a Hamiltonian $H\neq 0$ introduces diffusion in several directions. As in the classical rolling wheel example, this diffusion arises from iterated commutators between the single stochastic vector field $G_{L_1}$ and the deterministic term $-i\, [H, \cdot ]$.

In the middle panels of Fig.~\ref{fig:ill:1}, the two commuting measurement operators $L_1$ and $L_2$ generate two distinct vector fields $G_{L_k}$. Indeed, e.g.~at the maximally mixed state  $\rho=(\q{0}\qd{0}+\q{1}\qd{1}+\q{2]\qd{2}})/3$, one finds:
\begin{eqnarray*}
G_{L_1}(\rho) &=& \tfrac{-5.6}{9} \q{0}\qd{0} + \tfrac{0.4}{9} \q{1}\qd{1} + \tfrac{5.2}{9} \q{2}\qd{2} \; , \\
G_{L_2}(\rho) &=& \tfrac{-0.8}{3} \q{0}\qd{0} +\tfrac{1.2}{3} \q{1}\qd{1} + \tfrac{-0.4}{3} \q{2}\qd{2} \; ,
\end{eqnarray*}
which are clearly not collinear. However, commutators with those vector fields appear to generate no additional directions, since we observe that the dynamics remains confined to a two-dimensional surface. \hfill \textbf{R.e.(2)}~$\square$


\subsection{Computing low-dimensional deterministic manifolds in practice}\label{ssec:practice}

A first step in analyzing the algebra $\mathfrak{G}_F$ is computing commutators between the relevant vector fields. The identities below help systematize this computation. (Note the dual use of brackets: commutators of vector fields on the left, and operator commutators on the right.)
\begin{eqnarray}
\nonumber G_{L_1+L_2} &=& G_{L_1}+G_{L_2}  \; ;\\
\nonumber G_{\alpha I} &\equiv& 0 \text{ for $I=$ identity, } \alpha \in \mathbb{C}  \; ;\\
\nonumber G_{-iH}(\cdot) & = &  -i[H,\cdot]  \; ;\\
 \label{eq:Gcommute}
[G_{L_j},\,G_{L_k}] & = & G_{[L_j,L_k]} \; ;\\
\label{eq:Gcommute.a} [F_{L_j}+ D_{L_j},\,G_{L_k}](\rho) & = & \\ \nonumber
&& \hspace{-3cm} (1-\eta_j) \, \left([L_j,L_k]\rho L_j^\dagger  - \tr([L_j,L_k]\rho L_j^\dagger)\rho \quad + h.c. \right) \\
\nonumber  & & \hspace{-3cm}+ (1-\eta_j)\, G_{L'}(\rho) + \eta_j \,  G_{L''}(\rho) \\
\nonumber & & \hspace{-3cm}  + \eta_j \big(c_1(\rho)\, G_{[L_j,L_k]}(\rho) + c_2(\rho) G_{L_j}(\rho)\big) \; .
\end{eqnarray}
Here, $L' = \tfrac{1}{2}\, [L_k,L_j^\dagger L_j]$ and $ L'' = \tfrac{1}{2}\, \left[L_k\,,\, (L_j^\dagger+L_j)\, L_j \right]$, while $h.c.$ denotes the Hermitian conjugate. The coefficients $c_1(\rho)$ and $c_2(\rho)$ are real-valued functions of $\rho$ whose specific values are not needed, since the associated vector fields are already present in $\mathfrak{G}_F$. For $k=j$, the last identify specializes to:
\begin{equation}\label{eq:jisk}
[F_{L_j}+D_{L_j},\,G_{L_j}](\rho) = G_{L'}(\rho) + \eta_j c_2(\rho)\, G_{L_j}(\rho) \; .
\end{equation}
For further commutators involving such terms, it is useful to rewrite \eqref{eq:Gcommute.a} in a form where the nonlinear part is explicitly cast in the usual $G_L$ form:
\begin{eqnarray}
\label{eq:strangeterm2} 
&& [L_j,L_k]\rho L_j^\dagger  - \tr([L_j,L_k]\rho L_j^\dagger)\rho \quad + h.c. \\
\nonumber && =\left( [L_j,L_k]\rho L_j^\dagger - L_j^\dagger[L_j,L_k] \rho \quad + h.c. \right) + G_{L_j^\dagger [L_j,L_k]}(\rho) \; .
\end{eqnarray}
This structure often recurs under further commutators, though in the examples considered here, the expressions typically simplify before reaching that level.

A useful structural property follows from repeated application of the Jacobi identity: 
\begin{prop}\label{prop:Jacobi}
    Let $M_1$,$M_2$ and $M_3$ be vector fields. If $[M_1,M_3]=[M_2,M_3]\equiv 0$, then $M_3$ commutes with all elements of the Lie algebra generated by $M_1$ and $M_2$.
\end{prop}
This is particularly relevant when vector fields act on distinct subsystems of a composite quantum system.\\

A second important aspect is determining the dimension $M$ of $\mathfrak{G}_F$. Several considerations are worth noting:
\begin{itemize}
\item The dimension of $\mathfrak{G}_F$ is defined as the maximal dimension of $\mathfrak{G}_F(\rho)$ over all states $\rho$. Singular points --- where fewer directions are available --- can occur, but unlike in control theory, such singularities are not physically meaningful in this context.
\item The map $L_k \mapsto G_{L_k}$ is not injective. For instance, $\{ G_{i \sigma_x}, G_{i \sigma_y}, G_{i \sigma_z} \}$ are linearly dependent at any given $\rho$. Such dependencies can be detected numerically, by evaluating the vector fields at a random $\rho$, or identified analytically while searching for the manifold expressions. 
\item In some cases, the dimension of $\mathfrak{G}_F$ reduces for finely tuned relative amplitudes of the measurement operators $L_j$ in the deterministic drift $\sum_j  [F_{L_j}+D_{L_j}]$. Unless otherwise stated, we assume in our case studies that these rates are generic and not specially adjusted (See Section \ref{sec:Lewalle} for an exception.)\\
\end{itemize} 

Finally, to explicitly construct the confining manifolds, one must in principle integrate the vector field algebra. At each point $\rho$, the algebra $\mathfrak{G}_F$ defines the tangent space to a confining manifold. The full space of density operators on $\mathcal{H} \simeq \mathbb{C}^N$ is thus partitioned into a foliation of such manifolds. The state transits \emph{deterministically} from one manifold to another over time, while moving stochastically along the manifold. Two complementary approaches can be employed to characterize this reduced dynamics. 
\begin{enumerate}
\item \emph{Implicit representation via deterministic invariants.} The manifold is defined by a vector of $(N^2-1)-M$ variables --- nonlinear functions of $\rho$ ---  which evolve \emph{deterministically}:
\begin{equation}\label{eq:implicitsolution}
p_t = h(\rho_t)\quad \text{with}  \quad dp_t = f(p_t,t)\, dt \; .
\end{equation}
This is analogous to identifying conserved quantities in classical systems --- such as energy or angular momentum --- except that here the quantities evolve deterministically manner and are invariant only with respect to stochastic fluctuations. When performing coordinate transformations on \eqref{eq:SDE}, \Ito's Lemma must be applied. For a state $x$ with components $x_j$ satisfying $dx_j = f_j(x) dt + \sum_k g_{j,k}(x)\, dw_k$, and for transformed coordinates $x'_j = h_j(x)$, the \Ito~formula yields:
\begin{eqnarray}\label{eq:Itorule}
dx'_j &=& \sum_l\, \tfrac{\partial h_j}{\partial x_l} f_l dt + \sum_{l,k}\, \tfrac{\partial h_j}{\partial x_l} g_{l,k}\, dw_k\\
\nonumber && + \tfrac{1}{2} \sum_{l,m,k}\, \tfrac{\partial^2 h_j}{\partial x_l \partial x_m} g_{l,k}\, g_{m,k}\, dt \, .
\end{eqnarray}
The last sum is the \Ito~correction term. We thus seek functions $h(x)$ for which the stochastic terms vanish and the deterministic terms form a closed dynamical system. Note that by combining several such deterministically evolving variables, one may obtain quantities that are \emph{conserved}. effectively reducing the deterministic drift to a single variable.

\item \emph{Explicit parameterization.} Alternatively, the manifold can be described by an explicit solution of the form
\begin{equation}\label{eq:explicitsolution}
\rho_t =  f(\rho_0, r_t, \gamma_t) \; ,
\end{equation}
where $r_t$ is a vector of deterministic parameters that obey ordinary differential equations (ODEs) and are denoted by Latin letters in the case studies below; whereas $\gamma_t$ is an $M$-dimensional vector of irreducible stochastic parameters governed by the measurement backaction. Only the latter follow SDEs; they are represented by Greek letters in the case studies. This formulation is particularly useful for filtering and inference tasks. Such expressions are typically known only in special settings ----- e.g., linear quantum systems~\cite{LQS}, where Gaussianity is preserved. The algebraic analysis based on $\mathfrak{G}_F$ helps identify new scenarios where reduced forms like \eqref{eq:explicitsolution} may exist and thus be worth pursuing, and conversely, indicate when the state depends in a nontrivial way on the full measurement record.
\end{enumerate}
Whatever the procedure, intuitive insight is often necessary for identifying the most interpretable formulas. Symmetries of the system --- reflected as missing directions in the algebra $\mathfrak{G}_F$ --- can serve as a useful starting point for this task. Several of the case studies presented in Section \ref{sec:applications} illustrate such instances.

\vspace{2mm}

\noindent \textbf{Running example (3):} Let us compute the Lie brackets for the scenarios shown in Figure \ref{fig:ill:1}, using the identities \eqref{eq:Gcommute},\eqref{eq:Gcommute.a}, and \eqref{eq:jisk}.
\begin{itemize}
\item For the first two panels, we apply \eqref{eq:jisk} with $L_j=L_1$ Hermitian. Since $L_1^\dagger L_1$ commutes with $L_1$, the term $G_{L'}$ vanishes. The Lie bracket \eqref{eq:jisk} is then collinear with $G_{L_1}$ at any $\rho$, yielding no new directions. This is consistent with the manifold dimension $M=1$.
\item For the next two panels, \eqref{eq:Gcommute} shows that the two independent vector fields $G_{L_1}$ and $G_{L_2}$ commute. Their brackets with the deterministic drift involve \eqref{eq:jisk} as before, but also \eqref{eq:Gcommute.a}, where we have $[L_j,L_k]=L'=L''=0$. Hence, no additional directions arise, and $M=2$.
\item For the scenario of the last panel, there remains to compute repeated commutators with $G_{-iH}$ where $H \propto \q{0}\qd{1}+\q{1}\qd{0}$. Using \eqref{eq:Gcommute}, this amounts to computing operator commutators:
\begin{eqnarray*}
[L_1,iH] &\propto& i\q{0}\qd{1}-i\q{1}\qd{0} \\~
[L_1,[L_1,iH ] ] &\propto & i\q{0}\qd{1}+i\q{1}\qd{0} \\~
[H,[L_1,H ] ] &\propto & \q{0}\qd{0}-\q{1}\qd{1} \; .
\end{eqnarray*}
Evaluated at a generic $\rho$, these yield four linearly independent directions, while further commutators yield no others. This confirms $M=4$ (out of $3^2-1=8$ degrees of freedom in $\rho$).
\end{itemize}

Let us now describe the confining manifold for the first scenario. \vspace{2mm}
\newline - \emph{Explicit form:} Since $M=1$, the state $\rho_t$ should be fully determined by a single scalar driven by the measurement record. In this QND readout case, it is the integrated signal $y_{t,1} = \int_0^t dy_{t,1}$. One finds that the diagonal entries of $\rho_t$ evolve as:
$$\qd{b} \rho_t \q{b} = \qd{b} \rho_0 \q{b} \, e^{2 \lambda_b \sqrt{\eta_1} y_{t,1} - 2(\lambda_b)^2 \eta_1 t} \, \mu_t \; ,$$
where $\mu_t$ is a normalization factor independent of $b\in\{0,1,2\}$. For a given $y_{t,1}$, the expression in the exponential is maximized as a function of $\lambda_b \in \mathbb{R}$ when $y_{t,1} = 2 \sqrt{\eta_1} \lambda_b \, t$, which matches the mean signal if the system were in the eigenstate $\rho_t=\q{b}\qd{b}$. This confirms that measurement backaction favors the eigenstate of $L_1$ with eigenvalue closest to the observed signal.\vspace{2mm}
\newline - \emph{Implicit form}
\newline \emph{Populations:} The diagonal entries $\qd{b}\rho_t \q{b}$ are constrained by $3-M=2$ relations. The first is the normalization condition  $\sum_b \qd{b}\rho_t \q{b}=1$. The second is the quantity 
$$z_t := \frac{\qd{2} \rho_t \q{2}^{\lambda_1-\lambda_0}\; \qd{0} \rho_t \q{0}^{\lambda_2-\lambda_1}}{\qd{1} \rho_t \q{1}^{\lambda_2-\lambda_0}} \; ,$$
which evolves deterministically as
$$z_t = z_0 \, e^{-2 \eta_1 (\lambda_2-\lambda_1) (\lambda_2-\lambda_0) (\lambda_1-\lambda_0) \, t} \; .$$ (See Appendix \ref{app:ssec:QNDgeneral} for this derivation, which is a special case of Prop.~\ref{prop:3}.) This exponential decay constrains how the distinct populations evolve while only one population will remain nonzero at $t\rightarrow +\infty$.
\newline \emph{Coherences:} The coherences between populations are governed by the following deterministic quantities:
\begin{eqnarray*}
    \frac{\qd{a}\rho_t\q{b}^2}{\qd{a}\rho_t\q{a}\qd{b}\rho_t\q{b}} = \frac{\qd{a}\rho_0\q{b}^2}{\qd{a}\rho_0\q{a}\qd{b}\rho_0\q{b}} \; e^{-(1-\eta_1)(\lambda_a-\lambda_b)^2\,t} \; ,
\end{eqnarray*}
(see again Appendix \ref{app:ssec:QNDgeneral}.) For $\eta_1 < 1$, this implies exponential decay of all off-diagonal elements, with rates set by the eigenvalue differences. For $\eta_1=1$, the state remains pure at all times, so off-diagonal elements only decay insofar as the corresponding populations decay proportionally. 

From these variables, the following conserved quantities can be identified:
\begin{itemize}
\item The complex phase of $\qd{a}\rho_t\q{b}$ remains constant.
\item For any pairs $(a,b)$ and $(a',b')$, the ratio
$$
{\left(\frac{\qd{a}\rho_t\q{b}^2}{\qd{a}\rho_t\q{a}\qd{b}\rho_t\q{b}}\right)^{\frac{1}{(\lambda_a-\lambda_b)^2}}}\Big/{\left(\frac{\qd{a'}\rho_t\q{b'}^2}{\qd{a'}\rho_t\q{a'}\qd{b'}\rho_t\q{b'}}\right)^{\frac{1}{(\lambda_{a'}-\lambda_{b'})^2}}}
$$
is conserved.
\item The ratio
$${\left(\frac{\qd{a}\rho_t\q{b}^2}{\qd{a}\rho_t\q{a}\qd{b}\rho_t\q{b}}\right)^{\frac{1}{(1-\eta_1)(\lambda_a-\lambda_b)^2}}}\Big/{(z_t)^{\frac{1}{2 \eta_1 (\lambda_2-\lambda_1) (\lambda_2-\lambda_0) (\lambda_1-\lambda_0)}}}$$ is also conserved.
\end{itemize}
\hfill \textbf{R.e.(3)}~$\square$


\subsection{Clarifications on the technical framework}\label{ssec:Remarks}

\noindent 1. \emph{Special cases $\eta_j=0$ and $\eta_j=1$.} The cases $\eta_j=0$ and $\eta_j=1$ represent limiting regimes of measurement efficiency. When $\eta_j=0$, the output increment $dy_{t,j}$ carries no information, and the associated contribution to the SDE for $d\rho_t$ reduces to purely \emph{deterministic} decoherence. Conversely, when $\eta_j=1$ for all $j$, the measurement extracts maximal information, and any initially pure state remains pure over time. However, this does not imply that all the confining manifolds remain stationary. Counterexamples appear in the applications below and, for example, in \cite{SRqubit}, when measuring $L_1=\sigma_-$ on a qubit: here, the manifolds corresponding to mixed states exhibit an irreversible drift, despite $\eta=1$. From a technical perspective, when $\eta_j=1$, the non-standard term in \eqref{eq:Gcommute.a} vanishes, simplifying the computation of $\mathfrak{G}_F$. The distinctions related to these limiting cases are not systematically treated in the case studies below.\\

\noindent 2. \emph{On convex decomposition intepretations.} It may be tempting to interpret the dynamics at intermediate efficiencies $\eta_j \in (0,1)$ as a weighted interpolation between the ideal measurement case ($\eta_j=1$, where the state is ``confined'' to remain pure) and the no-measurement case ($\eta_j=0$, corresponding to deterministic evolution), e.g.,
\begin{equation}\label{eq:MazInterp}
\rho_t = (1-c_t) \, \rho_{\eta=1,t} + c_t \, \rho_{\eta=0,t} \; ,
\end{equation}
where only the pure state $\rho_{\eta=1,t}$ depends on the measurement record. However, such an interpretation generally fails, even in canonical examples. For instance: 
\begin{itemize}
\item The convex combination of a distribution of pure trajectories on one side and a deterministically evolving point on the other side, spans a manifold of dimension $N-1$ when $\mathcal{H} \equiv \mathbb{C}^N$. This is inconsistent with the manifold dimensions observed in most case studies. Already in the qubit study of \cite{SRqubit}, a single generic $L_1$ --- namely any operator not trivially reducible to $\alpha \sigma_-$ or $\alpha \sigma_z$ for $\alpha \in \mathbb{C}$ --- induces diffusion in all directions of the Bloch sphere.
\item Even when dimensions match, the decomposition \eqref{eq:MazInterp} leads to contradictions. Consider, for example, a qubit undergoing QND $\sigma_z$ measurement. The eigenstates $\rho= \q{g}\qd{g}$ and $\rho= \q{e}\qd{e}$ belong to all confining manifolds $\mathcal{M}_t$. Suppose that for a particular realization of the measurement process, the SDE yields $\rho_t \simeq \q{g}\qd{g}$ at some tim $t$. Then consistency would require either (a) both $\rho_{\eta=1,t}$ and $\rho_{\eta=0,t}$ to be close to $\simeq\q{g}\qd{g}$ --- which contradicts the possibility that another measurement realization leads to a trajectory near $\q{e}\qd{e}$, since $\rho_{\eta=0,t}$ should remain unchanged --- or (b) $c_t \simeq 0$, rendering the decomposition vacuous.
\end{itemize}
Other inconsistencies can be identified using similar reasoning. While a more refined interpretation along these lines is not excluded, it would require substantially more care than the naive form of \eqref{eq:MazInterp}.\\

\noindent 3. \emph{On normalized vs.~unnormalized dynamics.} In constructing the Lie algebra $\mathfrak{G}_F$, the nonlinear term in \eqref{eq:SDE} is typically manageable, as it appears as a scalar multiple of $\rho$. More substantial complications often arise from the accumulation of operators acting both on the left and right of $\rho$, a feature that persists even in the linear SDE governing the unnormalized state. While the unnormalized picture can simplify some algebraic manipulations, the benefits thus remain limited. Proposition \ref{prop:1} remains applicable to the linear (unnormalized) case, but it may count an additional direction corresponding to changes in the trace of $\rho$. When projecting back to normalized states, this direction becomes irrelevant, and should be excluded when determining the dimension of the confining manifold.


\section{Case studies}\label{sec:applications}

This section presentes the results of applying the methodology developed in Section \ref{sec:theory} to several scenarios of practical relevance. In some cases, we also highlight associated physical interpretations. Section \ref{sec:QND} focuses on QND measurements; 
Section \ref{sec:HO} addresses the continuous measurement of harmonic oscillator quadratures; and Section \ref{sec:bipar} explores bipartite quantum systems. General formulas for each setting, along with detailed derivations, are provided in the Appendix. An outline at the beginning of the Appendix indicates where each scenario is treated.

\subsection{Quantum non-demolition measurement}\label{sec:QND}

QND measurement \cite{Haroche,Braginsky} is the continuous-time analogue of projective measurement, aiming to extract information about the eigenstate populations of a Hermitian observable $Q=\sum_{n=1}^N q_n \q{d_n}\qd{d_n}$ in the orthonormal basis $\{\q{d_n}\}_{n=1}^N$, without disturbing any eigenstate $\q{d_n}$. A QND measurement is characterized by \eqref{eq:SDE} with (possibly several) $L_k$ that commute with $Q$. Continuous-time measurement of a single QND channel is now routinely performed in various experimental setups \cite{holland1991quantum,sayrin2011real,essig2021multiplexed}. The associated dynamics express progressive wavefunction collapse onto a specific eigenstate, as a function of the associated output signals. We now characterize the dynamics of this process from the viewpoint of confining manifolds. Both Fig.\ref{fig:ill:1} and Fig.\ref{fig:ill:2} represent scenarios of this type.
\begin{enumerate}
\item \emph{Algebraic criterion.} So-called homodyne QND measurement considers all the $L_k$ to be Hermitian, $L_k=\sum_{n=1}^N \lambda_k(n) \q{d_n}\qd{d_n}$, with eigenvalues $\lambda_k(n) \in \mathbb{R}$. We admit degenerate $L_k$, as often appears in practice. In optical field quadrature measurements, see e.g.~\cite{PhCManifolds}, it is natural to also encounter \emph{heterodyne} measurement, i.e.~adding channels $L_{k'} = i L_k$.
\begin{prop}
In both homodyne and heterodyne QND measurement, the vector fields in \eqref{eq:SDE} all commute. As a result, the system is confined to a manifold of dimension at most $\bar{k}$, the number of measurement channels.
\end{prop}
This property is a direct consequence of the fact that $[L_k,L_j] = 0$ and $[L_k,L_j^\dagger] = 0$ for all $j,k$.

\item \emph{Manifolds implicit form.} Let $\rho_t(a,b) = \qd{d_a} \rho_t \q{d_b}$ denote matrix components in the eigenbasis of the $L_k$. The confining manifold can be described by the following deterministically evolving variables.
\begin{prop}\label{prop:3a}
For homodyne measurement of $\bar{k}$ commuting Hermitian $L_k$:
\newline
$\bullet$ The \emph{complex phases} of all $\rho_t(a,b)$ are constant.
\newline $\bullet$ For $a\neq b$, the \emph{magnitudes} $|\rho_t(a,b)|$ decay deterministically, relative to the corresponding populations, at exponential rate $\sum_{k=1}^{\bar{k}} \; (1-\eta_k) (\lambda_k(a)-\lambda_k(b))^2$.
\newline 
$\bullet$ For \emph{populations}, $N-1-\bar{k}$ independent linear combinations of the $\log(\rho_t(a,a))$ evolve deterministically.
\newline
\indent For heterodyne detection, with $L_{\bar{k}/2+k}= i L_k$ in addition to commuting Hermitian operators $L_k$ for $k=1,2,...\bar{k}/2$, the phases are not constant, but correlated via $N(N-1)/2-\bar{k}/2$ deterministic quantities.
\end{prop}
The corresponding formulas and their derivation can be found in Appendix \ref{app:ssec:QNDgeneral}.

Let us briefly comment on Proposition \ref{prop:3a}.

- The decay rate of coherences $|\rho_t(a,b)|$ reflects information loss: for $\eta_k = 1$, the state remains pure; for $\eta_k < 1$, the coherences decay faster than the corresponding populations.

- The deterministically evolving combinations of populations $\rho_t(b,b)$ are entirely new with respect to the qubit case \cite{SRqubit}, and maybe surprising at first sight. For example, with $\bar{k} = 1$, knowing the trajectory $\rho_t(1,1)$ suffices (together with $\rho_0$) to reconstruct the full state trajectory.

- Interdependence of the phases of the $\rho_t(a,b)$ is an expected feature. Indeed, starting with an arbitrary positive hermitian matrix $\rho$, if the off-diagonals could pick up independent complex phases, then the state would not necessarily remain positive.

When some levels $a,b$ are degenerate in all $L_k$, the confinement formulas in Appendix \ref{app:ssec:QNDgeneral} properly capture that there is no dynamics within the corresponding eigenspace. In particular, the quantity $\log(\rho_t(a,a))-\log(\rho_t(b,b)) = \log\left(\frac{\rho_t(a,a)}{\rho_t(b,b)}\right)$ remains invariant over time.

Singularities arising from $\rho_{t}(a,a)=0$ can be removed by discarding the corresponding variables from the outset.

\item \emph{Explicit form.} The correlation of distinct $\rho_t(b,b)$ can be understood from the following expression.
\begin{prop}\label{prop:5}
The populations in the setting of Proposition \ref{prop:3a} evolve as:
\begin{equation}\label{eq:VinceRes}
\rho_t(b,b) = \rho_0(b,b) \cdot e^{2 \sum_{k=1}^{\bar k} \;( \lambda_k(b) \sqrt{\eta_k}\, {y^k_t} - \lambda_k(b)^2\eta_k\, t) } \cdot \mu(t) \; ,
\end{equation}
with a normalization factor $\mu(t)$ independent of $b$. Hence, the state at time $t$ is fully determined by the initial state and the integrated value of each measurement signal $y^k_t$.
\end{prop}
This reflects a core QND property: the ordering of measurement outcomes doesn't affect the final state --- only their accumulated values matter. The normalization factor $\mu(t)$ depends on the measurement record and on $\rho_0$; leaving it unspecified is a standard feature of Bayesian filtering, e.g.~in robotics too.
\\
\end{enumerate}

Concrete examples are presented in Appendix Sections \ref{app:ssec:QNDexampleN}--\ref{app:ssec:QNDexampleR}. Below, we highlight a few notable observations.\\

\paragraph*{Ex. Quantum number measurement:} With the single operator
$L_1 = \mathbf{N} = \sum_{n=0}^{n_{\max}} \, n\, \q{n}\qd{n} \, ,$ the combination of populations
$$z_t = \left(\frac{\rho_t(n_4,n_4)}{\rho_t(n_1,n_1)}\right)^{(n_3-n_2)} \left(\frac{\rho_t(n_2,n_2)}{\rho_t(n_3,n_3)}\right)^{(n_4-n_1)}$$
remains constant for $(n_2-n_1) = (n_4-n_3)$.\\
 
\paragraph*{Ex. Continuous-variable measurement:} With $L_1 = \mathbf{X} = \int_{\mathbb{R}} x\, \q{x}\qd{x} \, dx$ the position operator, coherences decay at a rate $(1-\eta)\, (x_1-x_2)^2$, naturally indicating that positions further apart are easier to distinguish. The population distribution $p(x) = \rho(x,x)$ satisfies, in particular:
\begin{eqnarray*}
\frac{d^2}{dx^2}[\log p_t(x)] &=& \frac{d^2}{dx^2}[\log p_0(x)] - 2 \eta t  \\
\frac{d^k}{dx^k}[\log  p_t(x)] &=& \frac{d^k}{dx^k}[\log p_0(x)] \text{ invariant for } k>2 \, ,
\end{eqnarray*}
implying preservation of shape features in the distribution. The explicit solution takes the form:
\begin{eqnarray}
p_t(x) &=& p_0(x) \cdot e^{- (x-\gamma_{t})^2/ (2 \sigma_t^2)} \, \cdot \mu_t \; ,
\end{eqnarray}
with deterministic width $\sigma_t = \frac{1}{\sqrt{2 \eta t}}$ and stochastic center $\gamma_{t} = \frac{y_t}{\sqrt{\eta} t}$. We show in Appendix \ref{app:ssec:QNDexampleC} how a heterodyne measurement, with $L_1 = \mathbf{X}$ and $L_2 = i \, \mathbf{X}$, additionally induces a particularly correlated evolution of the off-diagonal phases. In Section \ref{sec:HO}, we consider this example with additional dynamics, under the framework of linear quantum systems.\\

\paragraph*{Ex. 3-qubit repetition code:}  Quantum Error Correction (QEC; see e.g.~\cite[chapter 10]{NC},\cite{StabCode}) motivates the measurement of several commuting Hermitian $L_k$, each with eigenvalues $\pm 1$ detecting an error syndrome. Depending on the setting, \eqref{eq:SDE} may be an appropriate way to model the finite rate at which such information is acquired.

The simplest error-correcting code is the 3-qubit repetition code. This is the setting illustrated on Fig.\ref{fig:ill:2}. The ideal code subspace is $V_0 = \text{span}(\q{000},\q{111})$ and flipping one bit maps this to subspaces $V_1,V_2,V_3$ respectively. Each measurement operator $L_k$ compares two qubits along the canonical basis and is thus 4-fold degenerate. The manifolds framework retrieves the following points.
\begin{itemize}
\item The relative values of $\rho_t(a,a)$, $\rho_t(b,b)$ and $\rho_t(a,b)$ are conserved quantities when $\q{a},\q{b}$ belong to the same subspace $V_k$. This reflects how quantum information encoded inside a subspace is not affected by syndrome measurements.
\item With $L_1,L_2,L_3$ each comparing a different pair of qubits, we have $M=3$. Each $2\times2$ matrix block corresponding to a subspace $(\q{a},\q{b}) \subset V_j$, follows the explicit form $\; \rho_t(a,b) = \rho_0(a,b) \gamma_{t,j}  \;$, with the $\gamma_{t,j}$ evolving independently (up to overall normalization). The other matrix blocks all decay deterministically, at a rate $8(1-\eta)$ when $\eta_1=\eta_2=\eta_3$. This backaction indicates how the syndrome measurements help decide to which subspace $V_j$ the state belongs.
\item Only two $L_k$ are needed to distinguish the $V_j$. Omitting e.g.~the comparison of qubits 1 and 3, corresponding to $L_2$, gives a manifold dimension $M=2$ and thus introduces an extra deterministic quantity, such as:  
$$z_t = \frac{\rho_t(000,000) \rho_t(010,010)}{\rho_t(001,001)\rho_t(100,100)} = z_0 \; .$$
This conservation law indicates how, when e.g.~the state converges towards $V_0$, the population on $V_2$ must converge to zero quicker than the populations on $V_1$ and $V_3$. This reflects that confusing $V_0$ with $V_2$ would correspond to confusing the measurements of \emph{both} $L_1$ and $L_3$, whereas confusing e.g.~$V_0$ and $V_1$ implies confusion on a single measurement operator. The coherences among subspaces similarly follow different decay rates in absence of $L_2$.
\item When spurious bit flips are added to the model, the deterministic vector field interacts with the measurement-induced ones, and we have analytically constructed at least 15 independent direnctions in $\mathfrak{G}_F$. This highlights the difficulty of exactly capturing memory effects in this realistic setting and underscores the need for \emph{approximate} model reduction in this case. Appendix~\ref{app:ssec:QNDexampleR} further illustrates that a low-dimensional manifold persists in the presence of bit-flip on a single qubit. This is a direct consequence of Proposition~\ref{prop:Jacobi}, which states that local operations increase the manifold's dimension only locally. More generally, this suggests that highly non-local correlations typically correspond to higher-order Lie brackets, whose contributions remain negligible over moderate timescales. Neglecting the directions associated with these terms can thus yield efficient approximate filters, even though an exact filter would need to span the entire state space.
\end{itemize}


\subsection{Measuring harmonic oscillator quadratures}\label{sec:HO}

The quantum harmonic oscillator is a foundational model in quantum physics \cite{Haroche}, with an infinite-dimensional Hilbert space and equally spaced energy levels. Its dynamics are governed by the annihilation and creation operators ${\bf a}$ and ${\bf a}^\dag$, satisfying $[{\bf a},{\bf a}^\dagger] = {\bf I}$. The position and momentum operators, ${\bf X} = \tfrac{{\bf a}+{\bf a}^\dagger}{2}$ and ${\bf P} =  \tfrac{{\bf a}-{\bf a}^\dagger}{2i}$, serve as typical observables for interaction, measurement, and control.

Several experimentally relevant continuous-time measurement settings on this system feature confining deterministic manifolds $\mathcal{M}_t$, and some of those have no projective counterpart. These are characterized below in terms of the Wigner function $\mathcal{W}^{\{\rho\}}(x,p)$, a quasi-probability distribution in phase space (see Appendix \ref{app2}). The Gaussian kernels generalize known preservation properties of Gaussian states under quantum filtering \cite{GaussPreserved}. In the Heisenberg picture, these examples fall within the framework of linear quantum systems \cite{LQS}. 


\subsubsection{Monitoring the annihilation channel}\label{sec:aandia}

This setting, also known as fluorescence measurement \cite{QuadMeasRef1,QuadMeasRef2,QuadMeasRef3}, corresponds to monitoring $L_1={\bf a}$, and possibly $L_2=i{\bf a}$ in a heterodyne detection. The resulting outputs
\begin{eqnarray}\label{eq:dyforac}
dy_{t,1}  &=& 2\sqrt{\eta_1} \text{Trace}({\bf X} \rho_t)\, dt + dw_{t,1} \quad , \\ \nonumber
dy_{t,2} &=&  -2\sqrt{\eta_2} \text{Trace}({\bf P} \rho_t)\, dt + dw_{t,2} \; 
\end{eqnarray}
reveal information about position and momentum. Unlike measurements of $L_1={\bf X}$ or ${\bf P}$ however, this configurations causes the system to dissipate energy, asymptotically approaching the ground state $\q{0}$. This is directly analogous to $L_1=\sigma_-$ and $L_2 = i \sigma_-$ on a qubit, as explored experimentally in \cite{PhCManifolds}, which motivated this work. The following results are derived in detail in Appendix \ref{ssec:app:aandia}.

\begin{enumerate}
\item \emph{Algebraic criterion.} According to Proposition \ref{prop:1}, measuring $L_1$ and $L_2$ with a Hamiltonian $H_0 = \Delta {\bf a}^\dagger {\bf a}$ yields confining manifolds of dimension $M=2$. Adding a control Hamiltonian involving operators ${\bf X}$ and ${\bf P}$ does not increase this dimension. If the system also couples to a thermal bath with nonzero temperature ($n_{th} > 0$), modeled by an additional channel $L_3={\bf a}^\dag$ with $\eta_3=0$, the manifolds dimension increases to $M=4$. Under homodyne detection (i.e.~omitting~$L_2$) and with $\Delta=0$, all dimensions are halved.
\newline
\emph{Thus, for $n_{th} >0$, attempting to capture the state with a single integral per output signal is bound to fail. Yet, tracking two integrals per channel recovers a full and compact description of $\rho_t$.}
\newline
Figure \ref{fig:ill:3} illustrates the case $M=2$ under homodyne detection with $\Delta=0$ and $n_{th}> 0$.

\item \emph{Manifolds explicit form.} In general, for all those cases, the state evolution admits the following representation:
\begin{prop}\label{prop:HOmeas}
The quantum state evolution for fluorescence measurement writes, in Wigner representation:
\begin{multline}    
\mathcal{W}^{\{\rho\}}_t(x,p) = \label{eq:GenGaussKer} \\ \nonumber \int \int \mathcal{W}^{\{\rho\}}_0(x_0,p_0)\, K_{t}(x,x_0,p,p_0)\, dx_0\, dp_0\; \nu_t \; ,
\end{multline}
where $\nu_t$ is a normalization constant and $K_{t}(x,x_0,p,p_0)$ is a two-dimensional Gaussian involving both deterministically and stochastically evolving parameters.
\end{prop}
For instance, in the heterodyne case with equal detection rates, $\Delta = 0$ (without loss of generality), thermal noise $n_{\mathrm{th}} > 0$, and coherent control via ${\bf X}$ and ${\bf P}$, the state evolution admits the following representation:
\begin{prop}\label{prop:Ameas}
The quantum state evolution in this fluorescence measurement scenario, described by the SDE \eqref{eq:drhoHO} in Appendix \ref{ssec:app:aandia}, takes $K_{t}(x,x_0,p,p_0) = K_{x,t}(x,x_0)\, K_{p,t}(p,p_0)$ with
\begin{eqnarray*}
K_{x,t} &=& \exp\left( \frac{-(x-x_0 a_t - \xi_t)^2}{s_t} + (d_t \, x_0^2+\theta_{t}x_0) \right)\;\; , \\
K_{p,t} &=& \exp\left( \frac{-(p-p_0 a_t - \pi_t)^2}{s_t} + (d_t \, p_0^2+\phi_{t}p_0) \right)\;\;.
\end{eqnarray*}
The parameters $a_t, s_t, d_t$ evolve deterministically, independently of both measurement records and controls. For $n_{th}>0$, the variables $\xi_t, \pi_t, \theta_t, \phi_t$ explore their full 4-dimensional space as a function of the measurement realizations. For $n_{th}=0$, deterministic relations link $\theta_t$ to $\xi_t$ and $\phi_t$ to $\pi_t$.
\end{prop}
The full formulas, their derivation, and similar forms for alternative scenarios can be found in Appendix \ref{ssec:app:aandia}.

\item \emph{Special cases.} The above formulation applies for arbitrary $\rho_0$, but it simplifies in special cases. Specifically, when integrating the Gaussian kernel with particular initial states, e.g. Gaussian ones, the dependence on some parameters may vanish. For instance, with $n_{th}=0$ and an initial coherent state $\rho_0 = \q{\alpha}\qd{\alpha}$, $\alpha \in \mathbb{C}$, the state remains coherent and independent of measurement outcomes: $\rho_t = \q{\alpha_t}\qd{\alpha_t}$ for a deterministic evolution $\alpha_t$. This can also be seen by noting that the ensemble-averaged Lindblad equation maintains such a pure state, hence all quantum trajectories must coincide \cite[Ch.4,Lem.2]{PRCohTB}. The manifold dimension thus reduces to zero in this case. 
\end{enumerate}


\subsubsection{Measuring position and momentum simultaneously}\label{sec:XandP}

Taking $L_1={\bf X} = \frac{{\bf a}+{\bf a}^\dag}{2}$ and $L_2 = {\bf P} = \frac{{\bf a}-{\bf a}^\dag}{2i}$ yields the same form \eqref{eq:dyforac} for the output channels, but with a very different backaction on $\rho_t$. In this case, the backaction is closer to a QND behavior, where the mean values of ${\bf X}$ and ${\bf P}$ do not drift on average.

\begin{enumerate}
\item \emph{Algebraic criterion.} Although ${\bf X}$ and ${\bf P}$ do not commute, we find that $[G_{\bf X},\, G_{\bf P}] = 0$. According to the criterion of Proposition \ref{prop:1}, commutation with the deterministic dynamics yields two new vector fields $G_{i\bf X},\, G_{i\bf P} \in \mathfrak{G}_F$. Further commutations generate no new element. This structure remains unchanged when adding Hamiltonians in ${\bf X}$, ${\bf P}$, or ${\bf a}^\dag {\bf a}$, and even under dissipation in a thermal environment, modeled by $L_3=\sqrt{(1+n_{th})}{\bf a}$ and $L_4=\sqrt{n_{th}}{\bf a}^\dag$ with $\eta_3=\eta_4=0$. In all cases, the dynamics are confined to deterministic manifolds of dimension $M=4$.
\item \emph{Manifolds explicit form.} The Wigner function can be written using a Gaussian kernel, as in Proposition~\ref{prop:HOmeas}. More detailed expressions are provided in Appendix \ref{sssec:app:XandPmeas}.

\item Setting the measurement strength of $L_2={\bf P}$ to zero recovers a QND measurement of $L_1={\bf X}$ (see Section \ref{sec:QND}). However, unlike the purely QND case, the presence of additional terms in the deterministic dynamics causes the state to evolve within a larger manifold of dimension $M = 4$. This behavior is best captured by Proposition~\ref{prop:7} in Appendix \ref{sssec:app:XandPmeas}. When $\Delta=0$, the manifold dimension reduces to $M=2$; and if the coupling to the thermal environment also vanishes, the dynamics further collapse to a one-dimensional manifold ($M=1$), in agreement with the QND measurement case in Section \ref{sec:QND}. Thanks to $[G_{i {\bf X}}\, , \, G_{\bf X}] = [G_{i {\bf P}}\, , \, G_{\bf X}] = 0$, the property $M=1$ remains valid in presence of arbitrary drives in ${\bf X}$ and ${\bf P}$.
\end{enumerate}


\subsection{Bipartite quantum systems}\label{sec:bipar}

Identifying low-dimensional manifolds is especially valuable in multipartite quantum systems, whose state space dimension scales as the product of their subsystems' dimensions. While the general method remains applicable, we here focus on a few representative situations.

\subsubsection{Indirect Measurement}\label{sec:OneMTwo}

Consider a bipartite system on $\mathcal{H} = \mathcal{H}_A \otimes \mathcal{H}_B$, where subsystem A is coupled to B via a Hamiltonian, and only subsystem B is directly measured. The general dynamics read:
\begin{eqnarray}\nonumber
d\rho_t &=& \left(\, -i\left[\,H_A \otimes {\bf I} + {\bf I} \otimes H_B + {\textstyle \sum_{m=1}^{\bar{m}}} A_m \otimes B_m \, ,\; \rho_t\,\right] \, \right) \; dt \\
\nonumber && + \left( \sum_{\ell=1}^{\bar{\ell}} F_{M_\ell \otimes {\bf I}}(\rho_t) \, \right) \; dt + \left( \sum_{k=1}^{\bar{k}} F_{{\bf I} \otimes L_k}(\rho_t)\,  \right) \; dt \\
\label{eq:bipart} && + \sum_{k=1}^{\bar{k}} \,\sqrt{\eta_k} G_{{\bf I} \otimes L_k}(\rho_t) \; dw_{t,k}\; .
\end{eqnarray}
This setup, in which subsystem A is monitored via its coupling to B, is widely used in quantum experiments (see e.g.~\cite{holland1991quantum,essig2021multiplexed,sayrin2011real} and many more). 

\begin{enumerate}
\item \emph{Algebraic criterion.} Despite properties like Prop.~\ref{prop:Jacobi}, the bipartite structure alone does not guarantee the presence of low-dimensional manifolds. For instance, first setting$ B_m = 0$ one can characterize the subalgebra $\mathfrak{G}_B$ generated by the dynamics on subsystem B. Next, taking any $G_{I \otimes \tilde{B}_k} \in \mathfrak{G}_B$, together with the coupling term $A_m \otimes B_m$ yields a commutator of the form $ A_m \otimes [B_m, \tilde{B}_k] =: A_m \otimes \tilde{B}_{k,m}$. Further commutations compound A-side operators --- e.g.~$A_n A_m \otimes \tilde{B}_{k,m,n}$ at the next stage ---, thus possibly spanning a large subspace of A-side operators, even if the B-side operators keep cycling within a small algebra $\mathfrak{G}_B$.
\item \emph{Relation to Reservoir Engineering.} In the limit $\eta=0$, the structure \eqref{eq:bipart} is often used for engineering dissipation to stabilize subsystem A into some target subspace \cite{ResEng1,ResEng2}. In this regime, reduced models can be derived that describe the effective dynamics of A alone, while incorprating the influence of B. Such reductions may in principle be exact when based on invariant eigenspaces of the Lindbladian, as in adiabatic elimination techniques \cite{AdEl1,AdEl2,AdEl3}. \emph{When $\eta\neq0$, i.e.~when the dissipation channel is actively monitored, the algebraic criterion shows that measurement backaction may significantly obstruct efficient model reduction.} It also provides a systematic way to identify cases where significant model reduction remains possible.
\item \emph{Special cases.} A stronger reduction occurs, for example, when all coupling terms commute with the measurement dynamics on B:
$\; [B_m\, , \, L] = 0$ for all $m$ and all $G_L \in \mathfrak{G}_B \; $. In this case, $\mathfrak{G}_F = \mathfrak{G}_B$ and the measurement-induced diffusion is unaffected by the dynamics on A. However, such a setting typically prevents extracting useful information about subsystem A. For instance, it it obtained with a QND measurement of B's populations in the diagonal basis of $L$, when these populations remain unaffected by A (since $[B_m,L]=0$).
\newline The following examples illustrate common experimental settings where strong dimensional reductions based on deterministic manifold structures do occur.\\ 
\end{enumerate}


\paragraph*{Ex. monitored harmonic oscillator, dispersively coupled to a qudit:} Consider subsystem B as a harmonic oscillator with annihilation operator ${\bf b}$, and subsystem A as a finite-dimensional system on $\mathcal{H}_A \simeq \mathbb{C}^d$. The dynamics are given by:
\begin{eqnarray}
\label{eq:couplex1}
d\rho_t &=& \left(\, -i[\chi Q_A \otimes ({\bf b}^\dag {\bf b}),\, \rho_t ]  + 2F_{{\bf I} \otimes {\bf b}}(\rho_t) \, \right) dt \;\; \\
\nonumber && -i[{\bf I} \otimes (u_t ({\bf b}+{\bf b}^\dagger) - iv_t ({\bf b}-{\bf b}^\dagger))\,,\, \rho_t \,]\; dt\\
\nonumber && + \sqrt{\eta}\, G_{{\bf I} \otimes {\bf b}}(\rho_t)\, dw_{t,1} + \sqrt{\eta}\, G_{{\bf I} \otimes i{\bf b}}(\rho_t)\, dw_{t,2} \; .
\end{eqnarray}
Here, $Q_A$ is a Hermitian operator (e.g., $\sigma_z$ for a qubit), and $u_t+i v_t$ is a control signal on B, which is monitored through its annihilation channel. This type of model is typical in idealized ``dispersive readout'' schemes for superconducting qubits~\cite{QubitIndirectQND,GambMeas}. It can indeed be viewed as an indirect QND measurement of $Q_A = \sum_{s=1}^d \, \lambda_s \, \q{s}\qd{s}$, considered non-degenerate here for simplicity of notation. Previous studies have derived explicit solutions of \eqref{eq:couplex1} for special initial conditions, such as a separable initial state with B in a coherent state \cite{GambMeas}. Appendix \ref{ssec:app:iqm} generalizes such solutions to arbitrary initial states. 
\begin{enumerate}
\item \emph{Algebraic criterion.} When $u_t=v_t=0$, the dynamics are confined to a manifold of dimension $M=2d$. With general drives, this increases to $M=4d-2$. Moreover, under all measurement realizations and control signals, the reachable states remain within a time-dependent manifold of dimension $\tilde{M} = 3 d^2 +2d - 1$.

This upper bound $\tilde{M}$ is relevant from a control-theoretic perspective, as it constrains the reachable set under measurement-induced feedback \cite{mirrahimi2004controllability}.

\item \emph{Manifolds explicit form.} The solution takes the form $\rho_t = \sum_{j,k=1}^{d} \q{j}\qd{k} \otimes \rho_{t,(j,k)} \; ,$ with each $\rho_{t,(j,k)}$ described by a Gaussian Wigner function. \emph{For $u_t=v_t=0$, all the parameters are deterministically tied to the Gaussian means of the diagonal blocks $\rho_{t,(k,k)}$.} The algebraic criterion has been instrumental in guiding the search for dependencies among those parameters.
\item \emph{Interpretation.}  At long times, each $\rho_{t,(k,k)}$ converges toward a coherent state $\q{\alpha_k}\qd{\alpha_k}$, with amplitude $\alpha_k$ conditioned on qudit level $k$. These amplitudes may evolve under control drives but remain independent of the measurement outcomes. When the $\rho_{t,(k,k)}$ have reached this form, the qudit dynamics resemble a standard QND measurement as in Section \ref{sec:QND}, with the coherent state amplitudes $\alpha_k$ playing the role of effective measurement eigenvalues.\\
\end{enumerate}

\paragraph*{Ex. two resonantly coupled harmonic oscillators:}  Now let both  subsystems A and B be harmonic oscillators with annihilation operators ${\bf a}$ and ${\bf b}$, respectively. Their joint dynamics, under continuous monitoring of the annihilation channel of B, are:
\begin{eqnarray}
\nonumber
d\rho &=& \left(\, -i g\, [ {\bf a} \otimes {\bf b}^\dag + {\bf a}^\dag \otimes {\bf b}\,,\; \rho\, ] 
-i[ \Delta {\bf a}^\dag  {\bf a}\otimes {\bf I}\,,\; \rho \,]\, \right)\; dt \\ && \label{eq:couplex2} -i[u_t ({\bf a}+{\bf a}^\dagger) - iv_t ({\bf a}-{\bf a}^\dagger) ) \otimes {\bf I}\,,\; \rho \,]\; dt\\ 
\nonumber &&+ 2(1+n_{th})F_{{\bf I} \otimes {\bf b}}(\rho)\, dt + 2n_{th} F_{{\bf I} \otimes {\bf b}^\dag}(\rho)\, dt \\
\nonumber && + \sqrt{\eta}\, G_{{\bf I} \otimes {\bf b}}(\rho)\, dw_{t,1} + \sqrt{\eta}\, G_{{\bf I} \otimes i{\bf b}}(\rho)\, dw_{t,2} \; .
\end{eqnarray}
This is a standard model obtained e.g.~after performing Rotating Wave Approximation on two quasi-resonant oscillators with dipolar coupling \cite[Chap.3.2]{Haroche}, with complex drives $u_t+i v_t$ on A, and with additional measurement and thermal noise on subsystem B.
\begin{enumerate}
\item \emph{Algebraic criterion.} As shown in Appendix \ref{app:add:algcrit2cavs}, the dynamics are confined to a manifold of dimension $M=8$ when $n_{th}>0$, and reduce to $M=4$ for $n_{th}=0$. Drives on B (e.g.,~$({\bf b}+{\bf b}^\dagger)$ or ${\bf b}^\dagger {\bf b}$) do not change this structure.
\item \emph{Manifolds explicit form.} System \eqref{eq:couplex2} is again a linear quantum system \cite{LQS}, and its state is fully described by a Gaussian kernel on the Wigner function of the joint phase space. This structure extends naturally to networks of coupled oscillators. 
\end{enumerate}


\subsubsection{Indistinguishable emission}\label{sec:Lewalle}

Detecting photons emitted by multiple quantum systems without distinguishing their origin is a standard technique, commonly used to generate entanglement between distant qubits \cite{EntCond1,EntCond2,EntCond3}.

Here, we examine a diffusive variant of this setting --- motivated both by the methods developed in Section \ref{sec:theory} and by the fact that, for instance in superconducting circuit platforms, field-quadrature measurements were experimentally available well before photodetection (see e.g.~\cite{QQuadRef} vs.~\cite{ManuPhotoDet}. The system consists of two qubits, A and B, with no Hamiltonian dynamics, and two measurement channels
\begin{eqnarray}\label{eqL:sde}
L_1 &=& (\sigma_{-A} + \sigma_{-B})/\sqrt{2} \;  ,\\ \nonumber  L_2 &=& i (\sigma_{-A} - \sigma_{-B})/\sqrt{2} \; ,
\end{eqnarray}
where $\sigma_{-k}$ is the lowering operator (from excited to ground state) for qubit $k$. In practice, $L_1$ and $L_2$ arise by interfering the emitted signals from the two qubits on a balanced beamsplitter \cite{EntCond3}, which naturally ensures equal measurement rates for the two resulting channels~\footnote{Inhomogeneous decay rates of $\sigma_{-A}$ and $\sigma_{-B}$ would yield different effective measurement operators $L_1$ and $L_2$, i.e.~involving other superpositions than the symmetric/antisymmetric ones targeted here. This case is not treated further.}.
\begin{enumerate}
\item \emph{Algebraic criterion.} The vector fields $G_{L_1}$ and $G_{L_2}$ commute. For arbitrary measurement rates $\kappa_1, \kappa_2$, their commutators with the corresponding drift terms $\kappa_1\, (F_{L_1}+D_{L_1})$ and $\kappa_2\, (F_{L_2}+D_{L_2})$ --- explicitly computed in Appendix \ref{app:ssec:Lewalle}--- generate an expanding algebra $\mathfrak{G}_F$, precluding efficient model reduction. However, when the natural symmetry condition $\kappa_1 = \kappa_2$ is imposed (while still allowing $\eta_1 \neq \eta_2$), the commutators with the \emph{total} deterministic evolution produce no new diffusion directions. The full 15-dimensional state space of two qubits is then confined to a 2-dimensional manifold. Additional variations on this setting are discussed in Appendix \ref{app:ssec:Lewalle}.
\item \emph{Manifolds {\bf im}plicit form.} In the standard Pauli basis, we express the two-qubit state as
$\rho_t = \sum_{j,k \in \{I,X,Y,Z\}} \, r(jk)_t \; \frac{{\bf \sigma}_{j} \otimes {\bf \sigma}_k}{4} \; ,$
where $\sigma_I = {\bf I}$ is the identity, and the other $\sigma_j$ are the Pauli matrices. The normalization condition $\tr(\rho_t)=1$ implies $r(II)_t=1$ for all $t$, leaving 15 free coordinates. In Appendix \ref{app:ssec:Lewalle}, we define normalized variables $B_{jk,t} = \frac{r(jk)_t}{r(ZZ)_t-r(IZ)_t-r(ZI)_t}$, and progressively construct $13$ linearly independent polynomials in the $B_{jk,t}$ --- each of degree at most $3$ --- that evolve deterministically in time, independently of the measurement records $dy_{1,t}$ and $dy_{2,t}$. 
\end{enumerate}


\subsubsection{A qudit monitored through $n$ harmonic oscillators}\label{sec:PRmodel}

The example of Section~\ref{sec:OneMTwo} can be generalized to a setting where a qudit is coupled to $n>1$ harmonic oscillators, whose fluorescence is measured through $m\geq 1$ detection channels. In \cite{PRmodelNote}, inspired by \cite{criger2016multi}, the system dynamics were analyzed in the case $m=1$ and with the oscillators initialized in coherent states. In that regime, the oscillators remain in coherent states at all times, with amplitudes evolving deterministically --- independent of the measurement outcomes. The stochastic dynamics then reduce to at most $d^2-1$ variables associated with the qudit alone. The present framework allows us to go beyond these assumptions and handle arbitrary initial conditions.

We consider a composite system of $n$ harmonic oscillators and one $d$-level qudit. Working in a rotating frame for the qudit’s free Hamiltonian (and possibly after Rotating Wave Approximation), the dynamics are governed by the following dispersive coupling model:
\begin{eqnarray}\label{eq:PRmodel0}
d\rho &=&-i\; { \sum_{k=1}^n} \; [ (\Delta_k {\bf a}_k^\dag  {\bf a}_k \\ \nonumber && \qquad + u_{t,k} ({\bf a}_k+{\bf a}_k^\dagger) - iv_{t,k} ({\bf a}_k-{\bf a}_k^\dagger) )\,,\; \rho \,] \; dt \\  \nonumber
&& -i { \sum_{k=1}^n \sum_{j=1}^d}\; \chi_{j,k}\; [\q{j}\qd{j} \otimes {\bf a}_k^\dag  {\bf a}_k\,,\; \rho]\; dt\\  \nonumber
&& + { \sum_{l=1}^m} \; F_{{\bf b}_l}(\rho)\, dt + \sqrt{\eta_l}\, G_{{\bf b}_l}(\rho)\, dw_{t,l} \; ,
\end{eqnarray}
where ${\bf a}_k$ is the annihilation operator for oscillator $k$. 
Measurement is performed on collective modes defined by:
\begin{equation}\label{eq:PRmodel1}
{\bf b}_{l} = { \sum_{k=1}^n} \; \beta_{l,k} \; {\bf a}_k \;, \quad l=1,2,...,m \; ,
\end{equation}
with arbitrary complex coefficients $\beta_{l,k} \in \mathbb{C}$.
\begin{enumerate}
\item \emph{Algebraic criterion.}  The vector fields $G_{{\bf b}_l}$ all commute. As shown in Appendix \ref{app:ssec:PRmod}, their commutators with the deterministic terms generally generate vector fields spanning all combinations of $\q{j}\qd{j}$ with complex linear combinations of the ${\bf a}_k$ and the identity. This yields a confining manifold of dimension $M=2 d (n+1)-2$. 

When $n$ is small compared to $d$, this is smaller than the $d^2-1$ qudit variables considered in \cite{PRmodelNote,criger2016multi}, revealing additional deterministic structure. Conversely, when $n$ is large compared to $d$, the manifold dimension exceeds $d^2 - 1$, reflecting the more complex dynamics arising from generic initial conditions beyond coherent states.

\item \emph{Manifolds explicit form.} Like in the single-resonator case, the formulas in Appendix \ref{app:ssec:PRmod} express $\rho_t$ in terms of a Wigner function with a Gaussian kernel. The evolving parameters of this kernel fully characterize the system. The stochastic degrees of freedom essentially correspond to the Gaussian means and to the relative weights of the diagonal qudit blocks $\qd{j} \rho_t \q{j}$, with $\q{j}$ a qudit computational state. This matches the $O(nd)$ scaling predicted by the algebraic criterion. All off-diagonal blocks $\qd{j} \rho_t \q{j'}$ for $j'\neq j$ are deterministically determined by the diagonal blocks, except for their global phases,  which --- perhaps unexpectedly --- introduce additional stochastic variables.

Under ``sufficiently rich observation'' conditions, i.e.~ensuring convergence of the Gaussian profile, the $n$-resonator system asymptotically approaches a product of coherent states $\bigotimes_{k=1}^n \, \q{\alpha^{(j,k)}}$, entangled with the qudit level $j$. In thsi regime, we recover the setting of \cite{PRmodelNote,criger2016multi}: \emph{Each coherent amplitude $\alpha^{(j,k)}_t \in \mathbb{C}$ evolves deterministically, independently of the measurement realizations. When the resonators are initialized in such states, the qudit populations $\text{Tr}(\qd{j} \rho_t \q{j})$ evolve as in a standard QND measurement of the qudit, with effective time-dependent measurement operators $L_{l,t} = \sum_{j=1}^d \; \lambda^{(j)}_{l,t} \; \q{j}\qd{j} \, $, where
$\lambda^{(j)}_{l,t} = 
 {\textstyle \sum_{k=1}^n}\mathfrak{R}\text{eal}(\beta_{l,k}^* \; \alpha^{(j,k)}_t) \; , $ for $l=1,2,...,m$.}
\end{enumerate}


\section{Conclusion}

This note provides low-dimensional formulas for the explicit solution of continuously monitored quantum systems, modeled by nonlinear stochastic differential equations driven by Wiener processes capturing the randomness of measurement outcomes. Across several representative settings, we show that the quantum state often remains confined to a low-dimensional, deterministically evolving nonlinear manifold, whose governing equations we compute explicitly. In these settings, the remaining stochasticity --- due to measurement outcomes --- spans a low-dimensional space, making the system more amenable to analysis. This framework can be seen as extending the principle of Linear Quantum Systems, which are reducible to Gaussian-state representations of harmonic oscillators. Here, we derive low-dimensional descriptions for arbitrary quantum states --- still leveraging Gaussian kernels in oscillator settings, although the equations are nonlinear; and extending them to various finite-dimensional quantum systems. In essence, while a single measurement-induced Wiener process could in principle cause full diffusion across the state space, we identify many common settings where this is not the case.

From the perspective of representation theory, our results characterize the minimal memory required to express the statistics of future measurement outcomes as functions of past measurement records. Beyond its analytical value, this reduction has practical implications for efficient parameter estimation, experiment design guided by measurement statistics \cite{JordanMLT1}, and real-time quantum filtering in control applications. The approach could be extended to other filters, such as past quantum state models, to reduce computational complexity in broader tasks.\\ 

A key tool enabling this framework is a simple algebraic criterion from control theory, which identifies whether compact parameterizations exist. This criterion also reveals how small modifications (e.g., adding decoherence or Hamiltonian terms) impact the potential for model reduction. For instance, it helped derive the reduced parameterization of system \eqref{eq:couplex1}, by indicating both the presence of a confining structure and when further search for dependencies should cease.

Conversely, the same criterion can serve as a no-go theorem: it identifies memory effects that make exact model reduction impossible. In the 3-qubit repetition code with bit-flip channels, for example, no reduced filter based on a few integrated signals exists. Likewise, in indistinguishable emission scenarios, confinement depends on specific parameter conditions. For bipartite systems where only one subsystem is monitored, reduced deterministic manifolds should not generally be expected.\\

Confining manifolds have been observed in experiments \cite{PhCManifolds,MMmanifolds}, although from a mathematical viewpoint they are not structurally robust. Even small Hamiltonian imperfections typically induce diffusion in all directions under a single stochastic channel. However, this diffusion is often slower in directions involving higher-order commutators with the measurement-induced vector field. This opens paths toward studying \emph{approximate} confinement and exploiting it for approximate model reduction. Starting from the opposite extreme --- idealized projective measurements --- adding any refinement in terms of memory effects should already improve model approximations. In this sense, the exploration of coupled systems, error correction scenarios, or more generally timescale-separated dynamics, has only been initiated in this note.\\

Another promising direction is the interaction between measurement backaction and feedback control. On one hand, as shown for system \eqref{eq:couplex1}, the algebraic criterion quickly reveals the dimension of the reachable state space \cite{mirrahimi2004controllability} when both measurement backaction and control are treated as actuation resources. On the other hand, since designing highly efficient quantum feedback controllers --- numerically or analytically --- remains a complex challenge, any principled method to reduce the system dimension is a highly valuable tool. In particular, a first step towards feedback controllers featuring less memory than a full state model, could involve applying feedback control directly to the reduced parameters identified through this framework.

Feedback strategies that fully counteract stochastic effects have been proposed, e.g., in \cite{PhysRevA.102.062418}. It remains an open question whether actively \emph{designing} a system or control scheme to confine the dynamics to a low-dimensional manifold --- independent of measurement realizations --- offers operational benefits in its own right.\\

Finally, while this work focuses on continuous-time measurements modeled by Wiener processes (e.g., field amplitude measurements), an analogous study could be envisioned for Poisson-type measurements (\cite{belavkin1989continuous,belavkin2006nondemolition}, e.g., photodetection). In this case, although the number of jumps yields a discrete set of options, the timing of jumps could still give rise to continuous, low-dimensional structure.

\section*{Acknowledgments}
The authors thank Mazyar Mirrahimi, Benjamin Huard, Philippe Campagne-Ibarcq, Vincent Martin, Stefaan De Kesel, Zibo Miao, Paolo Forni, Francesca Chittaro, Philippe Lewalle and Birgitta Whaley for useful discussions. This work was partially supported by the ANR project HAMROQS and by Plan France 2030 through the project ANR-22-PETQ-0006. This work was partially supported by the European Research Council (ERC) under the European Union’s Horizon 2020 research and innovation program (grant agreement No.884762).

\bibliography{sections/bibliography}
\bibliographystyle{plain}

\newpage
\onecolumn

\appendix

\section*{Appendix}

Here, we give the detailed settings and expressions of the results summarized in the main text, as well as the associated proofs. 

Section \ref{app1} considers the scenarios with Quantum Non-Demolition Measurement, first general expressions then each of the 3 examples. Section \ref{app2} addresses the cases of harmonic oscillator quadrature measurements, in the same order as in the main text. Section \ref{app3} addresses the scenarios with bi-partite quantum systems.\\

\begin{itemize}[noitemsep]
\item[] \hyperref[app1]{A.~Appendix 1: Quantum Non-Demolition Measurement} \dotfill \pageref{app1}\\[2mm]
\item[] \hyperref[app:ssec:QNDgeneral]{A.1.~General QND expressions} \dotfill \pageref{app:ssec:QNDgeneral}
\item[] \hyperref[prop:3]{$*$ formulas: homodyne QND} \dotfill \pageref{prop:3}
\item[] \hyperref[prop:4]{$*$ formulas: heterodyne QND}
\dotfill \pageref{prop:4}
\item[] \hyperref[app:ssec:QNDexampleN]{A.2.~QND example: Quantum number measurement} \dotfill \pageref{app:ssec:QNDexampleN}
\item[] \hyperref[eq:Nphase]{$*$ formulas: quantum number phases} \dotfill \pageref{eq:Nphase}
\item[] \hyperref[app:ssec:QNDexampleC]{A.3.~QND example: continuous-variable measurement} \dotfill \pageref{app:ssec:QNDexampleC}
\item[] \hyperref[eq:Xphase]{$*$ formulas: position coherence phases} \dotfill \pageref{eq:Xphase}
\item[] \hyperref[app:ssec:QNDexampleR]{A.4.~QND example: repetition code} \dotfill \pageref{app:ssec:QNDexampleR}\\[2mm]
\item[] \hyperref[app2]{B.~Appendix 2: harmonic oscillator quadratures} \dotfill \pageref{app2}\\[2mm]
\item[] \hyperref[app2details]{B.1.~Full models, Algebraic criterion and Wigner function} \dotfill \pageref{app2details}
\item[] \hyperref[ssec:app:aandia]{B.2.~Full solution for measurement of ${\bf a}$ and $i {\bf a}$} \dotfill \pageref{ssec:app:aandia}
\item[] \hyperref[prop:6]{$*$ formulas: general case} \dotfill \pageref{prop:6}
\item[] \hyperref[prop:6specialcase]{$*$ formulas: special case $n_{th}=0$} \dotfill \pageref{prop:6specialcase}
\item[] \hyperref[sssec:app:XandPmeas]{B.3.~Full solution for measurement of ${\bf X}$ and ${\bf P}$} \dotfill \pageref{sssec:app:XandPmeas}
\item[] \hyperref[prop:7]{$*$ formulas: general case} \dotfill \pageref{prop:7}
\item[] \hyperref[cor:71]{$*$ formulas: special case $\Delta=0$} \dotfill \pageref{cor:71}
\item[] \hyperref[cor:72]{$*$ formulas: measuring ${\bf X}$ with drives, $n_{th} \neq 0$} \dotfill \pageref{cor:72}\\[2mm]
\item[] \hyperref[app3]{C.~Appendix 3: bipartite quantum systems} \dotfill \pageref{app3}\\[2mm]
\item[] \hyperref[ssec:app:iqm]{C.1.~Indirect qudit measurement} \dotfill \pageref{ssec:app:iqm}
\item[] \hyperref[prop:12]{$*$ formulas: general case} \dotfill \pageref{prop:12}
\item[] \hyperref[prop:13]{$*$ formulas: special case with no drive} \dotfill \pageref{prop:13}
\item[] \hyperref[app:add:algcrit2cavs]{C.2.~Two resonantly coupled harmonic oscillators} \dotfill \pageref{app:add:algcrit2cavs}
\item[] \hyperref[app:ssec:Lewalle]{C.3.~Indistinguishable emission} \dotfill \pageref{app:ssec:Lewalle}
\item[] \hyperref[prop:LewRes]{$*$ formulas: indistinguishable emission} \dotfill \pageref{prop:LewRes}
\item[] \hyperref[app:ssec:PRmod]{C.4.~A qudit monitored through $n$ harmonic oscillators} \dotfill \pageref{app:ssec:PRmod}
\item[] \hyperref[prop:LLast]{$*$ formulas: qudit with $n$ oscillators} \dotfill \pageref{prop:LLast}
\end{itemize}
 

\section{~~~~Appendix 1:  Quantum Non-Demolition Measurement}\label{app1}

This section contains the details associated to Section \ref{sec:QND}.


\subsection{~~~~General QND expressions}\label{app:ssec:QNDgeneral}

We start with the general expressions, first the case of homodyne measurement.

\begin{prop}\label{prop:3}
Consider a quantum master SDE \eqref{eq:SDE} on $\mathcal{H} = \mathbb{C}^N$, with $L_k=\sum_{n=1}^N \lambda_k(n) \q{d_n}\qd{d_n}$ in the orthonormal basis $\{\q{d_n}\}_{n=1}^N$ and all $\lambda_k(n)$ real for $k=1,2,...,\bar{k}$. Then the state $\rho_t$ is restricted to a deterministically evolving manifold of dimension $\leq \bar{k}$, characterized by:
\begin{eqnarray*}
\phi^{(a,b)}_t & = & \phi^{(a,b)}_0 \text{ where } \phi^{(a,b)}_t = \mathrm{phase}(\rho_t(a,b)) \quad \text{for all } a,b=1,2,...,N  \;\;\;\; ;\\[2mm]
c^{(a,b)}_t & = & c^{(a,b)}_0 \, e^{\big( - \overline{(1-\eta)(Q(a)-Q(b))^2}\, t \big)} \quad  \text{for all } a,b=1,2,...,N  \;\;, \\
& & \text{where } \;\;\; c^{(a,b)}_t \; = \; \frac{\vert \rho_t(a,b) \vert^2}{\rho_t(a,a)\rho_t(b,b)} \\ 
&& \text{ and } \;\; \overline{(1-\eta)(Q(a)-Q(b))^2} = {\textstyle \sum_{k=1}^{\bar{k}}} \; (1-\eta_k) (\lambda_k(a)-\lambda_k(b))^2 \;\; ;\\[2mm]
z^{(\alpha)}_t & = &  z^{(\alpha)}_0 \, e^{\big( -2 \sigma^2_\alpha \; t \big)} \\
& & \text{where } z^{(\alpha)}_t = \prod_{b=1}^N \, (\rho_t(b,b))^{\alpha_b} \;\; \text{ and } \;\; 
\sigma^2_\alpha = \sum_{b=1}^N \sum_{k=1}^{\bar k} \, \alpha_b \, (\lambda_k(b))^2 \, \eta_k \\
& & \text{with any } \alpha\in\mathbb{R}^N \text{ solving }  \;\;\; \textstyle \sum_{b=1}^N  \alpha_b = 0 \; ,\\
&& \phantom{\text{with any } \alpha\in\mathbb{R}^N \text{ solving }  \;\;\;} \;\;\; \eta_k \; \textstyle \sum_{b=1}^N  \alpha_b \, \lambda_k(b) = 0 \;\; \text{for all } k=1,2,...,\bar{k} \, \; ,\\
&& \phantom{\text{with any } \alpha\in\mathbb{R}^N \text{ solving }  \;\;\;}  \;\;\; \text{ and } \sigma^2_\alpha \geq 0 \, .
\end{eqnarray*}
\end{prop}

\noindent \emph{Technical remarks:} This is a more refined version of Proposition \ref{prop:3a}, whose third bullet is obtained by taking the logarithm of the variables $z^{(\alpha)}_t$. (Note that arbitrary functions of deterministically evolving variables also evolve deterministically, and thus follow the usual rules under coordinate transformation.) Since $z^{(-\alpha)} = 1/z^{(\alpha)}$ and $\sigma^2_{-\alpha} = -\sigma^2_{\alpha}$, the inequality constraint just selects variables tending to $0$ rather than to $\infty$. When $\eta_k=0$ for some $k$, there are more linearly independent choices for $z^{(\alpha)}$, yielding a manifold of correspondingly lower dimension. When some levels $a,b$ are degenerate in all measurement operators, taking all components of $\alpha$ to be zero except $\alpha_a=-\alpha_b$ is a valid choice and yields $\sigma_\alpha^2 = 0$. This properly expresses the conservation of $\rho(a,a)/\rho(b,b)$ for these indistinguishable populations.\\

\noindent \underline{Proof of Proposition \ref{prop:3}:} We first write the dynamics of each component $\rho_t(a,b)$, for instance for the diagonal elements:
\begin{eqnarray}
\label{eq:dpt}
d\rho_t(b,b) & = & \sum_{k=1}^{\bar{k}} \; \sqrt{\eta}_k \left(2 \lambda_k(b) \rho_t(b,b) - ({\textstyle \sum_{a=1}^N \lambda_k(a) \rho_t(a,a)})\, \rho_t(b,b) \right)  \, dw_{t,k} \; .
\end{eqnarray}
We then look for combinations in which the vector field multiplying each $dw_k$ cancels. One can quickly check that this is the case for the $\phi^{(a,b)}$, $c^{(a,b)}$ and $z^{(\alpha)}$ of the statement. Next, the key step towards obtaining the form \eqref{eq:implicitsolution} is to ensure that this set of variables also evolves \emph{autonomously}: after having established that their dynamics takes the form e.g.
$$dc^{(a,b)}_t = f(\rho_t) \, dt \; ,$$
one must further ensure that $f(.)$ depends on $\rho_t$ through only the variables $\phi^{(a,b)},c^{(a,b)}$ and $z^{(\alpha)}$. We have obtained even more, namely independent equations for each variable, e.g.~ $dc^{(a,b)}_t = f\, dt$ with $f$ depending only on $c^{(a,b)}_t$. This is not difficult to check a posteriori. Note that the computation of $f(\rho_t)$ must use the \Ito~rule \eqref{eq:Itorule}. Integration of $dc^{(a,b)}_t = f(c^{(a,b)}_t) dt$ then yields the result, and similarly for the other variables.
\hfill $\square$\\

We next generalize to the case of a heterodyne measurement. For Hermitian $L_k$, the output signal $dy_{t,k}$ associated to $i L_k$ contains only Wiener noise, independent of the system state. The practical interest of this output channel is thus questionable, but since it is a natural experimental setting for general $L_k$, we can quickly treat it for completeness.
\vspace{2mm}

\begin{prop}\label{prop:4}
Consider a quantum master SDE \eqref{eq:SDE} on $\mathcal{H} = \mathbb{C}^N$, with $L_k=\sum_{n=1}^N \lambda_k(n) \q{d_n}\qd{d_n}$ in the orthonormal basis $\{\q{d_n}\}_{n=1}^N$ and all $\lambda_k(n)$ real for $k=1,2,...,\bar{k}/2$, whereas $L_{\bar{k}/2+k} = i L_{k}$. Then the state $\rho_t$ is restricted to a deterministically evolving manifold of dimension $\leq \bar{k}$, characterized by:
\begin{eqnarray*}
\phi^{(\beta)}_t & = &  \phi^{(\beta)}_0 \quad \text{where } \phi^{(\beta)} = {\textstyle \sum_{a,b=1}^N} \, \beta_{a,b} \, \mathrm{phase}(\rho(a,b)) \\
& & \text{with any } \beta\in\mathbb{R}^{N^2} \text{ solving }\\
& & \qquad  \eta_{\bar{k}/2 + k}\textstyle \sum_{a,b=1}^N  \beta_{a,b} \, (\lambda_k(a)-\lambda_k(b)) = 0 \;\; \text{for all } k=1,2,...,\bar{k}/2 \; ;\\[2mm]
c^{(a,b)}_t & = & c^{(a,b)}_0 \, e^{\big( - \overline{(1-\eta)Q_{a,b}^2} \, t \big)}  \quad \text{for all } a,b=1,2,...,N  \;\;, \\
& & \text{where } \;\;\; c^{(a,b)}_t \; = \; \frac{\vert \rho(a,b) \vert^2}{\rho_t(a,a)\rho_t(b,b)} \\ &&\text{ and } \;\; \overline{(1-\eta)Q_{a,b}^2} =  {\textstyle \sum_{k=1}^{\bar{k}}} \; (1-\eta_k) |\lambda_k(a)-\lambda_k(b)|^2 \;\; ; \\[2mm]
z^{(\alpha)}_t & = &  z^{(\alpha)}_0 \, e^{\big( -2 \sigma^2_\alpha \; t \big)} \\
& & \text{where } z^{(\alpha)}_t = \prod_{b=1}^N \, (\rho_t(b,b))^{\alpha_b} \;\; \text{ and } \;\; 
\sigma^2_\alpha = \sum_{b=1}^N \sum_{k=1}^{\bar k/2} \, \alpha_b \, (\lambda_k(b))^2 \, \eta_k \\
& & \text{with any } \alpha\in\mathbb{R}^N \text{ solving } \textstyle \sum_{b=1}^N  \alpha_b = 0 \; ,\\
& & \phantom{ \text{with any } \alpha\in\mathbb{R}^N \text{ solving } } \eta_k \; \textstyle \sum_{b=1}^N  \alpha_b \, 
\lambda_k(b) = 0 \;\; \text{for all } k=1,2,...,\bar{k}/2 \, \; ,\\
& & \phantom{ \text{with any } \alpha\in\mathbb{R}^N \text{ solving } } \text{ and } \sigma^2_\alpha \geq 0 \, .
\end{eqnarray*}
\end{prop}

\noindent \underline{Proof:} The proof is similar to the homodyne case.\\

The set of SDEs governing the diagonal components $\rho_t(b,b)$ is unaffected by the $i\, L_k$, hence the $z^{(\alpha)}_t$ of Proposition \ref{prop:3} remain valid.

The dynamics of $x^{a,b} := (\rho(a,b)+\rho(b,a))/2$ and $y^{a,b} := (\rho(a,b)-\rho(b,a))/2i$ are subject to a new stochastic process, which is associated to a vector field of components $y^{a,b}_t$ and $-x^{a,b}_t$ respectively.
\begin{itemize}
\item For any two given levels $a,b$, such a vector field expresses stochastic rotation on the equator of the Bloch sphere. In other words, the phases of $\rho_t(a,b)$ are not preserved anymore but instead follow the random walks:
$$d\phi^{(a,b)}_t \; = \; {\textstyle \sum_{k=1}^{\bar{k}/2} }\; (\lambda_k(a)-\lambda_k(b)) \, \sqrt{\eta_k} \, dw_{k+\bar{k}/2\; , \; t} \, .$$
Those walks are correlated since the system still stays confined in a $\bar{k}$-dimensional manifold, and this correlation is expressed by the $(N^2 - N - \bar{k})/2$ independent variables of type $\phi^{(\beta)}$. (Indeed, thanks to $\rho$ being Hermitian, one checks that the values of $(\beta_{a,b}+\beta_{b,a})$, including $\beta_{a,a}$ for $a=b$, have no influence on $\phi^{(\beta)}$.)
\item On the other hand, when $\eta_k<1$ for some $k>\bar{k}/2$, the lack of knowledge on phase random walk with respect to someone having access to the full output ($\eta_k=1$), is reflected as a faster decay of off-diagonal elements. More precisely, the deterministic decay still happens on $c^{(a,b)}$ like in the homodyne case, but with contributions from both parts of the heterodyne channel to the decay rate.
\end{itemize}
\hfill $\square$\\

We finally prove the simple explicit expression of the state as a function of the measurement record, as stated in Proposition \ref{prop:5}.\\

\noindent \underline{Proof of Proposition \ref{prop:5}:}  From \Ito~computations on \eqref{eq:dpt} and recalling the link between $dw_{k,t}$ and output signal $dy_{t,k}$, we obtain the following dynamics for $r(a,b) := \log(\rho(a,a)/\rho(b,b))$:
$$dr_t(a,b) = 2 \sum_{k=1}^{\bar k} \; (\lambda_k(a)-\lambda_k(b)) \sqrt{\eta_k}\, dy_{t,k} - (\lambda_k(a)^2-\lambda_k(b)^2) \eta_k\, dt \; .$$
The right hand side can be trivially integrated; taking its exponential, we get:
$$\left(\frac{\rho(a,a)}{\rho(b,b)}\right)_t = \left(\frac{\rho(a,a)}{\rho(b,b)}\right)_0 \; e^{2 \sum_{k=1}^{\bar k} \; (\lambda_k(a)-\lambda_k(b)) \sqrt{\eta_k}\, y_{t,k} - (\lambda_k(a)^2-\lambda_k(b)^2) \eta_k\, t } \; .$$
This explicit expression is compatible with the stated result.\hfill $\square$\\
 
 
\subsection{~~~~QND example: Quantum number measurement}\label{app:ssec:QNDexampleN}
 
Particularizing the above expression to the single measurement channel $L_1=\sum_{n=0}^{n_{\max}} n \, \q{n}\qd{n}$, we get the following deterministic variables. We speak of ``photon number'' for brevity, although the results of course hold for any physical system with this mathematical description. For the homodyne case:
\begin{itemize}
\item $\phi^{(n_1,n_2)}_t =\phi^{(n_1,n_2)}_0$ i.e.~the relative phase between coefficients of all photon number states $\q{n_1},\q{n_2}$ is conserved.
\item $\dfrac{|\rho_t(n_1,n_2)|^2}{\rho_t(n_1,n_1)\, \rho_t(n_2,n_2)} \, = \, \dfrac{|\rho_0(n_1,n_2)|^2}{\rho_0(n_1,n_1)\, \rho_0(n_2,n_2)}\cdot e^{-(1-\eta)(n_1-n_2)^2\, t}$, thus the coherence between two different photon numbers decays faster if their numerical difference is larger. This stands in one-to-one correspondence with being more distinguishable in the measurement output.
\item For the diagonal components, we can obtain invariant variables 
\begin{multline*}
z^{(\alpha)}_t = z^{(\alpha)}_0 \; \text{ for } \; z^{(\alpha)}=\rho(n_1,n_1)^{\frac{1}{(n_2-n_1)(n_3-n_1)(n_4-n_1)}}\, \rho(n_2,n_2)^{\frac{1}{(n_1-n_2)(n_3-n_2)(n_4-n_2)}} \\ \rho(n_3,n_3)^{\frac{1}{(n_1-n_3)(n_2-n_3)(n_4-n_3)}} \, \rho(n_4,n_4)^{\frac{1}{(n_1-n_4)(n_2-n_4)(n_3-n_4)}}\; .
\end{multline*}
This gives the expression reported in the main text when $(n_2-n_1) = (n_4-n_3)$.
\end{itemize}
The characterization  through Proposition \ref{prop:5} speaks for itself.

For the heterodyne case, thus $L_1=\sum_{n=0}^{n_{\max}} n \, \q{n}\qd{n}$ and  $L_2 = i  L_1$, the main novelty is that the $\phi^{(n_1,n_2)}_t$ now undergo correlated random walks. We can write some invariant phase combinations:
\begin{eqnarray*}
&& \phi^{(\beta)}_t = \phi^{(\beta)}_0 \;\; \text{ for } \\
&& \quad \phi^\beta = \text{phase}(\rho(a,b)) - \text{phase}(\rho(a+m,b+m)) \;\; \text{ and for } \\
&& \quad \phi^\beta = \tfrac{1}{a+1-b}\, \text{phase}(\rho(a+1,b)) - \tfrac{1}{a-b}\, \text{phase}(\rho(a,b))\; .
\end{eqnarray*}
In the form \eqref{eq:explicitsolution}, we can write the phase evolution with a single stochastic variable $\gamma_t$ as:
\begin{equation}\label{eq:Nphase}
\text{phase}(\rho_t(a+m,a)) = \text{phase}(\rho_0(a+m,a)) + m\, \gamma_t \;\;\; \text{ for all } a,m\; .
\end{equation}
Thus, the further from the diagonal, the more the component is affected by the stochastic phase $\gamma_t$. 

 
\subsection{~~~~QND example: continuous-variable measurement}\label{app:ssec:QNDexampleC}

The main text considers a homodyne measurement, we here consider the heterodyne case, thus $L_1 = \mathbf{X}$ and $L_2 = i \mathbf{X}$ where $\mathbf{X} = \int_{\mathbb{R}} x\, \q{x}\qd{x} \, dx$ represents the position operator. The results are obtained by repeating computations similar to the quantum number measurement and taking the continuum limit. 
\begin{itemize}
\item The phases of off-diagonal components feature invariant functions $\phi^{(\beta)}_t =  \phi^{(\beta)}_0$, including
\begin{eqnarray*}
\phi^{(\beta)} = \tfrac{d}{ds} \mathrm{phase}(\rho(x_1+s,x_2+s)) \;\;& \text{ and } &\;\; \phi^{(\beta)} = \tfrac{d}{dx_1}\left( \frac{1}{x_1-x_2} \mathrm{phase}(\rho(x_1,x_2)) \right) \; .
\end{eqnarray*}
\item The off-diagonal amplitudes still follow, for all $x_1,x_2 \in \mathbb{R}$:
\begin{eqnarray*}
c_t(x_1,x_2) =  c_0(x_1,x_2) \, e^{-2(1-\eta)\, (x_1-x_2)^2 \, t} &\;\; \text{ where } \;\; & c(x_1,x_2)\; = \; \frac{\vert \rho(x_1,x_2) \vert ^2}{\rho(x_1,x_1)\rho(x_2,x_2)} \; .
\end{eqnarray*}
Thus coherences between positions further apart decay faster. The factor 2 appears because $L_1$ and $L_2$ each contribute.
\item Regarding the $z_t^{(\alpha)}$, which govern correlations among the diagonal components of $\rho$, we obtain the same expressions as in the main text (homodyne case), as $L_2= i \mathbf{X}$ gives no information on QND level populations.
\end{itemize}
In the explicit form \eqref{eq:explicitsolution}, we can write the phases as:
\begin{equation}\label{eq:Xphase}
\mathrm{phase}(\rho_t(x+s,x)) = \mathrm{phase}(\rho_0(x+s,x)) + s\, \gamma_{t,2}
\end{equation}
for all $x,s \in \mathbb{R}$, with a single stochastic variable $\gamma_{t,2}$ independent of $x,s$. The distribution $p(x) = \rho(x,x)$ evolves as for the homodyne case.

 
\subsection{~~~~QND example: repetition code}\label{app:ssec:QNDexampleR}

We give more details about the scenario of 3-qubit repetition code. The Hilbert space is $\mathcal{H}=\text{span}(\{\q{0},\q{1}\} \otimes \{\q{0},\q{1}\} \otimes \{\q{0},\q{1}\}) \simeq \mathbb{C}^8$. 

In a first setting, we just consider \eqref{eq:SDE} with three commuting measurement operator. The measurements each compare whether two qubits have the same or a different value in the canonical basis $\{\q{0},\q{1}\}$. Denoting $P_{\{\q{v_1},\q{v_2},...\}}$ the orthonormal projector onto the subspace spanned by $v_1,v_2,...$, we thus have:
\begin{eqnarray*}
L_1 &=& P_{\{\q{0\; 00},\q{1\; 00},\q{0\; 11},\q{1\; 11} \}} - P_{\{\q{0\; 10},\q{1\; 10},\q{0\; 01},\q{1\; 01} \}} = \sigma_{z,2}\,\sigma_{z,3}\\
L_2 &=& P_{\{\q{0\, 0 \, 0},\q{0\, 1\, 0},\q{1\, 0\, 1},\q{1\, 1 \, 1} \}} - P_{\{\q{1\, 0\, 0},\q{1\,1\, 0},\q{0\, 0 \, 1},\q{0 \, 1\, 1} \}}  = \sigma_{z,1}\,\sigma_{z,3}\\
L_3 &=& P_{\{\q{00 \; 0},\q{00 \; 1},\q{11 \; 0},\q{11 \; 1} \}} - P_{\{\q{10 \; 0},\q{10 \; 1},\q{01 \; 0},\q{01 \; 1} \}}  = \sigma_{z,1}\,\sigma_{z,2}\; ,
\end{eqnarray*}
where $\sigma_{z,k}$ applies the Pauli $z$ operator on qubit $k$.
We consider the same measurement rate, and also the same measurement efficiency $\eta_1=\eta_2=\eta_3$, for simplicity of notation only. Together, these $L_k$ distinguish four eigenspaces, each one twofold degenerate: 
\begin{align*}
 (\lambda_1(.),\lambda_2(.),\lambda_3(.)) &\; =(+1,+1,+1) & \text{ on } \text{span}(\q{000},\q{111})=:V_0 \; ;  \\ & \; =(+1,-1,-1) & \text{ on } \text{span}(\q{100},\q{011})=:V_1 \; ; \\
 &\; =(-1,+1,-1) & \text{ on } \text{span}(\q{010},\q{101})=:V_2 \; ;  \\ & \; =(-1,-1,+1) & \text{ on } \text{span}(\q{001},\q{110})=:V_3 \; .
\end{align*}
The twofold degeneracy gives the possibility to encode a qubit in the codespace $V_0$ as $\q{\psi} = c_0 \q{000} + c_1 \q{111}$; the syndrome measurements further allow to identify and correct a single spurious bit-flip $0 \leftrightarrow 1$ without any loss of the quantum information encoded through $c_0,c_1$.

The confining manifolds for this case feature no more peculiarities than explained in the main text. The decay rate 
$8\, (1-\eta)$ for off-diagonal components is obtained according to Proposition \ref{prop:3}: each pair $V_j$, $V_{j'\neq j}$ is distinguished by two of the three syndromes $L_k$, with $|\lambda_k(a)-\lambda_k(b)|=2$.\\

The main purpose in this appendix is to discuss the addition of unmonitored, spurious bit flips to these dynamics. In \eqref{eq:SDE} this means adding $$ L_{3+k}=\sqrt{\Gamma_k} \sigma_{x,k}  \text{ with } \eta_{3+k}=0 \text{ for } k=1,2,3 \; .$$ 
Thus the SDE is driven by no new stochastic processes. However, like in the classical example with the rolling wheel, the addition of a deterministic drift to few stochastic vector fields can induce diffusion in many more directions. This turns out to be the case here. We provide two technical results and include one comment.\\
\begin{itemize}
    \item We first consider $\Gamma_1=\Gamma_2=\Gamma_3 \neq 0$ and sketch how to generate several independent vector fields in $\mathfrak{G}_F$, which shows how the presence of deterministic dynamics due to spurious bit flips significantly complicates the expression of an explicit solution.
    \begin{itemize}
\item We start with the 3 independent and commuting vector fields $G_{L_1},G_{L_2},G_{L_3} \in \mathfrak{G}_F$ corresponding to syndrome measurements.
\item The commutator with $F+D$ reduces to the commutator with  $\sum_{k=4}^6 F_{L_k}$ and generates three new independent vector fields. Using the form \eqref{eq:strangeterm2} and noting that $L_j^\dagger L_k L_j = \pm L_k$ for all operators present --- a property inherited from the fundamental identity $\sigma_{x} \sigma_{z} = - \sigma_{z} \sigma_{x}$ for the situations yielding a minus sign, and from commutation of $L_j$ and $L_k$ for the other cases ---, we get:
$$ J_1(\rho) = L_1\, ( \sigma_{x,2} \rho \sigma_{x,2} + \sigma_{x,3} \rho \sigma_{x,3} + 2 \rho) + ( \sigma_{x,2} \rho \sigma_{x,2} + \sigma_{x,3} \rho \sigma_{x,3} + 2 \rho)\, L_1 \; ,$$
and $J_2,J_3$ with circular permutation of the indices.
\item Since $\mathfrak{G}_F$ contains the $J_k$, it must contain their commutators. This gives 3 new vector fields of the form:
\begin{eqnarray*}
\tilde{J}_{1}(\rho) &=&  \sigma_{x,1} \sigma_{x,2} \xi \sigma_{x,1} \sigma_{x,2} - \sigma_{x,1} \sigma_{x,3}\xi \sigma_{x,1} \sigma_{x,3} \\
&& \text{with } \xi = L_1 \rho + L_2 \rho L_3 + L_3 \rho L_2 + \rho L_1 \; , 
\end{eqnarray*}
and circular permutation of the indices. The commutators with the $G_{L_k}$ give vector fields:
\begin{eqnarray*}
JG_{1,1}(\rho) &=& (\sigma_{x,2} \rho \sigma_{x,2} + \sigma_{x,3} \rho \sigma_{x,3}) + L_1\, ( \sigma_{x,2} \rho \sigma_{x,2} + \sigma_{x,3} \rho \sigma_{x,3}) \, L_1 - 4 \rho \; , \\
JG_{2,3}(\rho) &=& L_1 \xi + L_2 \xi L_3 + L_3 \xi L_2 + \xi L_1 - 4 \text{Trace}(L_1 \rho) \rho \\
&& \text{with } \xi = \sigma_{x,1} \rho \sigma_{x,1} \; ,
\end{eqnarray*}
for commutation of $G_1$ with $J_1$ and of $G_3$ with $J_2$ respectively; to this we add the permutation of the indices.
\end{itemize}
Checking numerically at various $\rho$ values, we see that these indeed all provide independent directions, such that the manifold has dimension 15 at least, which is the dimension of two coupled qubits. At this point the model reduction appears quite inefficient at best, so we give up the computation of further commutators. \\
\item To describe the above situation in other words: The 3 values of integrated measurement outcomes $y_{t,1}, y_{t,2}, y_{t,3}$ would not be enough anymore to identify the state. Instead, to exactly follow the state, the associated memory effect would have to be modeled with at least 15 variables.

An idea towards efficient \emph{approximate} model reduction, requiring little online adaptation \cite{ZFfilter}, could be to typically drop terms corresponding to higher-order commutators. Indeed, at least on the short run, dissipation can be expected slower on these directions, as they result from accumulating a sequence of motions with different components of the dynamics.\\
\item We conclude by illustrating how the algebra $\mathfrak{G}_F$ is generated \emph{locally}, by considering spurious bit-flips on the first qubit only. Concretely, we consider syndrome measurement channels $L_1$ and  $L_3$ as described in the general model, as well as spurious bit-flips with $L_4 = \sqrt{\Gamma_4} \sigma_{x,1}$ associated to $\eta_4=0$; however, in this toy model, we let $\Gamma_5=\Gamma_6=0$. 
\begin{itemize}
\item We start with the two commuting vector fields $G_{L_1}$ and $G_{L_3}$. There remains to compute their iterated commutators with the deterministic drift.
\item Since $L_1$ commutes with $L_3$ and with any operator on the first qubit, by Proposition \ref{prop:Jacobi}, it will generate no new vector fields.
\item Similarly to $J_1(\rho)$ in the model with spurious bit-flips everywhere, the commutator of $F+D$ with $G_{L_3}$ reduces to
$$J(\rho) = L_3 (L_4 \rho L_4 + \rho) + {\color{gray} G_{L_3}(\rho)} \;\; + h.c. \; . $$
The shaded term is already in $\mathfrak{G}_F$ and can thus be dropped, leaving a linear vector field.
\item By using the properties $L_4 L_3 L_4 = - L_3$ and $L_k^\dagger L_k = (L_k)^2=$ identity, we quickly see that $[F_{L_4}\,,\,J]$ and $[D_{L_3}+F_{L_3}\,,\,J]$ yield no new vector fields. The commutator of $J$ with $G_{L_3}$ yields a new vector field
$$JG(\rho) = L_4 \rho L_4 + L_3 L_4 \rho L_4 L_3 - 2 \rho \; .$$
\item The commutators of $JG$ with all the vector fields encountered so far produces no new vector field, so the algebra is closed at this point and we have $\mathfrak{G}_F=\text{span}\{G_{L_1}\,,\,G_{L_3}\,,\,J\,,\,JG\}$, implying a confining manifold of dimension $M=4$.
\end{itemize}
\end{itemize}


\section{~~~~Appendix 2:  harmonic oscillator quadratures}\label{app2}

This section contains the details associated to Section \ref{sec:HO}.

\subsection{~~~~Full models, Algebraic criterion and Wigner function}\label{app2details}

We first investigate fluorescence measurement, thus with $L_1=\sqrt{\Gamma_1}\,{\bf a}$ and $L_2=i \sqrt{\Gamma_2}\,{\bf a}$. The corresponding SDE, including drives and decoherence in a thermal environment, writes:
\begin{eqnarray}\label{eq:drhoHO}
d\rho_t &=& -i\, [u_t ({\bf a}+{\bf a}^\dag) - iv_t ({\bf a}-{\bf a}^\dag) + \Delta\, {\bf a}^\dagger{\bf a} \, , \; \rho_t ] \, dt \\
\nonumber	&& + 2\, (1+n_{th})\, ({\bf a} \rho_t {\bf a}^\dag - \tfrac{1}{2} {\bf a}^\dag {\bf a} \rho_t - \tfrac{1}{2} \rho_t {\bf a}^\dag {\bf a})\, dt \;\; + 2\, n_{th} ({\bf a}^\dag \rho_t {\bf a} - \tfrac{1}{2}{\bf a}{\bf a}^\dag\, \rho_t - \tfrac{1}{2}\rho_t \, {\bf a}{\bf a}^\dag ) \, dt \\
\nonumber	&& + \sqrt{\eta_1\Gamma_1}({\bf a} \rho_t + \rho_t {\bf a}^\dag - \tr({\bf a}\rho_t + \rho_t {\bf a}^\dag)\rho_t )\, dw_{t,1} \;\; + i\sqrt{\eta_2\Gamma_2}({\bf a} \rho_t - \rho_t {\bf a}^\dag - \tr({\bf a}\rho_t - \rho_t {\bf a}^\dag)\rho_t )\, dw_{t,2}  \; .
\end{eqnarray}
Here $u_t$ and $v_t$ are real control signals and $n_th$ denotes the effective thermal photon number. For greater generality, we have introduced different rates and efficiencies ($\Gamma_k$ and $\eta_k$) for the two channels; assuming units are taken such that $\Gamma_1+\Gamma_2 \leq 2$, we allow for cavity decay beyond the two measurement channels.

Applying the algebraic criterion of Proposition \ref{prop:1} to this setting proceeds as follows.
\begin{itemize}
\item We have $[F_{\bf a}+ D_{\bf a},\, G_{\bf a}] = \frac{1}{2} G_{\bf a}$, so having only the measurement channels should yield one- and two-dimensional manifolds, for the homodyne and heterodyne cases respectively.
\item Including a detuning Hamiltonian proportional to ${\bf N} = {\bf a}^\dag {\bf a}$, its commutator with $G_{\bf a}$ yields $G_{i{\bf a}}$, since $[{\bf a},{\bf N}] = {\bf a}$.  This is understood easily from a physical viewpoint, as a rotation in phase space with Hamiltonian ${\bf N}$ transforms ${\bf a}$ into $i{\bf a}$. Detuning thus has no effect on the dimension $M$ of $\mathcal{M}_t$ in the heterodyne case, while in the homodyne case it adds one dimension.
\item When adding $n_{th}>0$, the corresponding commutation of $F_{{\bf a}^\dagger}$ with $G_{\bf a}$ adds a vector field $G_{{\bf a}^\dag}$  to the algebra $\mathfrak{G}_F$, and commutation with $G_{i {\bf a}}$ yields the vector field $G_{i{\bf a}^\dag}$. Those vector fields yield no new terms under further commutation. I.e., the manifold dimension is doubled and no more.
\item Adding a Hamiltonian proportional to ${\bf X}$ or ${\bf P}$ has no effect on the manifold dimension, in either case. Indeed, e.g.~$[{\bf a}, 2 {\bf X}] = [{\bf a},{\bf a}^\dag]={\bf I}$ and $G_{\alpha {\bf I}}(\rho) = 0$ for all $\rho$ and all $\alpha \in \mathbb{C}$. The same holds when commuting with $G_{i {\bf a}}$, $G_{{\bf a}^\dag}$, or $G_{i{\bf a}^\dag}$, if those are present in $\mathfrak{G}_F$.
\item Incidentally, adding Hamiltonians in ${\bf X}^2$, ${\bf P}^2$ or ${\bf X\, P}$ would preserve the manifold dimension $M=4$.
\end{itemize}
Thus, the measurement outcomes associated to the photon loss channel, in presence of arbitrary drives in ${\bf X}$ or ${\bf P}$, can stochastically move the system in 1 dimension ($\Gamma_2=0$ i.e.~homodyne measurement, and $\Delta=n_{th}=0$); in 2 dimensions ($\Gamma_2=\Delta=0$, with $n_{th}$ nonzero; or $\Gamma_2$ or $\Delta$ nonzero, with $n_{th}=0$);  or in 4 dimensions ($\Gamma_2$ or $\Delta$ nonzero, and $n_{th}$ nonzero).\\

We next investigate simultaneous weak measurement of $L_1=\sqrt{\Gamma_x}\,{\bf X}$ and $L_2=\sqrt{\Gamma_p}\,{\bf P}$, optionally with the same additional dynamics as in \eqref{eq:drhoHO}. Applying the algebraic criterion yields the following results.
\begin{itemize}
\item We have $[G_{\bf X},\, G_{\bf P}] = 0$ because $G_{[{\bf X},\,{\bf P}]} = G_{\alpha {\bf I}}=0$ for any $\alpha \in \mathbb{C}$. From computations done for the QND case, we already know that $[F_{L}+\eta D_{L},\, G_{L}]=0$ for both $L={\bf X}$ or $L={\bf P}$. But we have 
\begin{eqnarray*}
[F_{\bf X}+D_{\bf X},\, G_{\bf P}] =  G_{i{\bf X}} \;\;&,&\;\; [F_{\bf P}+D_{\bf P},\, G_{\bf X}] =  -G_{i{\bf P}} \;\; ,
\end{eqnarray*}
giving two new vector fields in the Lie algebra. The resulting four vector fields commute and further commutation with $\Gamma_x(F_{\bf X}+D_{\bf X}) + \Gamma_p(F_{\bf P}+ D_{\bf P})$ gives no new contributions, so in absence of further dynamics, we have found a four-dimensional $\mathfrak{G}_F$, accounting for a 4-dimensional manifold.
\item Commutators with the Hamiltonian vector fields involving ${\bf X}$, ${\bf P}$, or ${\bf a}^\dagger {\bf a}$, as modeled on the first row of \eqref{eq:drhoHO}, returns the same algebra $\mathfrak{G}_F$.
\item When adding relaxation in a thermal environment, the commutation of $F_{{\bf a}}$ with e.g.~$G_{\bf X}$ yields a linear combination of $G_{\bf X}$ and $G_{\bf a}$, which in turn can be written as a linear combination of $G_{\bf X}$ and $G_{i{\bf P}}$. A similar reasoning holds with $F_{{\bf a}^\dagger}$, which is present when $n_{th}>0$. 
\item Incidentally, adding Hamiltonians in ${\bf X}^2$, ${\bf P}^2$ or ${\bf X\, P}$ would also preserve this $\mathfrak{G}_F$.
\end{itemize}
All these effects can thus be included without pushing the manifold dimension beyond $M=4$.\\[3mm]

In all the models involving harmonic oscillators, we describe the confined state explicitly using its Wigner function representation:
$$\mathcal{W}^{\{\rho\}}(x,p) = \tfrac{2}{\pi}\,\tr((-1)^{\bf a^\dagger a}\; {\bf D}_{-(x+ip)}\,\rho\, {\bf D}_{x+ip} )$$
where ${\bf D}_\alpha = \exp(\alpha {\bf a}^\dag - \alpha^* a)$ is the so-called displacement operator \cite{Haroche}, satisfying ${\bf D}(\alpha)\, {\bf a}\, {\bf D}(\alpha)^\dagger = {\bf a} - \alpha$. The Wigner function is a quasi-distribution; it satisfies 
$\int\int \mathcal{W}^{\{\rho\}}(x,p) \,dx\,dp= 1$ and, although it is not necessarily positive at each $(x,p)$, its marginal $\; P^{\{\rho\}}(x)= \int \mathcal{W}^{\{\rho\}}(x,p)\,dp \; $ gives the probability distribution associated to $\rho$ for the operator ${\bf X} = \tfrac{{\bf a}+{\bf a}^\dag}{2}$, and similarly in rotated bases. The SDE on $\rho$ translates into a stochastic partial differential equation (SPDE) on $\mathcal{W}^{\{\rho\}}(x,p)$, with first and second-oder derivatives with respect to $x$ and $p$. Explicitly, to translate a quantum SDE on $\rho$ to the Wigner function representation, we use the standard properties:
\begin{eqnarray*}
\mathcal{W}^{\{{\bf a}^\dag \rho\}} = \left( (x-ip) - \tfrac{1}{4}(\tfrac{\partial}{\partial x}-i\tfrac{\partial}{\partial p})\right) \mathcal{W}^{\{\rho\}} &,&\;\; \mathcal{W}^{\{{\bf a} \rho\}} = \left( (x+ip) + \tfrac{1}{4}(\tfrac{\partial}{\partial x}+i\tfrac{\partial}{\partial p})\right) \mathcal{W}^{\{\rho\}}\; ,\\
\mathcal{W}^{\{\rho {\bf a}\}} = \left( (x+ip) - \tfrac{1}{4}(\tfrac{\partial}{\partial x}+i\tfrac{\partial}{\partial p})\right) \mathcal{W}^{\{\rho\}} &,& \;\; \mathcal{W}^{\{\rho {\bf a}^\dag\}} = \left( (x-ip) + \tfrac{1}{4}(\tfrac{\partial}{\partial x}-i\tfrac{\partial}{\partial p})\right) \mathcal{W}^{\{\rho\}}\; .
\end{eqnarray*}
For $\eta=0$, the SPDE becomes a linear deterministic PDE that can be solved with a Gaussian, time-varying Green function. Under measurements, the same representation stays rather efficient, although the Gaussian kernel undergoes a somewhat more complicated evolution.


\subsection{~~~~Full solution for measurement of ${\bf a}$ and $i {\bf a}$}\label{ssec:app:aandia}

We now characterize the deterministic manifold corresponding to the model \eqref{eq:drhoHO}, in the slightly particular case $\Gamma_1=\Gamma_2$. This setting also features $\Delta=0$ without loss of generality, since rewriting the model in a frame rotating with the unitary ${\bf U}_t = e^{-i \Delta {\bf a}^\dagger {\bf a}}$ yields back a model of the same type. This implies that the Gaussian kernel is circularly symmetric and separable in $x$ and $p$. We first re-state Proposition \ref{prop:Ameas} with all details.\\

\begin{prop}\label{prop:6}
The quantum SDE \eqref{eq:drhoHO}, with $\Gamma_1=\Gamma_2=1$ and $\Delta=0$, admits the following explicit solution:
$$\mathcal{W}^{ \{\rho\} }_t(x,p) = \int \int \mathcal{W}^{\{\rho\}}_0(x_0,p_0)\, K_{x,t}(x,x_0)\, K_{p,t}(p,p_0) \, dx_0\, dp_0\; \nu_t$$
where $\nu_t$ is a normalization constant and 
\begin{eqnarray*}
K_{x,t}(x,x_0) &=& \exp\left( \frac{-(x-x_0 a_t - \xi_t)^2}{s_t} + (d_t\, x_0^2+\theta_{t}x_0) \right)\;\; , \\
K_{p,t}(p,p_0) &=& \exp\left( \frac{-(p-p_0 a_t - \pi_t)^2}{s_t} + (d_t\, p_0^2+\phi_{t}p_0) \right)\;\;.
\end{eqnarray*}
These functions involve the deterministic time-dependent variables:
\begin{eqnarray*}
a_t &=& \frac{2 \kappa e^{-\kappa t}}{(\kappa+1-\eta) + (\kappa-1+\eta)e^{-2\kappa t}} \\
s_t &=& \frac{-(1-\eta)}{2\eta} + \frac{\kappa}{2\eta} \frac{1-\frac{\kappa-1+\eta}{\kappa+1-\eta} e^{-2\kappa t}}{1+\frac{\kappa-1+\eta}{\kappa+1-\eta} e^{-2\kappa t}}\\
d_{t} &=& \frac{2\eta(e^{-2\kappa t}-1)}{(\kappa+1-\eta) + (\kappa-1+\eta)e^{-2\kappa t}}
\end{eqnarray*}
where $\kappa = \sqrt{1 + 4 \eta\, n_{th}}$. Thus $\mathcal{W}^{\{\rho\}}_t(x,p)$ depends on the measurement record only through the four parameters:
\begin{eqnarray}\label{eq:AiAstochs}
d\xi_t &=& \sqrt{\eta}(s_t-\frac{1}{2})\, dy_{t,1} + \left(v_t - \xi_t - 2 \eta(s_t-\tfrac{1}{2})\, \xi_t \right) \, dt \\
\nonumber d\theta_t &=& 2\sqrt{\eta} a_t\, dy_{t,1} - 4 \eta\, a_t \xi_t \, dt\\
\nonumber d\pi_t &=& \sqrt{\eta}(s_t-\frac{1}{2})\, dy_{t,2} + \left(-u_t - \pi_t - 2 \eta(s_t-\tfrac{1}{2})\, \pi_t \right) \, dt \\
\nonumber d\phi_t &=& 2\sqrt{\eta} a_t\, dy_{t,2} - 4 \eta\, a_t \pi_t \, dt \, ,
\end{eqnarray}
initialized with $\xi_0=\theta_0=\pi_0=\phi_0 = 0$.
\end{prop}
\vspace{4mm}

\noindent \underline{Proof:} We start by converting \eqref{eq:drhoHO} to an SPDE on the Wigner representation:
\begin{eqnarray}
\label{eq:pSDE}
d\mathcal{W} &=& 2 \sqrt{\eta} \left( x + \tfrac{1}{4} \frac{\partial}{\partial x} - \bar{x}_{\mathcal{W}} \right)\, \mathcal{W}\, dw^1_t \;\;\; - 2 \sqrt{\eta} \left(p + \tfrac{1}{4} \frac{\partial}{\partial p} - \bar{p}_{\mathcal{W}} \right)\, \mathcal{W}\, dw^2_t \\
\nonumber && + \left(1 + (x-v_t)\frac{\partial}{\partial x} + \frac{1+2\, n_{th}}{4}\frac{\partial^2}{\partial x^2}\ \right) \, \mathcal{W}\, dt \;\;\; + \left(1 + (p + u_t)\frac{\partial}{\partial p} + \frac{1+2\, n_{th}}{4}\frac{\partial^2}{\partial p^2}\ \right) \, \mathcal{W}\, dt \; ,
\end{eqnarray}
where $\bar{x}_{\mathcal{W}} = \int\int x \mathcal{W}(x,p) \, dx\, dp$ and  $\bar{p}_{\mathcal{W}} = \int\int p \mathcal{W}(x,p) \, dx\, dp$. 

We observe that this equation splits into one part in $x$ and one part in $p$; since those two parts are subject to independent noises, there is no cross-term in the \Ito~correction. We thus try to express the solution with a factorized Green function $K_{x,t} \; K_{p,t}$ as in the statement, and search for an expression of $K_{x,t}$ that satisfies the $x$-part of the equation, for any $p$ and any $V_p(x,x_0) = \int \mathcal{W}^{\{\rho\}}_0(x_0,p_0)\, K_{p,t}(p,p_0)\, dp_0$; the expression of $K_{p,t}$ is then deduced by symmetry.

Using the Gaussian parameterization of $K_{x,t} \sqrt{\nu_t}$ as stated in the Proposition, we compute an expression for the right-hand side of \eqref{eq:pSDE} as a function of the parameters $a_t,s_t,d_t,\xi_t,\theta_t$ and $\sqrt{\nu_t}$; note that $\bar{x}(\mathcal{W})$ here becomes just a particular function of the parameters and of $\mathcal{W}^{\{\rho\}}_0$.
The left-hand side is expressed with the time variations $da_t,...,\, d\theta_t,\, d\nu_t$, including their squares as required by \Ito~calculus. Equating the same powers of $x$ and $x_0$ inside the integrals on the left- and right-hand side, and the terms in $dt$ or $dw^1_t$ respectively, yields a set of equations which turn out to have a solution. 

One first checks, equating the terms proportional to $dw^1_t$ and to various powers of $x,x_0$, that $da_t,\, ds_t,\, dd_t$ should involve no noise term, while $d\xi_t$, $d\phi_t$ and $d\nu_t$ include a noise term. Next, from the $x^2\, dt$ term, we get a nonlinear ODE for $s_t$:
$$\frac{ds_t}{dt} + 2 \eta (s_t-\tfrac{1}{2})^2 = -2 s_t + (1+2 n_{th})\, .$$
This is in Ricatti form and can be integrated with standard methods, e.g.~relating $s_t$ to $\frac{\dot{u}_t}{u_t}$ where $u_t$ is a mathematical auxiliary signal. After thus solving this equation for $s_t$, we get similar first-order ODEs for $a_t$ (containing $s_t$) and for $d_t$ (containing $a_t$) respectively from the term in $x x_0$ and from the term in $x_0^2$. Expressing the ODE for $a_t$ with the Ricatti auxiliary signal $u_t$ from the $s_t$ equation yields a quick solution, while the $d_t$ equation admits explicit integration. Finally, the terms in $x$ and $x_0$ yield the SDEs for $\xi_t$ and $\theta_t$ given in the Proposition statement. We do not need to compute the SDE for $\nu_t$ since we know beforehand that, as a normalization constant, it will have an expression as a function of the other parameters and of $\mathcal{W}^{\{\rho\}}_0(x_0,p_0)$. It was important however to include it in the equations in order to compute the \Ito~correction correctly. The requirement $\mathcal{W}^{\{\rho\}}_0(x,p) = \int \int \mathcal{W}^{\{\rho\}}_0(x_0,p_0)\, K_{x,0}(x,x_0)\, K_{p,0}(p,p_0) \, dx_0\, dp_0\; \nu_0$ allows to fix the initial values of all the parameters' differential equations.
 
Concerning $K_{p,t}(p,p_0)$, since the equations for the deterministic variables are exactly the same, we can use the same parameters $a_t,s_t,d_t$; the stochastic variables of course must be taken different, since they will evolve according to different Wiener processes. \hfill $\square$\\

Let us discuss a few alternative scenarios.
\begin{itemize}
\item For $\Gamma_1 \neq \Gamma_2$ but $\Delta=0$, the kernel remains separable in $x$ and $p$, but with different parameters, which can be easily computed along the same lines.
\item For $\Gamma_1 \neq \Gamma_2$ and $\Delta \neq 0$, the manifold dimension remains $M=4$ but $K_t(x,x_0,p,p_0)$ is not separable in $x$ and $p$ anymore. We did not compute it in full here, but its expressions for the case $n_{th}=0$ can be directly deduced from the solutions provided in Section \ref{app:ssec:PRmod}.
\item For $n_{th}=0$ in the setting of Proposition \ref{prop:6}, the manifold dimensions reduces to $M=2$, with the following simplifications in the explicit solution.
\begin{cor}\label{prop:6specialcase}
For $n_{th}=0$ in the setting of Proposition \ref{prop:6}, the set of equations \eqref{eq:AiAstochs} becomes equivalent to:
\begin{eqnarray*}
d\xi_t &=& \sqrt{\eta}(s_t-\frac{1}{2})\, dy_{t,1} + \left(v_t - \xi_t - 2 \eta(s_t-\tfrac{1}{2})\, \xi_t \right) \, dt \\
d\pi_t &=& \sqrt{\eta}(s_t-\frac{1}{2})\, dy_{t,2} + \left(-u_t - \pi_t - 2 \eta(s_t-\tfrac{1}{2})\, \pi_t \right) \, dt \\
\frac{dz_t}{dt}&=& v_t - z_t  \quad \text{ for } z_t= \xi_t + \tfrac{e^{-t}}{4}\, \theta_t \\
\frac{dh_t}{dt}&=& - u_t - h_t \quad \text{ for } h_t= \pi_t + \tfrac{e^{-t}}{4}\, \phi_t \, .
\end{eqnarray*}
This leaves only two independent stochastic variables.\\
\end{cor}
\noindent \underline{Proof:} In the Wigner representation kernel $K_{x,t}(x,x_0)$, the number of stochastic variables is thus supposed to decrease from 2 to 1. 

Like for any SDE, this can be directly checked with the algebraic criterion. We thus apply the algebraic criterion to the classical SDE governing the evolution of the 3-dimensional system $\xi_t,\theta_t$ and $t$ in \eqref{eq:AiAstochs}. The variable $t$ must be explicitly included since the expressions are all explicitly time-dependent through $a_t,s_t,d_t$. The Stratonovich form is identical to the \Ito~form, because the Wiener vector field only depends on $t$, which itself has no stochastic evolution ($dt = 1 \cdot dt + 0 \cdot dw^1_t$).
Then indeed, the commutator of the deterministic vector field with the Wiener vector field turns out to be proportional to the Wiener vector field itself (no new direction), if and only if $n_{th}=0$. In other words, the variables $\xi_t,\theta_t$ remain confined to a one-dimensional manifold. For $n_{th}>0$ a new component appears in the commutator, confirming that in this case the pair of variables $(\xi_t,\theta_t)$ driven by a single Wiener process $dw^1_t$, can diffuse to span the entire plane at any given time $t$, and can thus not be further reduced.

We can parameterize the time-dependent curve to which $\xi_t,\theta_t$ are confined for $n_{th}=0$, by saying that a combination $z_t(\xi_t,\theta_t,t)$ of the variables (not reduced to $z_t=t$) must evolve deterministically. Canceling the contribution of $dw^1_t$ in $dz$ leads to the constraint $\tfrac{\partial z_t}{\partial \xi} = 4 e^t\, \tfrac{\partial z_t}{\partial \theta}$.  This suggests the solution $z_t = 4 \xi_t + e^{-t} \theta_t$. We further check that the evolution of such $z_t$ is deterministic, i.e. the right-hand side of
$$\tfrac{d}{dt}z_t(\xi_t,\theta_t,t) = 4 e^t\, \tfrac{\partial z_t}{\partial \theta}\, (v_t - \xi_t) + \tfrac{\partial z_t}{\partial t}\; ,$$
depends only on $z_t$ and on $t$. Of course this solution is far from unique as any function of $z_t$ and $t$ would also be a solution.
\hfill $\square$\\
\end{itemize}
\vspace{4mm}


\subsection{~~~~Full solution for measurement of ${\bf X}$ and ${\bf P}$}
\label{sssec:app:XandPmeas}

We consider a general setting, with unequal rates on the two measurement channels, a detuning Hamiltonian $\Delta {\bf a}^\dagger {\bf a}$, control drives on Hamiltonians ${\bf X}$ and ${\bf P}$, and relaxation in a thermal environment.  Defining $\tilde{\bf X}_t = \cos(\Delta t)\, {\bf X} + \sin(\Delta t)\, {\bf P}$ and $\tilde{\bf P}_t = \cos(\Delta t)\, {\bf P} - \sin(\Delta t)\, {\bf X}$ in a rotating frame with respect to $H_0 = \Delta {\bf a}^\dagger {\bf a}$, and modulo a similar redefinition of the two control signals and the two Wiener processes, we then get the model:
\begin{eqnarray}\label{eq:drhoXP}
d\rho_t &=& -i\, [u_t ({\bf a}+{\bf a}^\dag) - iv_t ({\bf a}-{\bf a}^\dag) \, , \; \rho_t ] \, dt \\
\nonumber	&& + \Gamma_\ell\, (1+n_{th})\, ({\bf a} \rho_t {\bf a}^\dag - \tfrac{1}{2} {\bf N} \rho_t - \tfrac{1}{2} \rho_t {\bf N})\, dt \;\;  + \Gamma_\ell\, n_{th} ({\bf a}^\dag \rho_t {\bf a} - \tfrac{1}{2} ({\bf N+I})\rho_t - \tfrac{1}{2}\rho_t({\bf N+I})) \, dt \\
\nonumber	&& + \Gamma_x (\tilde{\bf X}_t \rho_t \tilde{\bf X}_t - \tfrac{1}{2} (\tilde{\bf X}_t)^2\, \rho_t - \tfrac{1}{2}\rho_t\, (\tilde{\bf X}_t)^2) \, dt \;\; + \Gamma_p (\tilde{\bf P}_t \rho_t \tilde{\bf P}_t - \tfrac{1}{2} (\tilde{\bf P}_t)^2\, \rho_t - \tfrac{1}{2}\rho_t\, (\tilde{\bf P}_t)^2) \, dt \\
\nonumber	&& + \sqrt{\Gamma_x \eta_x}(\tilde{\bf X}_t \rho_t + \rho_t \tilde{\bf X}_t - \tr(2 \tilde{\bf X}_t\rho_t )\rho_t )\, dw_{t,1} \;\; + \sqrt{\Gamma_p \eta_p}(\tilde{\bf P}_t \rho_t - \rho_t \tilde{\bf P}_t - \tr(2 \tilde{\bf P}_t\rho_t )\rho_t )\, dw_{t,2}  \; .
\end{eqnarray}
The deterministic manifold corresponding to this model is characterized as follows.\\

\begin{prop}\label{prop:7}
The quantum SDE \eqref{eq:drhoXP} admits the following explicit solution:
$$\mathcal{W}^{\{\rho\}}_t(x,p) = \int \int \mathcal{W}^{\{\rho\}}_0(x_0,p_0)\, K_{t}(x,x_0,p,p_0)\, dx_0\, dp_0\; \nu_t$$
where $\nu_t$ is a normalization constant and 
\begin{eqnarray*}
K_{t}(x,x_0,p,p_0) &=& \exp\big[ -J_t^T (S_t)^{-1} J_t + (q_0)^T Z_t q_0 + (\Lambda_t)^T (\, A_t \;,\; R_t \,)\, q_0 \big]\; ,
\end{eqnarray*}
with column vectors $q = (x \,; \; p)$, $q_0 = (x_0 \,; \; p_0)$ and $ J_t = (q-A_t x_0 - R_t p_0 - \Theta_t) \in \mathbb{R}^2$. The symmetric matrices $S_t, F_t \in \mathbb{R}^{2\times 2}$ evolve deterministically as:
\begin{eqnarray*}
\tfrac{d}{dt}S_t &=& -\Gamma_\ell\, S_t + \tfrac{(1+2n_{th}) \Gamma_\ell}{2} I - 2 S_t \, M_t \, S_t + \tfrac{1}{2}N_t \\
\tfrac{d}{dt}Z_t &=& - 2 \, \left(\begin{array}{c}
(A_t)^T \\ (R_t)^T
\end{array} \right) \, M_t \, \left(\begin{array}{cc}
A_t \, ,& R_t
\end{array} \right)
\\
&& \text{with }\;\; M_t = \left(\begin{array}{cc}
\eta_x \Gamma_x \cos^2(\Delta t) + \eta_p \Gamma_p \sin^2(\Delta t) & (\eta_x \Gamma_x - \eta_p \Gamma_p) \sin\cos(\Delta t) \\
(\eta_x \Gamma_x - \eta_p \Gamma_p) \sin\cos(\Delta t) & \eta_p \Gamma_p \cos^2(\Delta t) + \eta_x \Gamma_x \sin^2(\Delta t)
\end{array}\right) \\
&& \text{and } \;\;\;\; N_t = \left(\begin{array}{cc}
\Gamma_p \cos^2(\Delta t) + \Gamma_x \sin^2(\Delta t) & (\Gamma_p - \Gamma_x) \sin\cos(\Delta t) \\
(\Gamma_p - \Gamma_x) \sin\cos(\Delta t) & \Gamma_x \cos^2(\Delta t) + \Gamma_p \sin^2(\Delta t)
\end{array}\right) \; ,
\end{eqnarray*}
starting from initial conditions $S_0 = Z_0 = 0$. The column vectors $A_t,\; R_t \in \mathbb{R}^2$ follow the independent linear ODEs:
\begin{eqnarray*}
\tfrac{d}{dt} A_t & = & \tfrac{-\Gamma_l}{2} A_t - 2 S_t \, M_t \, A_t \\
\tfrac{d}{dt} R_t & = & \tfrac{-\Gamma_l}{2} R_t - 2 S_t \, M_t \, R_t \; ,
\end{eqnarray*}
starting from initial conditions $A_0 = (\, 1 \; ; \; 0 \,)$ and $R_0 = ( \, 0 \; ; \; 1 \, )$. Thus $\mathcal{W}^{ \{ \rho\} }_t(x,p)$ depends on the measurement record only through 2 column vectors in $\mathbb{R}^2$, which follow the linear, block-triangular set of SDEs:
\begin{eqnarray*}
d\Theta_t &=& \left( \tfrac{-\Gamma_\ell}{2} \Theta_t + \left(\begin{array}{c}
v_t \\ -u_t
\end{array} \right) - 2 S_t\, M_t \, \Theta_t \right) \, dt \\
&& + S_t \, \left(\begin{array}{c}
\sqrt{\eta_x \Gamma_x} \cos(\Delta t)\, dy_{t,1} - \sqrt{\eta_p \Gamma_p} \sin(\Delta t)\, dy_{t,2} \\
\sqrt{\eta_x \Gamma_x} \sin(\Delta t)\, dy_{t,1} + \sqrt{\eta_p \Gamma_p} \cos(\Delta t)\, dy_{t,2}
\end{array} \right) \\[2mm]
d\Lambda_t &=& \left( \tfrac{\Gamma_\ell}{2} \Lambda_t + 2 M_t \, S_t \Lambda_t - 4 M_t\, \Theta_t \right) \, dt \\
&& + 2\, \left(\begin{array}{c}
\sqrt{\eta_x \Gamma_x} \cos(\Delta t)\, dy_{t,1} - \sqrt{\eta_p \Gamma_p} \sin(\Delta t)\, dy_{t,2} \\
\sqrt{\eta_x \Gamma_x} \sin(\Delta t)\, dy_{t,1} + \sqrt{\eta_p \Gamma_p} \cos(\Delta t)\, dy_{t,2}
\end{array} \right)
\; ,
\end{eqnarray*}
starting from initial conditions $\Theta_t = \Lambda_t = 0$. This SDE on $\mathbb{R}^4$, driven by 2 noise processes, leads in general to diffusion in all 4 dimensions at any given time $t>0$.
\end{prop}\vspace{2mm}

\noindent \underline{Proof:}  The dynamics \eqref{eq:drhoXP} translate to the Wigner function format as follows:
\begin{eqnarray}
\label{eq:pSDE-XP}
d\mathcal{W} &=& 2 \sqrt{\eta_x\Gamma_x} \left( 2(x - \bar{x}(\mathcal{W}))\cos(\Delta t) + 2(p - \bar{p}(\mathcal{W}))\sin(\Delta t) \right) \mathcal{W}\, dw^1_t \\
\nonumber && + \sqrt{\eta_p\Gamma_p} \left( 2(p - \bar{p}(\mathcal{W}))\cos(\Delta t) - 2(x - \bar{x}(\mathcal{W}))\sin(\Delta t) \right) \mathcal{W}\, dw^2_t\\
\nonumber && + \frac{\Gamma_\ell}{2} \left(2 + x \tfrac{\partial}{\partial x} + p\tfrac{\partial}{\partial p} + \tfrac{1 + 2 n_{th}}{4} (\tfrac{\partial^2}{\partial x^2} + \tfrac{\partial^2}{\partial p^2})  \right) \, \mathcal{W} \, dt \\
\nonumber && + \left( -v_t\, \tfrac{\partial}{\partial x} + u_t\, \tfrac{\partial}{\partial p} \right) \, \mathcal{W} \, dt \\
\nonumber && + (\tfrac{\Gamma_x \cos^2(\Delta t)}{8} + \tfrac{\Gamma_p \sin^2(\Delta t)}{8}) \tfrac{\partial^2}{\partial p^2} \mathcal{W} \, dt + (\tfrac{\Gamma_x \sin^2(\Delta t)}{8} + \tfrac{\Gamma_p \cos^2(\Delta t)}{8}) \tfrac{\partial^2}{\partial x^2} \mathcal{W} \, dt \\
\nonumber && + \tfrac{\Gamma_p-\Gamma_x}{4} \sin\cos(\Delta t) \tfrac{\partial^2}{\partial x \, \partial p} \mathcal{W} \, dt \; ,
\end{eqnarray}
where again $\bar{x}(\mathcal{W}) = \int\int x \mathcal{W}(x,p) \, dx\, dp\;$ and  $\; \bar{p}(\mathcal{W}) = \int\int p \mathcal{W}(x,p) \, dx\, dp$. 

This SPDE contains a coupling between $x$ and $p$ variables through the term proportional to $\tfrac{\partial^2}{\partial x \partial p}$, expressing that there is a rotation at rate $\Delta$ between the diffusion principal axes and our coordinate frame. As a consequence, the equation does not separate in $x$ and $p$, so we introduce the joint Green function $K_t(x,x_0,p,p_0)$. Similarly to Proposition \ref{prop:6}, we then express $d\mathcal{W}$ on the left-hand side of \eqref{eq:pSDE-XP} through variations in the parameters of $K_t(x,x_0,p,p_0)$, and compute the right-hand side of \eqref{eq:pSDE-XP} explicitly with the Ansatz postulated in Proposition \ref{prop:7}. For the left-hand side, we must go up to second order (\Ito~correction) for the parameters whose evolution would involve a noise term. Equating the same powers of $x,\,p,\,p_0$ and $x_0$ inside the integrals on the left and right side, and the terms in $dw^1_t,\, dw^2_t$ or $dt$ respectively, leads to the set of equations that we have to solve.

We first check the matching of terms in $dw^1_t,\, dw^2_t$. Starting with the $x^2$, $p^2$ and $xp$ terms we see that $dS_t$ must be proportional to $dt$, with no component proportional to $dw^1_t,\, dw^2_t$. This also simplifies the other equations, as there is no second-order \Ito~term from $dS_t$. Then considering terms in $xx_0$, $xp_0$, $px_0$ and $pp_0$ yields the same conclusion for $A_t, R_t$. After this, the terms in $x_0^2$, $p_0^2$ and $x_0 p_0$ give the same conclusion for $Z_t$. Next, the terms proportional to $dy_{1,t}$ and $dy_{2,t}$ in the equations for $\Theta_t$ are obtained from the terms in $x,\,p$. From there the contributions to $\Lambda_t$ are obtained with the terms in $x_0,\,p_0$. Finally, the term proportional to $1$ yields the contribution to the normalizing constant $\nu_t$, which must just be taken into account for the \Ito~correction.

After this, we consider the terms proportional to $dt$, to obtain the deterministic contributions to the equations. Only few parameters contribute to the second-order \Ito~correction. We consider the coefficients in the same order as for the noise investigation, starting with $x^2$ and so on. It helps to use $Q_t = (S_t)^{-1}$ as an intermediate variable. One can check, using the algebraic criterion on the vector fields for the system $\Theta_t,\Lambda_t$, that in general it indeed yields diffusion in all 4 dimensions and thus cannot be reduced further. \hfill $\square$\\[3mm]

We can explicitly write down the result for some special cases where the equations simplify.\\

For $\Delta=0$, the $x$ and $p$ variables can again be separated. For $\Gamma_x = \Gamma_p$ and $\eta_x = \eta_p$, the model is invariant under a rotation of phase space; in this situation it is equivalent to set $\Delta=0$ as well. In these cases, we get the following simpler explicit solution, although the manifold remains of dimension $M=4$.

\begin{cor}\label{cor:71}
When $\Delta=0$ or $(\Gamma_x\, ,\, \eta_x) = (\Gamma_p\, ,\, \eta_p)$, 
the solution from Proposition \ref{prop:7} can be integrated as follows:
\begin{itemize}
\item[$\bullet$] $S_t=\text{diag}(\,s_{1,t}\; , \; s_{2,t}\,)$ and $Z_t=\text{diag}(\,z_{1,t}\; , \; z_{2,t}\,)$ are diagonal. Defining \linebreak $\kappa_1 = \sqrt{\tfrac{\Gamma_\ell^2}{4}+\tfrac{\Gamma_\ell(1+2n_{th}+\Gamma_p)}{2}}$ and $\kappa_2 = \sqrt{\tfrac{\Gamma_\ell^2}{4}+\tfrac{\Gamma_\ell(1+2n_{th}+\Gamma_x)}{2}}$, we have
\begin{eqnarray*}
s_{1,t} &=& \tfrac{\Gamma_\ell(1+2n_{th}+\Gamma_p)}{4 \eta_x \Gamma_x}\; \frac{e^{\kappa_1 t} - e^{-\kappa_1 t}}{(\kappa_1+\Gamma_\ell/2) e^{\kappa_1 t} + (\kappa_1-\Gamma_\ell/2) e^{-\kappa_1 t}}\\
s_{2,t} &=& \tfrac{\Gamma_\ell(1+2n_{th}+\Gamma_x)}{4 \eta_p \Gamma_p}\; \frac{e^{\kappa_2 t} - e^{-\kappa_2 t}}{(\kappa_2+\Gamma_\ell/2) e^{\kappa_2 t} + (\kappa_2-\Gamma_\ell/2) e^{-\kappa_2 t}} \;\; ,  \\
z_{1,t} &=& \frac{2 \eta_x \Gamma_x (e^{-2 \kappa_1 t}-1)}{(\kappa_1+\Gamma_\ell/2) + (\kappa_1-\Gamma_\ell/2) e^{-2\kappa_1 t}} \\
z_{2,t} &=& \frac{2 \eta_p \Gamma_p (e^{-2 \kappa_2 t}-1)}{(\kappa_2+\Gamma_\ell/2) + (\kappa_2-\Gamma_\ell/2) e^{-2\kappa_2 t}}  \; .
\end{eqnarray*}
\item[$\bullet$] The vectors $A_t$ and $R_t$ $\in \mathbb{R}^2$ feature $a_{2,t} = r_{1,t} = 0$, while
\begin{eqnarray*}
a_{1,t} &=& \frac{2 \kappa_1\, e^{-\kappa_1 t}}{(\kappa_1+\Gamma_\ell/2) + (\kappa_1-\Gamma_\ell/2) e^{-2\kappa_1 t}}\\
r_{2,t} &=& \frac{2 \kappa_2\, e^{-\kappa_2 t}}{(\kappa_2+\Gamma_\ell/2) + (\kappa_2-\Gamma_\ell/2) e^{-2\kappa_2 t}} \, .
\end{eqnarray*}
\item[$\bullet$] The stochastic variables decouple into two independent sets of equations for $(\theta_{1,t}\;,\;\lambda_{1,t})$ and for $(\theta_{2,t}\;,\;\lambda_{2,t})$ respectively.
\end{itemize}
\end{cor}

\noindent \underline{Proof:} The term in $\tfrac{\partial^2}{\partial x \partial p}$ vanishes in the SPDE of the Wigner function, thus in fact the $x$ and $p$ variables can effectively be separated, like in Proposition \ref{prop:6}. In the expressoins of Proposition \ref{prop:7}, this translates into decoupled equations. We first note that the matrices $M_t$ and $N_t$ become diagonal. From this and the initial condition $S_0=0$, we have that $S_t$ is diagonal. For $\eta_x \Gamma_x\, \eta_p \Gamma_p \neq 0$, the equations for the diagonal elements of $S_t$ take a Ricatti form which can be integrated. As a consequence, the elements of $A_t$ and $R_t$ follow 4 independent equations. From the initial conditions, the matrix $(A_t \; ; \; R_t)$ is diagonal and the components can be integrated once knowing $S_t$. Then a similar procedure allows to integrate $Z_t$, leading to the result. \hfill $\square$\\

By setting $\Delta=0$ and $\Gamma_p=0$ in \eqref{eq:drhoXP}, we recover a connection with Section \ref{sec:QND}, since we obtain a QND measurement of ${\bf X}$, but now in the presence of thermal relaxation and drives. The corresponding particular case of Proposition \ref{prop:7} yields the explicit $M=2$ solution presented below. For $\Delta \neq 0$, the algebraic criterion confirms the physical intuition that $M$ returns to $4$.\\
\begin{cor}\label{cor:72}
For $\eta_p \Gamma_p = 0$ and $\Delta=0$, the dependence of the solution of Proposition \ref{prop:7} on the measurement signal reduces to a 2-dimensional SDE, whereas we get the following particular deterministic parameter evolutions:
\begin{eqnarray*}
s_{2,t} &=& \tfrac{\Gamma_x + \Gamma_\ell(1+2n_{th})}{2 \Gamma_\ell}\,(1 - e^{-\Gamma_\ell t}) \; , \qquad
r_{2,t} = e^{-\Gamma_\ell t/2} \; , \qquad z_{2,t} \;=\; 0 \; ,  \qquad
\theta_{2,t} = \int_0^t -u_s\, e^{-(t-s)\Gamma_\ell/2} \, ds \; , \quad \lambda_{2,t} = 0 \; .
\end{eqnarray*}
The other parameters are obtained by plugging $\eta_p \Gamma_p = 0$ into Corollary \ref{cor:71} or Proposition \ref{prop:7}.
\end{cor}

\noindent \underline{Proof:} The proof is similar to the previous cases. Compared to Corollary \ref{cor:71}, the equation for $s_{1,t}$ does not take a Ricatti form anymore but it becomes affine. This yields the different form for the explicit solution. \hfill $\square$\\


\section{~~~~Appendix 3:  bipartite quantum systems}\label{app3}

This section contains the details associated to Section \ref{sec:bipar}.

\subsection{~~~~Indirect qudit measurement}\label{ssec:app:iqm}

This section provides the details for the setting of \eqref{eq:couplex1}, where a qudit is measured indirectly through its dispersive coupling to a monitored harmonic oscillator. This example is further generalized to $n$ harmonic oscillators in Section \ref{app:ssec:PRmod}. We start by detailing the investigation with the algebraic criterion, before giving the full expression of the associated manifold. The results are given for a non-degenerate coupling operator $Q_A = \sum_{s=1}^d \, \lambda_s \, \q{s}\qd{s}$ ; generalizing to a degenerate one involves no particular complications.\\

\begin{prop}\label{prop:10}
The system \eqref{eq:couplex1} features deterministic manifolds $\mathcal{M}_t$ of dimension $4 d - 2$, characterized by the Abelian algebra:
$$\mathfrak{G}_F = \text{span}\{G_{\q{s}\qd{s} \otimes \delta {\bf b}}  : s=1,2,...,d;\; \delta =1,i \} \cup \{ G_{\q{s}\qd{s} \otimes \delta {\bf I}} : s=1,2,...,d-1;\; \delta =1,i \} \; .$$
In absence of drives, thus for $u_t=v_t=0$, the manifold further reduces to dimension 2d, characterized by the Abelian algebra:
$$\mathfrak{G}_F = \text{span}\{G_{\q{s}\qd{s} \otimes \delta {\bf b}}  : s=1,2,...,d;\; \delta =1,i \} \; .$$
\end{prop}\vspace{2mm}

\noindent \underline{Proof:} Note that we check commutation with each of the deterministic vector fields independently, as they involve a priori independent scales and should thus allow to span their respective directions independently.
\begin{itemize}
\item[-] The operators acting on B alone give an algebra 
$\mathfrak{G}_B = \text{span}\{\, G_{\bf b}\,,\; G_{i{\bf b}} \, \} \; .$
\item[-] The commutation with $G_{-i\,Q_A \otimes ({\bf b}^\dag {\bf b})}$ yields two new vector fields 
$$ G_{Q_A \otimes {\bf b}} \;\;\; \text{ and } \;\;\; G_{iQ_A \otimes {\bf b}} \; .$$
Those commute with the elements of $\mathfrak{G}_B$, we just have to check further commutation with the terms of the deterministic vector field. Commutation with $F_{\bf b} = F_{i{\bf b}}$ yields no new term. Commutation with the drives yields 
$$ G_{Q_A \otimes {\bf I}} \;\;\; \text{ and } \;\;\; G_{iQ_A \otimes {\bf I}} \; .$$
\item[-] From here we can proceed by induction. Assume that we have the Abelian algebra of vector fields:
$$\mathfrak{G}_{\bar{s}} := \text{span}\{ G_{Q_A^{s-1} \otimes {\bf b}} \; ,\; G_{i Q_A^{s-1} \otimes {\bf b}} \; ,\; G_{Q_A^{s-1} \otimes {\bf I}} \; ,\; G_{i Q_A^{s-1} \otimes {\bf I}} \;\; : \;\; s=1,2,...,\bar{s} \} \; . $$
Then one commutation with the vector fields associated to the drive Hamiltonian, the coupling Hamiltonian and $F_{\bf b} + D_{\bf b}$ yields the Abelian algebra $\; \mathfrak{G}_{\overline{s+1}} \; $. Similarly, in absence of drives, consider the Abelian algebra of vector fields:
$$\tilde{\mathfrak{G}}_{\bar{s}} := \text{span}\{ G_{Q_A^{s-1} \otimes {\bf b}} \; ,\; G_{i Q_A^{s-1} \otimes {\bf b}}  \;\; : \;\; s=1,2,...,\bar{s} \} \; . $$
Then one commutation with the vector field associated to the coupling Hamiltonian and $F_{\bf b} + D_{\bf b}$ yields the Abelian algebra $\; \tilde{\mathfrak{G}}_{\overline{s+1}} \; $. 

The iteration stops once $\mathfrak{G}_{\overline{s+1}} = \mathfrak{G}_{\bar{s}} \; $, which for a $d$-dimensional (non-degenerate) Hermitian matrix $Q_A$ holds once $\bar{s} \geq d$.
\end{itemize}
From here the result follows by equivalence of $\text{span}\{ G_{Q_A^s \otimes {\bf Q}} : s=1,2,...,d \}$ and $\text{span}\{ G_{\q{s}\qd{s} \otimes {\bf Q}} : s=1,2,...,d \}$ for any operator ${\bf Q}$. For ${\bf Q} = \alpha {\bf I}$ a complex multiple of identity, since $G_{\alpha I \otimes {\bf I}} = 0$, we further have $\text{span}\{ G_{\q{s}\qd{s} \otimes \alpha{\bf I}} : s=1,2,...,d \} = \text{span}\{ G_{\q{s}\qd{s} \otimes \alpha {\bf I}} : s=1,2,...,d-1 \}$.
\hfill $\square$\\

Although the control signals are deterministic, they can a priori take any forms, and therefore the following result will further motivate the Ansatz for writing the actual manifold equations.

\begin{prop}\label{prop:11}
The set of states reachable by system \eqref{eq:couplex1}, under all measurement realizations and all control signals $u_t,v_t$, is confined to a time-dependent manifold $\tilde{\mathcal{M}}_t$ of dimension $3 d^2 +2d - 1$, characterized by the algebra:
\begin{eqnarray}
\label{eq:bipar1contalgcrit}
\mathfrak{G}_F &=& \text{span}\{G_{\q{s}\qd{s} \otimes {\bf I}} : s=1,2,...,d-1 \} \\ 
\nonumber &&  \cup \{ G_{\q{s}\qd{s} \otimes {\bf b}},\; G_{\q{s}\qd{s} \otimes i{\bf b}} : s=1,2,...,d \} \\ 
\nonumber &&  \cup \{ G_{\q{s}\qd{s} \otimes {\bf b}^\dagger},\; G_{\q{s}\qd{s} \otimes i{\bf b}^\dagger} : s=1,2,...,d \} \\ 
\nonumber &&  \cup \{ (\q{s}\qd{s} \otimes {\bf I}) \rho (\q{j}\qd{j} \otimes {\bf I}) + (\q{j}\qd{j} \otimes {\bf I}) \rho (\q{s}\qd{s} \otimes {\bf I}) \;\; : s,j=1,2,...,d \;,\;\; j<s  \} \\
\nonumber &&  \cup \{ i(\q{s}\qd{s} \otimes {\bf I}) \rho (\q{j}\qd{j} \otimes {\bf I}) - i(\q{j}\qd{j} \otimes {\bf I}) \rho (\q{s}\qd{s} \otimes {\bf I}) \;\; : s,j=1,2,...,d \;,\;\; j<s  \} \\ 
\nonumber &&  \cup \{ (\q{s}\qd{s} \otimes {\bf b}) \rho (\q{j}\qd{j} \otimes {\bf I}) + (\q{j}\qd{j} \otimes {\bf I}) \rho (\q{s}\qd{s} \otimes {\bf b}^\dagger) \;\; : s,j=1,2,...,d \;,\;\; j \neq s \} \\
\nonumber &&  \cup \{ i(\q{s}\qd{s} \otimes {\bf b}) \rho (\q{j}\qd{j} \otimes {\bf I}) - i(\q{j}\qd{j} \otimes {\bf I}) \rho (\q{s}\qd{s} \otimes {\bf b}^\dagger) \;\; : s,j=1,2,...,d \;,\;\; j \neq s  \} \; .
\end{eqnarray}
\end{prop}

\noindent \underline{Proof:} We must find the smallest algebra $\tilde{\mathfrak{G}}_F$ generated by the vector fields $G_{{\bf I} \otimes {\bf b}}$, $G_{{\bf I} \otimes i{\bf b}}$ (measurements) and $G_{{\bf I} \otimes {\bf b}^\dagger}$, $G_{{\bf I} \otimes i{\bf b}^\dagger}$ (independent linear combination of measurement and control vector fields), and commuting with $F_{{\bf I} \otimes {\bf b}} + D_{{\bf I} \otimes {\bf b}}$ as well as with the coupling Hamiltonian.

The first three lines of \eqref{eq:bipar1contalgcrit} are obtained similarly to the proof of Prop.\ref{prop:10}. Note that we have dropped the vector fields of the form $G_{\q{s}\qd{s} \otimes i{\bf I}}$ in the first line, because they are included in the set described in the fifth line. The new contributions are obtained as follows.
\begin{itemize}
\item[-] Commuting $G_{Q_A^s \otimes \delta{\bf b}^\dagger}$ with $F_{{\bf I} \otimes {\bf b}} + D_{{\bf I} \otimes {\bf b}}$ yields a vector field of the form
$$(Q_A^s\otimes {\bf I}) \rho ({\bf I} \otimes \delta{\bf b}^\dagger) - \text{Trace}((Q_A^s\otimes {\bf I}) \rho ({\bf I} \otimes \delta{\bf b}^\dagger)) \;\; + \text{\emph{hermit.conj.}} \; .$$
Taking a linear combination with previously constructed vector fields, we can rewrite it as (dropping the ${\bf I}$ symbols to avoid cluttered notation):
$$Q_A^s \rho \delta{\bf b}^\dagger + \delta^* {\bf b} \rho Q_A^s - (Q_A^s \otimes \delta^*{\bf b}) \rho -\rho (Q_A^s \otimes  \delta {\bf b}^\dagger) \; .$$
\item[-] Further repeated commutations with the third line of \eqref{eq:bipar1contalgcrit}, the Hamiltonian coupling, and $F_{{\bf I} \otimes {\bf b}} + D_{{\bf I} \otimes {\bf b}}$ yields all the vector fields of the form
\begin{eqnarray*}
&& Q_A^m \rho Q_A^k + Q_A^k \rho Q_A^m - Q_A^{m+k} \rho - \rho Q_A^{m+k}\\
&\text{ or } & i (Q_A^m \rho Q_A^k - Q_A^k \rho Q_A^m)\\
&\text{ or } & {\bf b} Q_A^m \rho Q_A^k + Q_A^k \rho Q_A^m {\bf b}^\dagger - {\bf b} Q_A^{m+k} \rho - \rho Q_A^{m+k} {\bf b}^\dagger \; ,
\end{eqnarray*}
as well as the ones obtained by replacing ${\bf b}$ by $i{\bf b}$. 
\item[-] One can now check that these vector fields form a closed algebra under mutual commutation, and under further commutation with the coupling Hamiltonian or with $F+\eta D$. 
\end{itemize}
The statement then follows by matching the basis involving operators $Q_A^k$ to the basis involving eigenstate projectors $\q{s}\qd{s}$. \hfill $\square$\\

Proposition \ref{prop:11} is helpful towards characterizing the manifolds under the form \eqref{eq:explicitsolution}. In fact, this provides a fully general representation of the link between input and output signals of the system \eqref{eq:couplex1} via $3 d^2 + 2 d -1$ dynamic parameters. An Ansatz using the Wigner function of the cavity, like in Section \ref{sec:HO}, leads to the following result. The expressions may look somewhat long, but remember that they analytically and exactly describe the solution of a composite quantum system. Simplified expressions are provided afterwards.\\

\begin{prop}\label{prop:12}
The quantum SDE \eqref{eq:couplex1}, for a general initial condition, admits the explicit solution:
$$\rho_t = \sum_{j,k=1}^{d} \q{j}\qd{k} \otimes \rho_{t,(j,k)} \; ,$$
where each $\rho_{t,(j,k)}$ is an operator on the cavity Hilbert space evolving as follows. Denote $\omega_k = \chi (Q_A)_{k,k}$ and $\omega_{j,k}=\chi((Q_A)_{j,j}-(Q_A)_{k,k})$.\vspace{2mm}

The \textbf{diagonal-block} operators $\rho_{t,(k,k)}$ correspond to the unnormalized Wigner representation:
$$\mathcal{W}^{ \{\rho_{t,(k,k)}\} }(\,x\;,\;p\,) = \mu_{t,k} \, \tfrac{1}{\Pi\, s_t}\; \iint \mathcal{W}^{\{\rho_{0,(k,k)}/\mu_{0,k}\}}(x_0,p_0)\, K_{k,t}(x,x_0,p,p_0) \, dx_0\, dp_0 \; ,$$
\begin{eqnarray*}
K_{k,t}(...) &=& \exp{\left( \frac{-\left\Vert \left(\begin{array}{c}
x-\xi_{t,k} \\ p-\pi_{t,k}
\end{array}\right)- a_t R_{\omega_k\, t} \left(\begin{array}{c}
x_0 \\ p_0
\end{array}\right) \right\Vert^2}{s_t} + d_t \left\Vert \begin{array}{c}
x_0 \\ p_0
\end{array} \right\Vert^2 + \left(\begin{array}{c}
\theta_{t,k} \\ \zeta_{t,k}
\end{array}\right)^T R_{\omega_k\, t} \left(\begin{array}{c}
x_0 \\ p_0
\end{array}\right) \right)}\;\; ,
\end{eqnarray*}
where we have introduced the rotation matrix
$$R_{\omega_k\,t} = \left(\begin{array}{cc} \cos(\omega_k t) & -\sin(\omega_k t) \\ \sin(\omega_k t) & \cos(\omega_k t) \end{array}\right) \; .$$
The deterministic parameters involved in this expression evolve as:
\begin{eqnarray*}
a_t &=& \frac{2 e^{-t}}{(2-\eta) + \eta\, e^{-2 t}} \\
s_t &=& \tfrac{1}{2} \, (1-a_t e^{-t})\\
d_{t} &=& a_t \, \eta \, (e^{-t}-e^{t})\\
\frac{d}{dt}\left(\begin{array}{c}
z_{t,k} \\ y_{t,k}
\end{array}\right)
&=& \, \left(\begin{array}{c}
v_t \\ -u_t
\end{array}\right) - \left(\begin{array}{c}
z_{t,k} \\ y_{t,k}
\end{array}\right) + \omega_k \left(\begin{array}{cc}
-y_{t,k} \\ z_{t,k}
\end{array}\right) \quad , \quad \left(\begin{array}{c}
z_{0,k} \\ y_{0,k}
\end{array}\right) = 0 \; , \\
&&\hspace{-20mm} \text{characterizing } \left(\begin{array}{c}
\theta_{t,k} \\ \zeta_{t,k}
\end{array}\right) = 4\, e^t \, \left(\begin{array}{c}
z_{t,k}- \xi_{t,k} \\ y_{t,k} - \pi_{t,k}
\end{array}\right) \; .
\end{eqnarray*}
Thus the $\rho_{t,(k,k)}$ depend on the measurement record only through the parameters:
\begin{eqnarray}
\nonumber d\left(\begin{array}{c} \xi_{t,k} \\ \pi_{t,k} \end{array}\right)
& = & \frac{-a_t e^{-t}\sqrt{\eta}}{2}\, \left( \begin{array}{c} dy_{t,1}-2 \sqrt{\eta}\xi_{t,k}\, dt \\ dy_{t,2} -2 \sqrt{\eta} \pi_{t,k} \, dt \end{array} \right) \\
&& \quad
+ \left( \, \left(\begin{array}{c}
v_t \\ -u_t
\end{array}\right) - \left(\begin{array}{c}
\xi_{t,k} \\ \pi_{t,k}
\end{array}\right) 
+ \omega_k \left(\begin{array}{cc}
-\pi_{t,k} \\ \xi_{t,k}
\end{array}\right) \,
\right) dt \; , \label{eq:Bipar1-coherentdisp} \\
\label{eq:Bipar1-bpmu}
d(\mu_{t,k}) & = & -2 \sqrt{\eta} \mu_{t,k} \left( \left(\begin{array}{c} \bar{x}_t \\ \bar{p}_t \end{array}\right) -\left(\begin{array}{c} \xi_{t,k} \\ \pi_{t,k} \end{array}\right) \right)^T \left(\begin{array}{c} dy_{t,1}-2\sqrt{\eta}\bar{x}_t dt \\
dy_{t,2}-2\sqrt{\eta}\bar{p}_t dt
\end{array}\right) \\
\nonumber && \text{ with } 2\bar{x}_t = \text{Trace}(({\bf b} + {\bf b}^\dagger)\, \rho_t) \;\;\; \text{ and } 2\bar{p}_t = \text{Trace}(i({\bf b} - {\bf b}^\dagger)\, \rho_t) \; .
\end{eqnarray}
These variables are initialized with $\xi_{0,k} = \pi_{0,k}=0$, while $\mu_{0,k} = \text{Trace}(\rho_{0,(k,k)})$. The sum of the $\mu_{t,k}$ is fixed by normalization, leading to $3d-1$ independent stochastic variables for these block-diagonal components, and $2d$ additional variables which depend on the control inputs.\vspace{2mm}

The \textbf{off-diagonal-block} operators 
$$\rho_{t,(j,k)} = e^{-i \tfrac{\omega_{k}+\omega_j}{2}\, {\bf N}t}\, \tilde{\rho}_{t,(j,k)}\, e^{i\tfrac{\omega_{k}+\omega_j}{2}\, {\bf N}t} \, ,$$
with ${\bf N} =  {\bf b}^\dagger  {\bf b}$ and $j \neq k$, involve $\tilde{\rho}_{t,(j,k)}$ associated to the unnormalized Wigner function:
$$\mathcal{W}^{\{\tilde{\rho}_{t,(j,k)}\}}(x,p) = \tfrac{\mu_{t,(j,k)}}{\Pi\, s_{t,(j,k)}}\; \int \int \mathcal{W}^{\{\tilde{\rho}_{0,(j,k)}/ \mu_{0,(j,k)}\}}(x_0,p_0)\, K_{(j,k),x,t}(x,x_0)\, K_{(j,k),p,t}(p,p_0) \, dx_0\, dp_0 \; ,$$
\begin{eqnarray*}
K_{(j,k),x,t}(x,x_0) &=& \exp\left( \frac{-(x-x_0 a_{t,(j,k)} - \xi_{t,(j,k)})^2}{s_{t,(j,k)}} + (d_{t,(j,k)} x_0^2+\theta_{t,(j,k)}x_0) \right)\;\; , \\
K_{(j,k),p,t}(p,p_0) &=& \exp\left( \frac{-(p-p_0 a_{t,(j,k)} - \pi_{t,(j,k)})^2}{s_{t,(j,k)}} + (d_{t,(j,k)} p_0^2+\zeta_{t,(j,k)}p_0) \right)\;\;.
\end{eqnarray*}
The parameters in these expressions belong to $\mathbb{C}$.
The deterministic parameters involved in these expressions, and not controllable by the inputs $(u_t,v_t)$ independently of the diagonal blocks, evolve as follows:
\begin{eqnarray*}
a_{t,(j,k)} &=& \frac{(2+i\omega_{j,k}) e^{-(1+i\tfrac{\omega_{j,k}}{2})t}}{(2-\eta+i\tfrac{\omega_{j,k}}{2}) + (\eta+i\tfrac{\omega_{j,k}}{2})\, e^{-(2+i\omega_{j,k}) t}} \\
s_{t,(j,k)} &=& \tfrac{1}{2} \, (1-a_{t,(j,k)} e^{-(1+i\tfrac{\omega_{j,k}}{2})t})\\
d_{t,(j,k)} &=& a_{t,(j,k)} \, \frac{2\eta+i\omega_{j,k}}{2+i\omega_{j,k}} \, (e^{-(1+i\tfrac{\omega_{j,k}}{2})t}-e^{(1+i\tfrac{\omega_{j,k}}{2})t})\\
\left(\begin{array}{c}
z_{t,(j,k)} \\ y_{t,(j,k)}
\end{array}\right)
&=& \tfrac{1}{2} \left(\;
\left(\begin{array}{cc} 1 & i \\ - i & 1 \end{array}\right)
R_{\tfrac{\omega_{j,k}}{2}\,t}
\left(\begin{array}{c} z_{t,k} \\ y_{t,k} \end{array}\right)
+
\left(\begin{array}{cc} 1 & -i \\ i & 1 \end{array}\right)
R_{\tfrac{-\omega_{j,k}}{2}\,t}
\left(\begin{array}{c} z_{t,j} \\ y_{t,j} \end{array}\right)
 \;\right) \\
&&\hspace{-20mm}\text{characterizing } \left(\begin{array}{c}
\theta_{t,(j,k)} \\ \zeta_{t,(j,k)}
\end{array}\right) = 4 e^{(1+i\tfrac{\omega_{j,k}}{2})t} \, \left(\begin{array}{c}
z_{t,(j,k)}- \xi_{t,(j,k)} \\ y_{t,(j,k)} - \pi_{t,(j,k)}
\end{array}\right) \; .
\end{eqnarray*}
The $(3d-1)(d-1)$ deterministic variables, which can be driven to independent values by the input signals $(u_t,v_t)$, evolve as follows:
\begin{eqnarray*}
\tfrac{d}{dt}\left(\begin{array}{c} X_{t,(j,k)} \\ P_{t,(j,k)} \end{array}\right) &=&
\left(\frac{1}{a_{t,(j,k)}} - \frac{e^{-i\tfrac{\omega_{j,k}}{2}\,t}}{a_t}\right)\, R_{\tfrac{\omega_j+\omega_k}{2}t}\, \left(\begin{array}{c} v_t \\ -u_t \end{array}\right) \quad \text{ with } \left(\begin{array}{c} X_{0,(j,k)} \\ P_{0,(j,k)} \end{array}\right) =0 \\
&& \hspace{-20mm}\text{characterizing } \;\; \left(\begin{array}{c} \xi_{t,(j,k)} \\ \pi_{t,(j,k)} \end{array}\right)
\;=\;  a_{t,(j,k)} \, \left(\begin{array}{c} X_{t,(j,k)} \\ P_{t,(j,k)} \end{array}\right) \\
&& + \frac{a_{t,(j,k)}}{2 a_t}\, e^{-i\tfrac{\omega_{j,k}}{2}\,t}\,
{\scriptstyle \left(\;
\left(\begin{array}{cc} 1 & i \\ - i & 1 \end{array}\right)
R_{\tfrac{\omega_{j,k}}{2}\,t}
\left(\begin{array}{c} \xi_{t,k} \\ \pi_{t,k} \end{array}\right)
+
\left(\begin{array}{cc} 1 & -i \\ i & 1 \end{array}\right)
R_{\tfrac{-\omega_{j,k}}{2}\,t}
\left(\begin{array}{c} \xi_{t,j} \\ \pi_{t,j} \end{array}\right)
 \;\right)} \;  \\
\tfrac{d}{dt} c_{t,(j,k)} &=& \Re\left((\eta+i\tfrac{\omega_{jk}}{2})\,a_{t,(j,k)}\,e^{-(1+i\tfrac{\omega_{j,k}}{2})t}\right)
- \eta\, a_t\,e^{-t} \\
&& + 4 \Re\left(\, e^{(1+i\tfrac{\omega_{j,k}}{2})t} \left(\begin{array}{c} X_{t,(j,k)} \\ P_{t,(j,k)} \end{array}\right)^T \, R_{\tfrac{\omega_k+\omega_j}{2}t} \, \left(\begin{array}{c} v_t \\ -u_t \end{array}\right)\,  \right) \\
&& \hspace{-20mm}\text{characterizing } \;\; \log\left(\frac{|\mu_{t,(j,k)}|}{\sqrt{\mu_{t,j}\mu_{t,k}}}\right)
\;=\; c_{t,(j,k)} -\Re\left(\,\frac{2 e^{(1+i\tfrac{\omega_{j,k}}{2})t}}{a_{t,(j,k)}} (\xi_{t,(j,k)}^2 + \pi_{t,(j,k)}^2)\,\right) \\ && \qquad \qquad \qquad \qquad \qquad \qquad
 + \frac{e^t}{a_t} (\xi_{t,k}^2+\pi_{t,k}^2+\xi_{t,j}^2+\pi_{t,j}^2) \\
\tfrac{d}{dt} q_{t,(j,k,\ell)} &=& Q_{t,(j,k)} + Q_{t,(k,\ell)} + Q_{t,(\ell,j)} \\
&& \text{where } \;\;  Q_{t,(j,k)}\;=\; \Im\left((\eta+i\tfrac{\omega_{j,k}}{2})a_{t,(j,k)}\,e^{-(1+i\tfrac{\omega_{j,k}}{2})t}\right)\\
&& \qquad \qquad \qquad \qquad \qquad +4 \Im\left(\, e^{(1+i\tfrac{\omega_{j,k}}{2})t} \left(\begin{array}{c} X_{t,(j,k)} \\ P_{t,(j,k)} \end{array}\right)^T \, R_{\tfrac{\omega_k+\omega_j}{2}t} \, \left(\begin{array}{c} v_t \\ -u_t \end{array}\right)\,  \right) \;\; , 
\end{eqnarray*}
characterizing $\; \text{phase}(\mu_{t,(j,k)}\mu_{t,(k,\ell)}\mu_{t,(\ell,j)}) =$
\begin{eqnarray*}
 && q_{t,(j,k,\ell)} - 2 \Im\left( \frac{e^{(1+i\tfrac{\omega_{j,k}}{2})t}}{a_{t,(j,k)}}(\xi_{t,(j,k)}^2 + \pi_{t,(j,k)}^2) \right) \\
&& - 2 \Im\left(  \frac{e^{(1+i\tfrac{\omega_{k,\ell}}{2})t}}{a_{t,(k,\ell)}}(\xi_{t,(k,\ell)}^2 + \pi_{t,(k,\ell)}^2) \right) - 2 \Im\left(  \frac{e^{(1+i\tfrac{\omega_{\ell,j}}{2})t}}{a_{t,(\ell,j)}}(\xi_{t,(\ell,j)}^2 + \pi_{t,(\ell,j)}^2) \right) \; .
\end{eqnarray*}
The last variables are initialized from $\mu_{0,(j,k)} = \text{Trace}(\rho_{0,(j,k)})$. The measurement realizations add new degrees of freedom to the off-diagonal blocks, only through:
\begin{eqnarray*}
\tfrac{d}{dt} \delta_{t,(j,k)} &=& Q_{t,(j,k)} +\frac{2\,e^t}{a_t}\left( \left(\begin{array}{c} \xi_{t,k} \\ \pi_{t,k} \end{array}\right)^T R_{\omega_k t}  -  \left(\begin{array}{c} \xi_{t,j} \\ \pi_{t,j} \end{array}\right)^T R_{\omega_j t}\right)  \, \left(\begin{array}{c} u_t \\ v_t \end{array}\right) \\
&& \text{characterizing } \;\; \text{phase}(\mu_{t,(j,k)}) \;=\; \delta_{t,(j,k)} - 2 \Im\left( \frac{e^{(1+i\tfrac{\omega_{j,k}}{2})t}}{a_{t,(j,k)}}(\xi_{t,(j,k)}^2 + \pi_{t,(j,k)}^2) \right) \; .
\end{eqnarray*}
The latter are linked through the deterministic variables $q_{t,(j,k,\ell)}$, such that they can move only in a subspace of dimension $d-1$ as a function of the measurement realizations.
\end{prop}

\noindent \underline{Proof:} 
From \eqref{eq:couplex1}, each component $\qd{k} \rho \q{j}$ of the state, with $j,k$ two qudit levels, evolves independently --- up to the common variables $\bar{x}_t$ and $\bar{p}_t$. The latter only depend on $\{\qd{k} \rho \q{k} : k=1,2,..., d\}$, so these diagonal blocks undergo autonomous dynamics. Furthermore, the dependence of $d \qd{k} \rho \q{j}$ on $\bar{x}_t$ and $\bar{p}_t$ multiplies $\qd{k} \rho \q{j}$ itself, so it looks mainly as a normalization. It is important to note the detuning though: for each level $k$, the output and input analogous to Prop.\ref{prop:6} would be in a different rotating frame. In a Rotating Wave Approximation (RWA), we would say that each qudit level is associated to a cavity state driven by separate inputs and outputs, corresponding to a component at the resonance frequency only. This has motivated the form of the solution, analogous to one cavity per qudit level, with a Prop.\ref{prop:6} type solution except each level gets its own variables and the normalization of each cavity state is now explicitly computed.\\

For the diagonal blocks $\qd{k} \rho \q{k}$, the equations to be solved in Wigner representation are exactly the same as for Proposition \ref{prop:6}. We just have to rotate input and output into the correct frame. This time, we explicitly compute the equation for the normalization constants, and we find how they depend on the measurement signals. Note that the deterministic parameters do not depend on $k$.

Since we have assumed $n_{th}=0$, the component $\theta_k$ is deterministically linked to $\xi_k$ as in Prop.\ref{prop:6}, but it cannot be explicitly integrated since this link depends on the control inputs in a $k$-dependent way.
Also, for $u_t,v_t$ nonzero, a given state of the cavity cannot be linked uniquely to a normalization constant for the corresponding qudit level: there is a memory effect, implying a separate dependence on the output signals, so the $\mu_k$ must be integrated separately.\\

For the off-diagonal blocks $\qd{j} \rho \q{k}$ with $j\neq k$, the corresponding cavity density operator component is non-hermitian and cannot be considered anymore as an unnormalized cavity state. We can nevertheless write the Wigner representation, accepting $\mathcal{W}$ to be complex rather than real. The algebraic criterion, by giving the number of independent parameters that depend on $dy_1,dy_2,u_t$ and $v_t$ as well as the associated vector fields, has been an invaluable tool towards guessing which representation could work. The proof then comes down, to a big extent, to identifying the resulting equations and solving them exactly like for the cavity alone. We consider each off-diagonal block independent in a first approach, planning to look for remaining dependencies later.

The stochastic partial differential equation followed by the Wigner representation of $\qd{j} \rho \q{k}$ contains an additional contribution: 
\begin{multline*}
 d\rho_{t,(j,k)} = ...  - i \tfrac{\omega_j-\omega_k}{2}(\,{\bf b}^\dagger {\bf b} \rho_{t,(j,k)} + \rho_{t,(j,k)}{\bf b}^\dagger {\bf b}\, ) dt + .... \\ 
\leftrightarrow d\mathcal{W}^{\{\rho_{t,(j,k)} \}} = ... -i(\omega_j-\omega_k) \, \left( x^2+p^2 - \tfrac{1}{2} - \tfrac{1}{16}(\tfrac{\partial^2}{\partial x^2} +\tfrac{\partial^2}{\partial p^2}  ) \right) \, \mathcal{W} \, dt + ... \;\; .
\end{multline*}
The equation is still separable in $x$ and $p$, so we try the same Gaussian representation as for the cavity alone, but now allowing complex parameters. We also assume a priori that all the parameters could depend on the index $(j,k)$.

From there, we proceed like for the other cases. First, we see that the partial stochastic differential equation contains no term in $x^2 dw_{t,1}$ and therefore $s_{t,(j,k)}$ is deterministic. The corresponding ODE is:
$$\frac{ds_{t,(j,k)}}{dt} + 2 \eta (s_{t,(j,k)}-\tfrac{1}{2})^2 = -2 s_{t,(j,k)} + 1 - i (\omega_j-\omega_k)(s_{t,(j,k)}^2 - \tfrac{1}{4})\, .$$
The equation can again be solved as a Ricatti form, it just depends, indeed, on $(j,k)$. Going on like for the single cavity, yields the deterministic solutions for $a_{t,(j,k)}$ and $d_{t(j,k)}$. Those do not depend on outputs nor inputs. The equations for the remaining parameters $\xi_{t,(j,k)}, \theta_{t,(j,k)}$ and $\mu_{t,(j,k)}$ do involve inputs and outputs. Note that $\mu_{t,(j,k)}$ is split between $x$ and $p$ contributions, the overall equation is easily obtained by analogy.

The off-diagonal parameters computed in this way are not all independent. Indeed, from Proposition \ref{prop:11}, the overall solution should involve $3d^2+2d-1$ true degrees of freedom depending on outputs or inputs, among which $4d-2$ can depend on the outputs. The diagonal blocks already involve $5d-1$ degrees of freedom, among which $\xi_k,\pi_k,\mu_k$ are $3d-1$ variables depending on outputs, while $z_{k},y_{k}$ depend on the inputs. Thus there should remain $3d(d-1)$ independent variables, among which only $d-1$ can be driven independently by the outputs. Instead, in total, our complex parameterization involves $5d(d-1)$ real degrees of freedom through the $(\xi,\pi,\mu,\theta,\zeta)_{(j,k)}$. 

The parameters $(\theta,\zeta)_{(j,k)}$ can be written as functions of $(\xi,\pi)$ and of deterministic variables $z_{(j,k)},y_{(j,k)}$, like for the diagonal components. We can rotate back the vector $(z,y)$ to a common rotating frame, for indices $(k,k)$, $(j,j)$ and $(j,k)$. Then we separate the real and imaginary parts of the $(j,k)$ components, and we observe the resulting dynamics in $\mathbb{R}^8$: the vector fields span a 4-dimensional linear subspace, such that real and imaginary components of $z_{(j,k)},y_{(j,k)}$ are just static linear functions of $(z_k,y_k)$ and $(z_j,y_j)$, in any common rotating frame. This suppresses $2d(d-1)$ parameters and indeed leaves $3d(d-1)$ independent ones, among which only $d-1$ may depend on the output realization.

For the $(\xi,\pi)_{(j,k)}$, we can repeat a procedure similar to the $(z,y)_{(j,k)}$, linking real and imaginary parts to $(j,j)$ and $(k,k)$ components in rotating frame. The control vector field does not drop out of these equations, yielding thus $2d(d-1)$ parameters which can be driven independently by the control vector fields (only). 

We are thus left with the $\mu_{t,(j,k)}$, among which $d-1$ degrees of freedom can depend on measurement realizations, while all the remaining ones are provided by the input signals. We must thus find a variable where the noise term cancels, and the rest ``looks nice''. From the SDEs describing $\xi_{t,(j,k)},\pi_{t,(j,k)}$ and $\mu_{t,(j,k)}$, we consider the variables:
$$\tilde{m}_{(j,k)} := \log(\mu_{j,k}) - \frac{\xi_{t,(j,k)}^2+\pi_{t,(j,k)}^2}{s_{t,(j,k)}-1/2} \; ,$$
Indeed, in the corresponding SDEs, the noise term becomes independent of the indices $j,k$, namely:
\begin{eqnarray*}
d\tilde{m}_{t,(j,k)} &=& -2\sqrt{\eta} \left(\begin{array}{c} \bar{x}_t \\ \bar{p}_t \end{array}\right)^T
\left(\begin{array}{c} dy_{t,1} \\ dy_{t,2} \end{array}\right) 
- (2 \eta + i (\omega_j-\omega_k))(s_{t,(j,k)}-\tfrac{1}{2})\,  dt \\
&& - \tfrac{2}{s_{t,(j,k)}-\tfrac{1}{2}}\left(\begin{array}{c} \xi_{t,(j,k)} \\ \pi_{t,(j,k)} \end{array}\right)^T
 \, R_{\tfrac{\omega_j+\omega_k}{2}t} \, \left(\begin{array}{c} v_t \\ -u_t \end{array}\right) \, dt \; .
\end{eqnarray*}
Thus, taking linear combinations of such variables, with linear coefficients summing to zero, allows to cancel the noise term. In general, there would remain a dependence on $\xi,\pi$, through which the system might diffuse in many directions, like the rolling wheel. But we have just established the interdependence of the $(\xi,\pi)_{(j,k)}$ too, up to control signals. Knowing this, we consider just
$$m_{(j,k)} = \tilde{m}_{(j,k)}- \tfrac{1}{2}(\tilde{m}_{(j,j)}+\tilde{m}_{(k,k)})\; ,$$
i.e.~the coherence compared to its associated diagonals. Using the interdependence of the $\xi,\pi$, the corresponding differential equation writes:
\begin{eqnarray*}
dm_{t,(j,k)}&=& 2\eta(s_t-\tfrac{1}{2})\, dt - (2 \eta + i (\omega_j-\omega_k))(s_{t,(j,k)}-\tfrac{1}{2})\, dt\\
&& + 4 e^{(1+i\tfrac{\omega_j-\omega_k}{2})t} 
\, \left(\begin{array}{c} X_{t,(j,k)} \\ P_{t,(j,k)} \end{array}\right)^T
 \, R_{\tfrac{\omega_j+\omega_k}{2}t} \, \left(\begin{array}{c} v_t \\ -u_t \end{array}\right) \, dt \\
&& + \frac{i}{s_t-\tfrac{1}{2}} \left(\begin{array}{c} \xi_{t,j} \\ \pi_{t,j} \end{array}\right)^T
 \, R_{\omega_j t} \, \left(\begin{array}{c} u_t \\ v_t \end{array}\right) \, dt
 - \frac{i}{s_t-\tfrac{1}{2}} \left(\begin{array}{c} \xi_{t,k} \\ \pi_{t,k} \end{array}\right)^T
 \, R_{\omega_k t} \, \left(\begin{array}{c} u_t \\ v_t \end{array}\right) \, dt \; .
\end{eqnarray*}
In the real part of $dm_{(j,k)}$, all the variables $\xi,\pi$ disappear and we obtain an equation involving deterministic parameters and the input signals $u_t,v_t$. In the imaginary part of $m_{(j,k)}$, a dependence on $\xi_k,\pi_k$ and $\xi_j,\pi_k$ remains. This is expected, since $d-1$ degrees of freedom can still be driven by measurement realizations. The reduction from $d(d-1)/2$ to $d-1$ is made by noticing how any $\xi,\pi$ again disappear in linear combinations $m_{(j,k)} + m_{(k,l)} + m_{(l,j)}$.

With this we have treated all the parameters. Note how the algebraic criterion has been crucial in order to identify when interdependencies remained, and when we can stop because no further reduction is possible for a general initial state and general input signals.
 \hfill $\square$\\

The above result can be simplified by integrating out several parameters when the system features no control drives.\\
\begin{cor}\label{prop:13}
For $v_t=u_t=0$, i.e.~in absence of drives on the cavity, the solution of Proposition \ref{prop:12} simplifies as follows:
$$\begin{array}{rlrl}
\theta_{t,k} &=\; -4\, e^{t}\,\xi_{t,k} \;\; , \quad & \zeta_{t,k} &=\; -4\, e^{t}\, \pi_{t,k} \;\; ,\\
\theta_{t,(j,k)} &=\; -4\, e^{(1+i\tfrac{\omega_{j,k}}{2})\,t} \xi_{t,(j,k)} \;\; , \quad &
\zeta_{t,(j,k)} &=\; -4\, e^{(1+i\tfrac{\omega_{j,k}}{2})\,t} \pi_{t,(j,k)} \;\; ; 
\end{array}$$
$$
\left(\begin{array}{c} \xi_{t,(j,k)} \\ \pi_{t,(j,k)} \end{array}\right)
\;=\; \frac{a_{t,(j,k)}}{2 a_t}\, e^{-i\tfrac{\omega_{j,k}}{2}\,t}\,
\left(\;
\left(\begin{array}{cc} 1 & i \\ - i & 1 \end{array}\right)
R_{\tfrac{\omega_{j,k}}{2}\,t}
\left(\begin{array}{c} \xi_{t,k} \\ \pi_{t,k} \end{array}\right)
+
\left(\begin{array}{cc} 1 & -i \\ i & 1 \end{array}\right)
R_{\tfrac{-\omega_{j,k}}{2}\,t}
\left(\begin{array}{c} \xi_{t,j} \\ \pi_{t,j} \end{array}\right)
 \;\right)  \; ; 
$$
\begin{align*}
\mu_{t,k} &= \mu_{0,k}\, a_t \, \exp\left( t+ \frac{\xi_{t,k}^2+\pi_{t,k}^2}{s_t-1/2}\right) \, \nu_t \; \, ,\\
\mu_{t,(j,k)} &= \mu_{0,(j,k)}\, a_{t,(j,k)} \, \exp\left( (1+\tfrac{i\omega_{j,k}}{2})\,t + \frac{\xi_{t,(j,k)}^2+\pi_{t,(j,k)}^2}{s_{t,(j,k)}-1/2}\right) \, \nu_t \; \, ,
\end{align*}
with $\nu_t$ a common normalization constant.
\end{cor}

\noindent \underline{Proof:} According to the algebraic criterion, in the diagonal components already, the normalization constants $\mu_k$ must become deterministically linked to $\xi_k,\pi_k$. Like for the interdependence of the $\mu_{(j,k)}$ in the above proof, the variable
$$\frac{\xi_{t,k}^2+\pi_{t,k}^2}{s_t-1/2} - \frac{\xi_{t,j}^2+\pi_{t,j}^2}{s_t-1/2} - \log(\mu_{t,k}/\mu_{t,j})$$
appears like a good candidate. One then checks that indeed, it undergoes an autonomous deterministic equation, provided $u=v=0$. This evolution of $\mu_{t,k}/\mu_{t,j}$ is compatible with introducing a common multiplicative constant and stating the explicit expression of the $\mu_k$ as done in the statement.

In the off-diagonal parameterization, a similar treatment holds for the $\mu_{t,(j,k)}$ when $u=v=0$. In particular, this fixes the phase of each $\mu_{t,(j,k)}$ as a deterministic function of the $\xi_{t,(j,k)}$ and $\pi_{t,(j,k)}$. For the other variables, $z_{t,(j,k)},y_{t,(j,k)},\xi_{t,(j,k)},\pi_{t,(j,k)}$ which would be driven independently of the diagonal blocks only through control inputs $u_t,v_t$, the degree of freedom obviously disappears once $u_t=v_t=0$.\hfill $\square$\\

The result of Corollary \ref{prop:13} thus indicates that, for $u_t=v_t=0$, once the $2d$ values at time $t$ of the ``mean position parameters'' $\xi_{t,k}$ and $\pi_{t,k}$ in the Green function are known, this defines the full state completely. In presence of control, the situation as described in Proposition \ref{prop:12} is more complicated. In particular: (i) the diagonal normalization constants $\mu_{t,k}$ are not deterministically linked anymore to the conditional cavity state; and (ii) the global angles of each conditional cavity state, expressed through the off-diagonal normalization constants $\mu_{t,(j,k)}$, are not deterministically linked to the rest of the system. All the other variables are essentially deterministically linked to $\xi_{t,k}$ and $\pi_{t,k}$, but some of them can be driven to independent values by the $(u_t,v_t)$. This can be understood with the following standard physical argument: qudit level $\q{s}$ shifts the cavity frequency by $\chi {Q_A}_{s,s}$, and driving the cavity on resonance with this shift would mainly, independently, drive the part of the state conditioned on qudit level $\q{s}$. A few further remarks are in order.

\begin{itemize}
\item For a general initial state, the $\mu_{t,(j,k)}$ or $\mu_{t,k}$, are not necessarily equal to the trace of $\rho_{t,(j,k)}$ for $t>0$. The full normalization indeed involves the integral with $\mathcal{W}^{\{\tilde{\rho}_{0,(j,k)}\}}(x_0,p_0)$. Although some interesting features already appear --- e.g.~$\mu_k$ is less subject to Wiener noise if its $\xi_{t,k},\pi_{t,k}$ match well with $\bar{x}_t,\bar{p}_t$ --- more conclusive treatments have to assume a particular type of initial state for the joint system, since the general case cannot be further reduced. In particular, the normalization condition may be different from $\sum_k \mu_{t,k} = 1$.
\item If some qudit level is initially not populated, i.e.~$\mu_{0,k}=0$ for some $k \in \{1,2,...,d\}$, then $\mu_{t,k}=0$ for all times and the corresponding $\rho_{t,(j,k)}$ can be dropped.\\
\end{itemize}

Finally, in order to gain more concrete insight on the meaning of Proposition \ref{prop:12}, we consider its result as $t \rightarrow +\infty$. This will lead to the interpretation stated in the main text. We first observe that
\begin{eqnarray*}
a_0=1 \rightarrow a_{+\infty}=0 &\;\; \text{ and } \;\; & s_0=0 \rightarrow s_{+\infty}=1/2
\end{eqnarray*}
at a unit exponential rate. This indicates that, up to normalization, $\mathcal{W}^{ \{\rho_{t,(k,k)}\} }(x,p)$ becomes independent of the initial condition. It exponentially converges towards the Wigner function of a coherent state $\q{\alpha_k}$ with $\alpha_k =  \xi_{t,k} + i \pi_{t,k} \in \mathbb{C}$.

In the SDE \eqref{eq:Bipar1-coherentdisp} governing $\xi_{t,k}+i\pi_{t,k}$, the dependence on measurement outputs also decreases exponentially, such that asymptotically they follow the ordinary differential equation:
\begin{eqnarray*}
\tfrac{d}{dt} \, \alpha_{t,k} = (v_t-i\,u_t) + i\, \omega_k \alpha_{t,k} \, - \alpha_{t,k} \; .
\end{eqnarray*}
This is the standard level-dependent evolution of a coherent state in a ``readout cavity'' to be probed, see e.g.~\cite{GambMeas}.

The normalization, reflecting respective qudit populations, is then given by:
\begin{eqnarray*}
\tilde{\mu}_k & = & \mu_{t,k} \iint \mathcal{W}^{\{\rho_{0,(k,k)}\}}(x_0,p_0) \;
\exp\left( \tfrac{-2\eta}{2-\eta}\left\Vert\begin{array}{c} x_0 \\ p_0 \end{array} \right\Vert^2 +  \left(\begin{array}{c}
\theta_{t,k} \\ \zeta_{t,k}
\end{array}\right)^T R_k \left(\begin{array}{c}
x_0 \\ p_0
\end{array}\right)\,\right)
\, dx_0 dp_0 \; .
\end{eqnarray*}
Recalling from Section \ref{sec:aandia} that 
$$
d\left( \begin{array}{c}
\theta_{t,k} \\ \zeta_{t,k}
\end{array} \right)
= 2 \sqrt{\eta} a_t \left( \begin{array}{c}
dw_{t,1} \\ dw_{t,2}
\end{array} \right)
+ 4 \eta a_t \left( \begin{array}{c}
\bar{x}_t - \xi_{t,k} \\ \bar{p}_t - \pi_{t,k}
\end{array} \right) \, dt
+ \omega_k \left( \begin{array}{c}
-\zeta_{t,k} \\ \theta_{t,k}
\end{array} \right) \, dt 
$$
and considering (again) that $\bar{x},\bar{p},\xi_k,\pi_k$ would remain bounded with high probability, we see that $\tfrac{\tilde{\mu}_k}{\mu_k}$ asymptotically converges towards a time-independent value for each $k$. We then have in fact, asymptotically:
\begin{eqnarray}\label{eq:appfl:mugetsQND}
d\tilde{\mu}_{t,k} &=& 2 \sqrt{\eta} \left( \sum_{j=1}^d \, (\xi_{t,k}-\xi_{t,j}) \tilde{\mu}_{t,j} \right) \, \tilde{\mu}_{t,k} \, dw_{t,1}\\ \nonumber
&& + 2 \sqrt{\eta} \left( \sum_{j=1}^d \, (\pi_{t,k}-\pi_{t,j}) \tilde{\mu}_{t,j} \right) \, \tilde{\mu}_{t,k} \, dw_{t,2} \, .
\end{eqnarray}
This dynamics corresponds to the QND measurement of the populations $\tilde{\mu}_{t,k}$ of the qudit levels, with two measurement channels $L_1(t) = \sum_{j=1}^d \q{j}\qd{j}\, \xi_{t,j}$ and $L_2(t) = \sum_{j=1}^d \q{j}\qd{j}\, \pi_{t,j}$.

The latter dynamics is relevant if it remains significant after the convergence of the $\rho_{t,(k,k)}$ towards coherent states, or if the cavity state is initialized as such.


\subsection{~~~~Two resonantly coupled harmonic oscillators}\label{app:add:algcrit2cavs}

In this section, we detail the investigation of the algebraic criterion for the model \eqref{eq:couplex2}.

\begin{prop}
The system \eqref{eq:couplex2} with $n_{th}>0$ features confining manifolds of dimension $M=8$, characterized by the algebra:
$$\mathfrak{G}_F = \text{span}\{G_{\delta {\bf b}},\, G_{\delta {\bf b}^\dag},\, G_{\delta {\bf a}},\, G_{\delta {\bf a}^\dag};\; \delta =1,i \} \; .$$
\begin{itemize}
\item For $n_{th}=0$, the manifold further reduces to $M=4$ as all the $^\dag$ terms in $\mathfrak{G}_F$ disappear. 
\item Adding Hamiltonians on B similar to those on A does not change this manifold structure.\\
\end{itemize}
\end{prop}

\noindent \underline{Proof:} Operators acting on B alone generate the algebra 
$\mathfrak{G}_B = \text{span}\{\, G_{\bf b}\,,\; G_{i{\bf b}} \, \} \; $  when $n_{th}=0$,  adding $G_{{\bf b}^\dag},\; G_{i{\bf b}^\dag}$ for $n_{th}>0$. The commutator of $G_{\bf b}$ (respectively $G_{{\bf b}^\dag}$) with the coupling Hamiltonian just yields $G_{\bf a}$(respectively $G_{{\bf a}^\dag}$). Commutation of the latter two yields $G_{[{\bf a},{\bf a}^\dag]} = G_{\bf I} = 0$. Further commutation of these A-dynamics with any B-dynamics vanishes. Further commutation of these A-dynamics with the coupling Hamiltonian yields back $G_{\bf b}$ and $G_{{\bf b}^\dag}$. Hence the algebra is closed at this point. 

Adding detuning on B, i.e.~a Hamiltonian term ${\bf I} \otimes \tilde\Delta {\bf b}^\dag  {\bf b}$, is equivalent to modifying the detuning on A. Indeed, we can annihilate this term by going to a common rotating frame, thus performing a unitary change of frame with the operator ${\bf U}_t = e^{-i \tilde{\Delta}({\bf a}^\dag  {\bf a}+{\bf b}^\dag  {\bf b})}$. This change of frame commutes with the coupling Hamiltonian and with the dissipation on B; it only induces a time-dependent redefinition of the output components $dy_{t,1}, dy_{t,2}$ and control signals $u_t, v_t$.

Drives on B, i.e.~a Hamiltonian term of the form ${\bf I} \otimes (\tilde{u}_t ({\bf b}+{\bf b}^\dagger) - i\tilde{v}_t ({\bf b}-{\bf b}^\dagger) )$, can be equivalently transferred into effective drives on subsystem A. Indeed, note that
$$ g  ({\bf a} \otimes {\bf b}^\dag + {\bf a}^\dag \otimes {\bf b})  +  {\bf I} \otimes (\tilde{u}_t ({\bf b}+{\bf b}^\dagger) - i\tilde{v}_t ({\bf b}-{\bf b}^\dagger) )\\ =  g  (({\bf a}+\beta_t) \otimes {\bf b}^\dag + ({\bf a}+\beta_t)^\dag \otimes {\bf b})\; , $$
with $\beta_t = \frac{\tilde{u}_t + i \tilde{v}_t}{g}$.
Hence, consider the time-dependent unitary change of frame $\tilde{\rho}_t = {\bf U}_t \rho_t {\bf U}_t^\dagger$ with ${\bf U}_t = {\bf D}(\beta_t) \otimes {\bf I}$. Here ${\bf D}(\alpha) = \exp(\alpha {\bf a}^\dag - \alpha^* {\bf a})$ is the displacement operator \cite{Haroche} on harmonic oscillator mode A, satisfying ${\bf D}(\alpha)\, {\bf a}\, {\bf D}(\alpha)^\dagger = {\bf a} - \alpha$ and $\tfrac{d{\bf D}(\alpha_t) }{dt}\, {\bf D}(\alpha_t)^\dagger  = \tfrac{d\alpha}{dt} {\bf a}^\dagger - \tfrac{d\alpha^*}{dt} {\bf a} + (\tfrac{d\alpha}{dt}\alpha^* - \tfrac{d\alpha^*}{dt} \alpha)\, {\bf I}$. With these elements and since ${\bf U}_t$ commutes with subsystem B, one checks that $d\tilde{\rho}$ follows the same equation \eqref{eq:couplex2}, except $u_t$ and $v_t$ are replaced by $u_t - \tfrac{\Delta \tilde{u}_t}{g} + \tfrac{d\beta^*/dt-d\beta/dt}{2i}$ and $v_t - \tfrac{\Delta \tilde{v}_t}{g} + \tfrac{d\beta^*/dt+d\beta/dt}{2}$ respectively.
\hfill $\square$\\


\subsection{~~~~Indistinguishable emission}\label{app:ssec:Lewalle}

We now move to the setting of Section \ref{sec:Lewalle}. We treat it in quite some detail in order to illustrate our method for finding deterministic manifold equations, in other contexts than Gaussian kernels.\\

\paragraph*{Algebraic criterion:} We first give the detailed computations concerning the algebraic criterion of Prop.\ref{prop:1}. After noting that $[L_1,L_2] = 0$ and hence $[G_{L_1},\; G_{L_2}]=0$, we compute further commutation with the deterministic evolution. We compute $[L_j^\dagger L_j , \; L_1 ] = \frac{1}{2}( \sigma_{-A}\pm\sigma_{zA}\sigma_{-B} + \sigma_{-B}\pm \sigma_{zB}\sigma_{-A})$ with signs depending on $j \in \{1,2\}$, and similarly for $[L_2,\; L_j^\dagger L_j ] $. Taken separately, these correspond to two new vector fields $G_{L'_k}$ with $L'_1 = \sigma_{zA}\sigma_{-B} + \sigma_{zB}\sigma_{-A}$ and $L'_2 = i (\sigma_{zA}\sigma_{-B} - \sigma_{zB}\sigma_{-A})$, whose further commutations happen to generate yet other directions. When the individual channels $L_1$ and $L_2$ have arbitrary rates, $\mathfrak{G}_F$ contains all these vector fields and the model reduction would not be efficient. However, when the rates of $F_{L_1}$ and $F_{L_2}$ are equal (and thus, without requiring $\eta_1=\eta_2$ since they do not appear there), we see instead that the commutator with their sum yields no new diffusion directions; this occurs through the opposite signs in the above individual commutators.

One can further check that when adding nonzero drives in $\sigma_x$ and/or $\sigma_y$ on A and/or B, under any conditions, the diffusion spans manifolds of dimension 10 --- hence, we expect no easy explicit formula for the solution in these cases. Adding detuning, i.e.~a Hamiltonian in $\sigma_{z,A}$ or $\sigma_{z,B}$, would just double the manifold dimension to $M=4$.\\

\paragraph*{Stochastic differential equations:} As stated in the main text, towards writing the deterministic manifold equations, we translate the dynamics \eqref{eqL:sde} to the Pauli basis. This is an orthonormal basis for the Frobenius norm, with~$r(jk)_t = \tr(\rho_t ({\bf \sigma}_{j} \otimes {\bf \sigma}_k))$; in particular $r(II)_t=1$ for all $t$. The dynamics of the 15 other coordinates are obtained by applying the corresponding trace to \eqref{eqL:sde}. 
\begin{itemize}
\item In fact, as the ground state on any qubit should be invariant, we readily introduce displaced Z variables, replacing $r(Zq),\; r(qZ),\; r(ZZ)$ respectively by $r(Zq^*) = r(Zq)-r(Iq),\; r(qZ^*) = r(qZ)-r(qI),\; r(ZZ^*)=r(ZZ)-r(ZI)-r(IZ)$ for $q\in \{I,X,Y\}$. The corresponding equations follow systematically, e.g.~
\begin{eqnarray*}
dr(XI)_t &=& -r(XI)_t\, dt + \big( -r(ZI^*)_t + r(XX)_t - (r(XI)+r(IX))_t r(XI)_t \big) \, \sqrt{\eta_1}\, dw_{t,1}\\
&& + \big( -r(XY)_t - (r(YI)-r(IY))_t\, r(XI)_t \big) \, \sqrt{\eta_2}\, dw_{t,2} \; .
\end{eqnarray*}
The other equations are all similar. Note that the output signals in this basis write:
$$dy_{t,1}= \sqrt{\eta_1} (r(XI)+r(IX))_t \; dt + dw_{t,1}\;\; , \quad dy_{t,2}= \sqrt{\eta_2} (r(YI)-r(IY))_t \; dt + dw_{t,2} \; .$$
\item We next group the variables in sums and differences, e.g.~$X_s = r(XI)+r(IX)$, $XY_d = r(XY)-r(YX)$, $YZ_s = r(YZ)+r(ZY)$ and so on. The dynamics are obtained as sums and differences, with no \Ito~specificities. This puts the system into a block-triangular form, with 9 variables that influence, but are not influenced by, the remaining 6: $X_d,\, Y_s,\, Z_d,\, XY_s, \, XZ_d,\, YZ_s$. 
\item We finally write the ratio between each of these variables and $r(ZZ)_t$, computing the resulting dynamics with due \Ito~corrections. Using $dy_{t,1}$ and $dy_{t,2}$ to include the stochastic contributions, this yields the following rather compact dynamics:
\begin{eqnarray*}
dB_{1,t} &=& B_{1,t}\, dt - 2 \sqrt{\eta_2}\, dy_{t,2} \quad \text{for } B_1 = YZ_d / r(ZZ)\\
dB_{2,t} &=& B_{2,t}\, dt - 2 \sqrt{\eta_1}\, dy_{t,1} \quad \text{for } B_2 = XZ_s / r(ZZ)\\
dB_{3,t} &=& 2 B_{3,t}\, dt + B_{1,t} \sqrt{\eta_2}\, dy_{t,2} \quad \text{for } B_3 = r(YY) / r(ZZ)\\
dB_{4,t} &=& 2 B_{4,t}\, dt - B_{2,t} \sqrt{\eta_1}\, dy_{t,1} \quad \text{for } B_4 = r(XX) / r(ZZ)\\
dB_{5,t} &=& 2 B_{5,t}\, dt + B_{1,t} \sqrt{\eta_1}\, dy_{t,1} + B_{2,t} \sqrt{\eta_2}\, dy_{t,2} \quad \text{for } B_5 = XY_d / r(ZZ)\\
dB_{6,t} &=& 2 B_{6,t}\, dt + B_{2,t} \sqrt{\eta_1}\, dy_{t,1} + B_{1,t} \sqrt{\eta_2}\, dy_{t,2} \quad \text{for } B_6 = Z_s / r(ZZ)\\
dB_{7,t} &=& 3 B_{7,t}\, dt + (2B_4-B_6)_t \sqrt{\eta_1}\, dy_{t,1} - B_{5,t} \sqrt{\eta_2}\, dy_{t,2} \quad \text{for } B_7 = X_s / r(ZZ)\\
dB_{8,t} &=& 3 B_{8,t}\, dt - B_{5,t} \sqrt{\eta_1}\, dy_{t,1} - (2B_3+B_6)_t \sqrt{\eta_2}\, dy_{t,2} \quad \text{for } B_8 = Y_d / r(ZZ)\\
dB_{0,t} &=& 4 B_{0,t}\, dt + B_{7,t} \sqrt{\eta_1}\, dy_{t,1} + B_{8,t} \sqrt{\eta_2}\, dy_{t,2} \quad \text{for } B_0 = 1/ r(ZZ)
\end{eqnarray*}
concerning the first block of 9 variables, and
\begin{eqnarray*}
dR_{1,t} &=& R_{1,t}\, dt  \quad \text{for } R_1 = XZ_d / r(ZZ)\\ 
dR_{2,t} &=& R_{2,t}\, dt  \quad \text{for } R_2 = YZ_s / r(ZZ) \\ 
dR_{3,t} &=& 2 R_{3,t}\, dt - R_{1,t} \sqrt{\eta_1}\, dy_{t,1} - R_{2,t} \sqrt{\eta_2}\, dy_{t,2} \quad \text{for } R_3 = Z_d / r(ZZ)\\ 
dR_{4,t} &=& 2 R_{4,t}\, dt - R_{2,t} \sqrt{\eta_1}\, dy_{t,1} + R_{1,t} \sqrt{\eta_2}\, dy_{t,2} \quad \text{for } R_4 = XY_s / r(ZZ)\\
dR_{5,t} &=& 3 R_{5,t}\, dt - R_{3,t} \sqrt{\eta_1}\, dy_{t,1} - R_{4,t} \sqrt{\eta_2}\, dy_{t,2} \quad \text{for } R_5 = X_d / r(ZZ)\\
dR_{6,t} &=& 3 R_{6,t}\, dt + R_{4,t} \sqrt{\eta_1}\, dy_{t,1} - R_{3,t} \sqrt{\eta_2}\, dy_{t,2} \quad \text{for } R_6 = Y_s / r(ZZ) 
\end{eqnarray*}
for the remaining block of 6 variables. This forms the basis for identifying variables that evolve deterministically, such as $R_1$ and $R_2$.
\end{itemize}

\paragraph*{A set of deterministically evolving variables:} We need to identify 13 independent deterministic combinations among the 15 equations above. 
\begin{itemize}
\item Since $B_1$ and $B_2$ are influenced by independent measurement records $dy_2$ or $dy_1$, we would keep those as representing the 2 stochastic variables. 
\item We next try to eliminate the stochastic term involving just $B_1$ in the expression of $dB_{3,t}$. By \Ito~calculus, we note that $d(\frac{B_1^2}{2}) = \big(2 \frac{B_1^2}{2} + 2 \eta_2\big) \, dt - 2 B_1 \sqrt{\eta_2}\, dy_2$, such that we can deduce
$$d\tilde{B}_{3,t} = (2 \tilde{B}_{3,t} + \eta_2)\, dt \quad \text{for } \tilde{B}_3 = (\tfrac{B_1^2}{4}+B_3) \; .$$
A similar treatment with $d(\frac{B_2^2}{2})$ and $d(\frac{B_1 B_2}{2})$ yields 
\begin{eqnarray*}
d\tilde{B}_{4,t} &=& (2 \tilde{B}_{4,t} + \eta_1)\, dt \quad \text{for } \tilde{B}_4 = (\tfrac{B_2^2}{4}-B_4) \; ,\\
d\tilde{B}_{5,t} &=& 2 \tilde{B}_{5,t}\, dt \quad \text{for } \tilde{B}_5 = (\tfrac{B_1 B_2}{2}+B_5) \; .
\end{eqnarray*}
\item The stochastic terms in $dB_{6,t}$ must be eliminated with a combination of $d(B_1^2)$ and $d(B_2^2)$. For the stochastic terms in $dB_{7,t}$ and $dB_{8,t}$, first we use products like e.g.~$d(B_4 B_2)$ to cancel $B_{4,t} \sqrt{\eta_1}\, dy_{t,1}$, then we add cubic terms in $B_1,\, B_2$ to eliminate the remaining terms. For $dB_0$, after $d(B_7 B_2 + B_8 B_1)$, adding $d(B_3^2+B_4^2+B_5^2+B_6^2)$ cancels all stochastic terms.
\item A similar procedure is applied to the $R_k$, combining them in products with $B_1,B_2$. Overall, this yields the following final result.
\end{itemize}

\begin{prop}\label{prop:LewRes}
The system \eqref{eqL:sde} is confined to a two-dimensional manifold and its solution can be written as follows (see above for notation). The dependence on measurement outcomes is characterized by:
\begin{eqnarray*}
B_{1}(t) &=& B_{1}(0) \, e^t - 2\sqrt{\eta_2}\, \int_0^t \, e^{t-s} \, dy_{s,2}  \;\; , \quad \text{for } \;\; B_1 = \tfrac{YZ_d}{r(ZZ)} \; , \\
B_{2}(t) &=& B_{2}(0) \, e^t  -2\sqrt{\eta_1}\, \int_0^t \, e^{t-s} \, dy_{s,1} \;\; , \quad \text{for } \;\;B_2 = \tfrac{XZ_s}{r(ZZ)} \; . 
\end{eqnarray*}
The other variables evolve independently of the measurement record, according to:
\begin{eqnarray*}
\tilde{B}_3(t) &=& \tilde{B}_{3}(0) \, e^{2t} + \tfrac{\eta_2}{2}(e^{2t}-1)  \;\; , \qquad \text{for }\;\; \tilde{B}_3 = (\tfrac{B_1^2}{4}+\tfrac{r(YY)}{r(ZZ)})  \; , \\
\tilde{B}_4(t) &=& \tilde{B}_{4}(0) \, e^{2t} + \tfrac{\eta_1}{2}(e^{2t}-1)  \;\; , \qquad \text{for }\;\; \tilde{B}_4 = (\tfrac{B_2^2}{4}-\tfrac{r(XX)}{r(ZZ)})  \; , \\
\tilde{B}_5(t) &=& \tilde{B}_{5}(0) \, e^{2t}  \;\; , \qquad  \qquad  \qquad  \qquad \text{for } \;\;\tilde{B}_5 = (\tfrac{B_1 B_2}{2}+\tfrac{XY_d}{r(ZZ)})  \; , \\
\tilde{B}_6(t) &=& \tilde{B}_{6}(0) \, e^{2t} + \tfrac{\eta_1+\eta_2}{2}(e^{2t}-1)  \;\; , \quad \text{for }\;\; \tilde{B}_6 = (\tfrac{B_1^2+B_2^2}{4}+\tfrac{Z_s}{r(ZZ)})  \; , \\
\tilde{B}_7(t) &=& \tilde{B}_{7}(0) \, e^{3t}  \;\; , \qquad \text{for } \;\;\tilde{B}_7 = (B_4 B_2 - \tfrac{B_6 B_2}{2} - \tfrac{B_5 B_1}{2} - \tfrac{B_1^2 B_2}{4} - \tfrac{B_2^3}{4} +\tfrac{X_s}{r(ZZ)})  \; , \\
\tilde{B}_8(t) &=& \tilde{B}_{8}(0) \, e^{3t}  \;\; , \qquad \text{for }\;\; \tilde{B}_8 = (-B_3 B_1 - \tfrac{B_5 B_2}{2} - \tfrac{B_6 B_1}{2} - \tfrac{B_2^2 B_1}{4} - \tfrac{B_1^3}{4} +\tfrac{Y_d}{r(ZZ)})  \; , \\
\tilde{B}_0(t) &=& \tilde{B}_0(0) \, e^{4t} + \big({\scriptstyle{(\eta_1+\eta_2)\tilde{B}_6(0)+2\eta_1\tilde{B}_4(0)+2\eta_2\tilde{B}_3(0)+\tfrac{(\eta_1+\eta_2)^2}{2}+\eta_1^2+\eta_2^2}}\big)\, \tfrac{e^{4t}-e^{2t}}{2}\\ && \qquad\qquad + \big({\scriptstyle{\tfrac{(\eta_1+\eta_2)^2}{2}+\eta_1^2+\eta_2^2}}\big)\,  \tfrac{e^{4t}-1}{4} \; ,\\
&& \qquad\qquad\qquad\quad \text{for } \;\;\tilde{B}_0 = (\tfrac{B_7 B_2}{2} + \tfrac{B_8 B_1}{2} + \tfrac{B_3^2 +B_4^2+B_5^2+B_6^2}{2} +\tfrac{1}{r(ZZ)}) \; , \\
\end{eqnarray*}
\begin{eqnarray*}
R_1(t) &=& R_1(0) \, e^t  \;\; , \qquad \text{for } \;\; R_1 = \tfrac{XZ_d}{r(ZZ)} \; , \\
R_2(t) &=& R_2(0) \, e^t  \;\; , \qquad \text{for } \;\; R_2 = \tfrac{YZ_s}{r(ZZ)} \; , \\
\tilde{R}_3(t) &=& \tilde{R}_3(0) \, e^{2t}  \;\; , \qquad \text{for }\;\; \tilde{R}_3 =   (-\tfrac{R_1 B_2}{2} - \tfrac{R_2 B_1}{2} +\tfrac{Z_d}{r(ZZ)}) \; , \\
\tilde{R}_4(t) &=& \tilde{R}_4(0) \, e^{2t}  \;\; , \qquad \text{for }\;\; \tilde{R}_4 =   (\tfrac{R_1 B_1}{2} - \tfrac{R_2 B_2}{2} +\tfrac{XY_s}{r(ZZ)}) \; , \\
\tilde{R}_5(t) &=& \big( \tilde{R}_5(0)+\tfrac{\eta_2-\eta_1}{4}\,R_1(0) \big) \, e^{3t} - \tfrac{\eta_2-\eta_1}{4} R_1(0) \, e^t  \; , \\
&& \quad \text{for }\;\; \tilde{R}_5 = (-\tfrac{R_3 B_2}{2} - \tfrac{R_4 B_1}{2} + \tfrac{R_2 B_1 B_2}{4} + \tfrac{R_1 B_2^2}{8} - \tfrac{R_1 B_1^2}{8} +\tfrac{X_d}{r(ZZ)}) \; , \\
\tilde{R}_6(t) &=& \big( \tilde{R}_6(0)+\tfrac{\eta_1-\eta_2}{4}\,R_2(0) \big) \, e^{3t} - \tfrac{\eta_1-\eta_2}{4} R_2(0) \, e^t \; , \\
&& \quad \text{for }\;\; \tilde{R}_6 = (-\tfrac{R_3 B_1}{2} + \tfrac{R_4 B_2}{2} + \tfrac{R_1 B_1 B_2}{4} + \tfrac{R_2 B_1^2}{8} - \tfrac{R_2 B_2^2}{8} +\tfrac{Y_s}{r(ZZ)}) \; .
\end{eqnarray*}
\end{prop}

This is but one way to describe the dynamics, as combinations of these variables yield other deterministic expressions, which the reader may set up according to preferences. For instance, $R_1/R_2$ is constant.


\subsection{~~~~A qudit monitored through $n$ harmonic oscillators}\label{app:ssec:PRmod}

Finally, we consider the setting of Section \ref{sec:PRmodel}. The corresponding solutions extend those of Section \ref{ssec:app:iqm}, which considered coupling to a single harmonic oscillator subject to heterodyne fluorescence measurement with equal rates. If the qudit is replaced by a single level, we recover the situation of fluorescence measurement on a cavity, as in Section \ref{ssec:app:aandia}, now generalized to allow arbitrary rates on the two measurement channels and the presence of a detuning Hamiltonian $\Delta\, {\bf a}^\dagger {\bf a}$. However, thermal relaxation is not included in the analysis of this section.

We provide many details to illustrate a last time how the procedure with Wigner functions works.\\

\paragraph*{Algebraic criterion:}  The measurement vector fields $G_{{\bf b}_l}$ all commute. Commutation with the deterministic terms, for $u_t=v_t=0$, yield linear combinations of the ${\bf a}_k$, possibly premultiplied by $\q{j}\qd{j}$. Explicitly, writing ${\bf q} = \sum_j q_{j,k}\; \q{j}\qd{j} \otimes {\bf a}_k$, one commutation of $G_{{\bf q}}$ with the deterministic vector field amounts to the linear map:
\begin{equation}\label{eq:gammaupdate}
q_{j,k} \mapsto \left(\,  {\bf I}_d \otimes (\beta \beta^\dagger + i \Delta)  + i X \, \right) \; q_{j,k} \; ,
\end{equation}
where $q_{j,k}$ is considered a column vector, in tensor product structure of the $j$ and $k$ coordinates; column $l$ of the matrix $\beta$ contains the components $\beta_{l,k}$; matrix $\Delta$ is diagonal with components $\Delta_k$; and matrix $X$ is diagonal with components $\chi_{j,k}$ (according to the same tensor product structure).

For generic values of $\Delta, X, \beta$, independently of the number of measurement channels $m$, the map \eqref{eq:gammaupdate} generates all $2 d n$ independent real linear combinations of $\{\, \q{j}\qd{j} \otimes {\bf a}_k , \; i\q{j}\qd{j} \otimes {\bf a}_k : j=1,...,d ; k=1,...,n \, \}$. By commuting the latter with the drive terms, in $u_t$ and $v_t$, we obtain contributions in $G_{\q{j}\qd{j} \otimes {\bf I}}$ and $G_{i\q{j}\qd{j} \otimes {\bf I}}$. Thus, taking into account that $G_{i\mathbf{I}} = G_{\mathbf{I}}=0$, in total the dimension of the manifold on which measurement outputs can spread the state is $2 d (n+1)-2$, in  presence of control signals and for general initial conditions; for $u_t=v_t=0$ it reduces to dimension $2d n$.\\ 

\paragraph*{Stochastic differential equations:} Like for our other examples with harmonic oscillators, we consider their Wigner function representation, involving a Gaussian kernel to model the low stochastic dimension.
\begin{itemize}
\item Thanks to the dispersive coupling with the qudit, we can write
\begin{equation}\label{eq:appl:rhojj'}
\rho_{t} = \sum_{j,j'=1}^d\; \q{j}\qd{j'} \otimes \rho_{t,(j,j')} \; ,
\end{equation}
where each $\rho_{t,(j,j')}$ is an operator on the Hilbert space of the $n$ cavities. For $j=j'$, these are unnormalized density operators, while for $j \neq j'$ they need not be hermitian nor positive. Conditioned on the measurement records, the $\rho_{t,(j,j')}$ evolve independently of each other. Indeed, the only place where one qudit level can influence another, is when taking the trace to define the $dy_{t,l}$.
\item We represent each $\rho_{t,(j,j')}$ via its multi-variate Wigner function $\mathcal{W}^{\{\rho_{t,(j,j')}\}}(q)$. This is just the straightforward generalization of the single oscillator case, i.e.~a pseudo-probability distribution over $2n$ variables, corresponding to the column vector $q=(x_1,p_1,x_2,p_2,...,x_n,p_n)$. Taking its marginal with respect to e.g.~$(p_1,x_2,p_3,x_4,...)$ gives the (truly real) probability distribution for a measurement of $x_1 \otimes p_2 \otimes x_3 \otimes ...$ which is indeed a valid observable. By using the translations to Wigner space of annihilation and creation operators acting on $\rho$, as recalled right before Section \ref{ssec:app:aandia}, the stochastic master equation \eqref{eq:PRmodel0},\eqref{eq:PRmodel1} translates into:
\begin{eqnarray} \nonumber
d\mathcal{W}^{(j,j')}_t \!\!\!&=& \sum_{k=1}^n \; \left(u_{t,k} \frac{\partial}{\partial p_k} - v_{t,k} \frac{\partial}{\partial x_k} \right) \, \mathcal{W}^{(j,j')}_t\, dt + \delta_k^{(j,j')}\; \left(x_k \frac{\partial}{\partial p_k} - p_k \frac{\partial}{\partial x_k} \right) \, \mathcal{W}^{(j,j')}_t\,dt \\ \label{eq:appl:WigPDE}
&& + \sum_{k=1}^n \; - i\, \omega_k^{(j,j')} \; \left( 2\, x_k^2 + 2\, p_k^2 -1 -\frac{1}{8}\frac{\partial^2}{\partial x_k^2} - \frac{1}{8}\frac{\partial^2}{\partial p_k^2} \right)  \, \mathcal{W}^{(j,j')}_t\,dt \\ \nonumber
&& + \sum_{l,k,k'} \beta_{l,k} \bar{\beta}_{l,k'} \; \left( \mathbf{1}_{k,k'}+ \tfrac{1}{8}(\tfrac{\partial}{\partial x_k} + i \tfrac{\partial}{\partial p_k})(\tfrac{\partial}{\partial x_{k'}} - i \tfrac{\partial}{\partial p_{k'}})\right) \mathcal{W}^{(j,j')}_t\,dt \\ \nonumber
&& + \sum_{l,k,k'} \frac{\beta_{l,k} \bar{\beta}_{l,k'}}{4} \; \left(\, (x_k+i p_k )(\tfrac{\partial}{\partial x_{k'}} \text{-} i \tfrac{\partial}{\partial p_{k'}}) +(x_{k'}\text{-}i p_{k'}) (\tfrac{\partial}{\partial x_{k}} + i \tfrac{\partial}{\partial p_{k}}) \, \right)  \mathcal{W}^{(j,j')}_t\,dt \\ \nonumber
&& \!\!\! \!\!\! \!\!\! \!\!\! \!\!\! + \sum_{l=1}^m \, 2 \sqrt{\eta_l} \left(\sum_{k=1}^n \mathfrak{R}(\beta_{l,k})\left(x_k + \tfrac{1}{4}\tfrac{\partial}{\partial x_k}\right) - \mathfrak{I}(\beta_{l,k})\left(p_k + \tfrac{1}{4}\tfrac{\partial}{\partial p_k}\right) - s_l \right)\, \mathcal{W}^{(j,j')}_t \; dw_{t,l} \; .
\end{eqnarray}
Here $\delta_k^{(j,j')} = \Delta_k + \frac{\chi_{j,k}+\chi_{j',k}}{2}$; $\omega_k^{(j,j')} = \frac{\chi_{j,k}-\chi_{j',k}}{2}$; the indicator function $ \mathbf{1}_{k,k'}$ equals 1 if $k=k'$ and 0 otherwise; $\bar{a}$ denotes the complex conjugate of $a$ (without transpose, for vectors and matrices, i.e.~$\bar{a}^\dagger$ is the transpose of $a$); $\mathfrak{R}$ and $\mathfrak{I}$ respectively denote real and imaginary part; and the measurement outcomes are described by
\begin{equation}\label{eq:appl:dy}
dy_{t,l} = 2 \sqrt{\eta_l}\, s_l\, dt + dw_{t,l} \; = \; 2 \sqrt{\eta_l}\, \sum_k (\; \mathfrak{R}(\beta_{l,k}) \text{Tr}(\rho_t \, \mathbf{X}_k) - \mathfrak{I}(\beta_{l,k}) \text{Tr}(\rho_t \, \mathbf{P}_k)  \; ) \, dt + dw_{t,l} \; .
\end{equation}
In a more compact notation, we have:
\begin{eqnarray} \nonumber
d\mathcal{W}^{(j,j')}_t &=& \left(\, u^\dagger (\mathbf{I}_n \otimes J) + q^\dagger (\delta^{(j,j')} \otimes J) \,\right)\,  \text{grad}_q  \, \mathcal{W}^{(j,j')}_t\,dt \\ \nonumber
&& \; -i\,\left(\, q^\dagger (2\omega^{(j,j')} \otimes \mathbf{I}_2) q - \text{Tr}(\omega^{(j,j')}) - \tfrac{1}{8}\text{Tr}( (\omega^{(j,j')} \otimes \mathbf{I}_2)\, \text{hess}_{q} )\, \right) \, \mathcal{W}^{(j,j')}_t\,dt \\ \nonumber
&& + \left(  \frac{1}{2}\text{Tr}(\tilde{R})  + \frac{1}{2} q^\dagger \tilde{R}\, \text{grad}_q + \frac{1}{8}\text{Tr}( \tilde{R}\, \text{hess}_q) \right)   \, \mathcal{W}^{(j,j')}_t\,dt 
\\ \label{eq:appl:WigPDE2}
&&  + \sum_{l=1}^m \, 2 \sqrt{\eta_l} \left( \beta_{l,A}^\dagger q + \tfrac{1}{4}\beta_{l,A}^\dagger\, \text{grad}_q - s_l \right) \mathcal{W}^{(j,j')}_t \; dw_{t,l} \;  ; \\
dy_{t,l} &=& 2 \sqrt{\eta_l}\, \beta_{l,A}^\dagger \, Q_{t} \; dt + dw_{t,l} \; .
\end{eqnarray}
Here $J=[0 \; ,\; 1 \; ;\; -1 \; ,\; 0 ] \in \mathbb{R}^{2 \times 2}$; the diagonal matrices $\delta^{(j,j')}$ and $\omega^{(j,j')}$ $\in \mathbb{R}^{n \times n}$ contain e.g.~$\delta_k^{(j,j')}$ as component $k$; the vector $u \in \mathbb{R}^2$ contains the control inputs in natural order; $\text{grad}_q$ and $\text{hess}_q$ respectively denote the gradient and hessian with respect to $q$; and column vectors $\beta_{l,A} = [\mathfrak{R}(\beta_{l,1}),\; -\mathfrak{I}(\beta_{l,1}),\; \mathfrak{R}(\beta_{l,2}),\; -\mathfrak{I}(\beta_{l,2}),\; ...  ]$ , $Q_{t} = [\,\text{Tr}(\rho_t\, \mathbf{X}_1),\;\text{Tr}(\rho_t\, \mathbf{P}_1),\;\text{Tr}(\rho_t\, \mathbf{X}_2),\;\text{Tr}(\rho_t\, \mathbf{P}_2),\;... \,] \; \in \mathbb{R}^{2n}$. We have further defined
\begin{eqnarray*}
\tilde{R} &=& \sum_{l=1}^m  \; R^{(re)}_l \otimes I_2 + R^{(im)}_l \otimes J \quad  \in \mathbb{R}^{2n \times 2n}\\
&& \text{with } R^{(re)}_{l,(k,k')} = \mathfrak{R}(\beta_{l,k}\bar{\beta}_{l,k'}) \;\; \text{ and } \;\;  R^{(im)}_{l,(k,k')} = \mathfrak{I}(\beta_{l,k}\bar{\beta}_{l,k'}) \; .
\end{eqnarray*}
Note that $\tilde{R}$ is real symmetric and $\text{Tr}(\tilde{R}) = 2 \sum_l \beta_{l,A}^\dagger \beta_{l,A}$. For the single cavity setting of Section \ref{ssec:app:iqm}, we just have $\tilde{R}=2\, \mathbf{I}_{2n}$. For further compactness, we will use below the notation $\tilde{J} = (\mathbf{I}_n \otimes J)$, $\tilde\delta^{(j,j')} = (\delta^{(j,j')} \otimes J)$ and $\tilde\omega^{(j,j')} = -i(\omega^{(j,j')} \otimes \mathbf{I}_2)$. All these matrices belong to $\mathbb{C}^{2n \times 2n}$, the first two are real skew-symmetric and the last one is imaginary diagonal; they are thus all skew-hermitian, i.e.~generators of rotations, as one would guess from the parameters they involve.

\item Describing a distribution of $2n$ variables involves, in full generality, a number of parameters which is exponential in $n$. In the present case, thanks to the low-dimensional confinement, we can reduce this complexity essentially to the initial state only: from there, the evolution in time involves a Gaussian and thus a number of parameters at most quadratic in $n$. Moreover, in agreement with the algebraic criterion, the number of parameters influenced by the stochastic measurement outcomes is only linear in $n$. Explicitly, we write
\begin{eqnarray}\label{eq:appl:WtoG}
\mathcal{W}^{\{\rho_{t,(j,j')}\}}(q) &=& \frac{\mu_t^{(j,j')}}{\pi^{n}\text{det}(S_t^{(j,j')})^{1/2}}\int \mathcal{W}^{\{\rho_0^{(j,j')}/\mu_0^{(j,j')}\}}(q_0) \, K^{(j,j')}_{t}(q,q_0)\, dq_0 \; , \\ \nonumber
K^{(j,j')}_{t}(q,q_0) &=& \exp\Bigg[-\left(q\text{-}\bar{\gamma}^{(j,j')}_t\text{-}\bar{M}^{(j,j')}_t q_0\right)^\dagger\; (S_t^{(j,j')})^{-1}\; \left(q\text{-}\gamma^{(j,j')}_t\text{-}M^{(j,j')}_t q_0\right) \\ \nonumber
&& \phantom{KKKKKKKKK} \; + q_0^\dagger D_t^{(j,j')} q_0 + (\bar{\lambda}^{(j,j')})^\dagger q_0 \; \Bigg] \; .
\end{eqnarray}
The time-dependent parameters whose equations we seek are scalars $\mu_t^{(j,j')} \in \mathbb{C}$; vectors $\gamma_t^{(j,j')}$ and $\lambda_t^{(j,j')}$  $\in \mathbb{C}^{2n}$; and matrices $S_t^{(j,j')}$, $M_t^{(j,j')}$ and $D_t^{(j,j')}$ $\in \mathbb{C}^{2n \times 2n}$. For $j=j'$, these parameters are all real. For $j \neq j'$, they can be complex and we have introduced some (not transposed) conjugates e.g.~$\bar{\gamma}_t^{(j,j')}$ to simplify expressions below. By analogy with Section \ref{ssec:app:iqm}, we have anticipated that the variables in latin letter will evolve deterministically, while those in greek letters will involve a stochastic component.

In order to reach the correct dimension $2d(n+1)-2$ for the deterministically evolving manifold containing all trajectories, we can view the $\gamma^{(j,j)}$ as independent stochastic variables (thus $(2n)d$ real values), as well as part of the $\mu^{(j,j')}$ (covering the $2d-2$ remaining variables); all the other parameters, in particular the $\lambda^{(j,j')}$ and the $\gamma^{(j\neq j')}$, will be deterministically related to them.

\item We next express $d\mathcal{W}^{(j,j')}_t$ in two ways. On the left-hand side, viewing $\mathcal{W}^{(j,j')}_t$ as a function of the Gaussian parameters, we express it with $d\mu_{t,(j,j')}, d\gamma_t^{(j,j')}, ...$, going up to second order Taylor expansion of the Gaussian in order to cover the \Ito~correction. On the right-hand side, viewing $\mathcal{W}^{(j,j')}_t$ as a function of $q$, we compute the right-hand side of \eqref{eq:appl:WigPDE2} for the particular Gaussian form \eqref{eq:appl:WtoG}. By separating, under the integral, the terms involving various powers of $q, q_0$ and the deterministic and stochastic parts, we obtain the following set of equations. To reduce clutter, we drop the indices $^{(j,j')}$ and introduce $Z_t=-S_t^{-1}$:

\begin{align} \label{eq:appl:F1}
q^\dagger\, & dZ_t\, q  + \tfrac{1}{2} q^\dagger (Z_t+\bar{Z}_t^\dagger) d\gamma_t\, d\bar{\gamma}_t^\dagger  (Z_t+\bar{Z}_t^\dagger)^\dagger \, q \\ \nonumber
& \;\; = q^\dagger\, [\tilde{\delta},\, Z_t]\, q \, dt + \tfrac{1}{2} q^\dagger\, \{ \tilde{R},\, Z_t \} \, q \, dt + 2 q^\dagger \, \tilde{\omega} \, q \, dt + \tfrac{1}{8}  q^\dagger\, (Z_t+\bar{Z}_t^\dagger)\, (\tilde{R} - \tilde{\omega})\, (Z_t+\bar{Z}_t^\dagger) \, q \, dt \\[4mm]
\nonumber
q_0^\dagger\, & dD_t \, q_0  + q_0^\dagger\, (d\bar{M}^\dagger_t Z_t M_t + \bar{M}^\dagger_t Z_t dM_t) \, q_0  +  q_0^\dagger \, (\bar{M}^\dagger_t dZ_t M_t ) \, q_0 \\  \label{eq:appl:F2} 
+ \tfrac{1}{2}  & q_0^\dagger\, \left(d\lambda_t + \bar{M}_t^\dagger\, (Z_t+\bar{Z}_t^\dagger)\, d\gamma_t \right) \left(d\bar{\lambda}_t^\dagger + d\bar{\gamma}_t^\dagger (Z_t+\bar{Z}_t^\dagger) M_t \right)  \, q_0 \\  \nonumber
& \;\; = \tfrac{1}{8} \; q_0^\dagger\, \bar{M}^\dagger \, (Z_t+\bar{Z}_t^\dagger) \,  (\tilde{R} - \tilde{\omega}) \, (Z_t+\bar{Z}_t^\dagger) \, M \, q_0 \, dt\\[4mm] \nonumber
q_0^\dagger\, & \bar{M}_t^\dagger (dZ_t+d\bar{Z}_t^\dagger) \, q +  q_0^\dagger\, d\bar{M}_t^\dagger (Z_t+\bar{Z}_t^\dagger) \, q +  q_0^\dagger\, \left(d\lambda_t + \bar{M}_t^\dagger\, (Z_t+\bar{Z}_t^\dagger)\, d\gamma_t \right)d\bar{\gamma}_t^\dagger (Z_t+\bar{Z}_t^\dagger) \, q \\  \label{eq:appl:F3}
& \;\; = -q_0^\dagger\, \bar{M}_t^\dagger (Z_t+\bar{Z}_t^\dagger)\, (\tilde{\delta}-\tfrac{1}{2}\tilde{R} \, q \, dt + \tfrac{1}{4} q_0^\dagger\, \bar{M}_t^\dagger (Z_t+\bar{Z}_t^\dagger)\,(\tilde{R}-\tilde{\omega})\,(Z_t+\bar{Z}_t^\dagger)\, q \, dt \\[4mm] \nonumber
q^\dagger\, & (Z_t+\bar{Z}_t^\dagger)\, d\gamma_{t,\text{stochastic}} \\ \label{eq:appl:F4}
& \;\; = -2 q^\dagger \, (\mathbf{I}_{2n}+\tfrac{Z_t+\bar{Z}_t^\dagger}{4})\; {\textstyle \sum_{l=1}^m} \; \sqrt{\eta_l} \, \beta_{l,A} \; dw_{t,l}\\[4mm] \nonumber
q_0^\dagger\, & \left(\, d\lambda_{t,\text{stochastic}}+\bar{M}^\dagger_t\, (Z_t+\bar{Z}_t^\dagger)\, d\gamma_{t,\text{stochastic}}\, \right) \\
\label{eq:appl:F5}
& \;\; = -2 q_0^\dagger \, \bar{M}^\dagger \, \tfrac{Z_t+\bar{Z}_t^\dagger}{4} \; {\textstyle \sum_{l=1}^m} \; \sqrt{\eta_l} \, \beta_{l,A} \; dw_{t,l}\\[4mm] \nonumber
&\!\!\!\!\!\! \frac{d\mu_{t,\text{stochastic}}}{\mu_t}\, + \bar{\gamma}^\dagger_t \, (Z_t+\bar{Z}_t^\dagger)\, d\gamma_{t,\text{stochastic}}  \\
\label{eq:appl:F6}
& \;\; = -2 {\textstyle \sum_{l=1}^m} \; \sqrt{\eta_l} \, \left( \bar{\gamma}_t^\dagger \, \tfrac{Z_t+\bar{Z}_t^\dagger}{4} \, \beta_{l,A} + s_l \right) \; dw_{t,l}\\[4mm] \nonumber
q^\dagger\, & (dZ_t+d\bar{Z}_t^\dagger)\, \gamma_t + q^\dagger\,(Z_t+\bar{Z}_t^\dagger)\, d\gamma_{t,\text{det}} \\ \nonumber
+& q^\dagger \, (Z_t+\bar{Z}_t^\dagger)\, d\gamma_t \frac{d\mu_t}{\mu_t} + q^\dagger\, (Z_t+\bar{Z}_t^\dagger)\, d\gamma_t\, \bar{\gamma}^\dagger_t \,  (Z_t+\bar{Z}_t^\dagger)\, d\gamma_t \\ \label{eq:appl:F7}
& \;\; = q^\dagger\,(Z_t+\bar{Z}_t^\dagger)\, \tilde{J} \, u_t \, dt + q^\dagger\, (\tilde{\delta}+ \tfrac{1}{2}\tilde{R})\, (Z_t+\bar{Z}_t^\dagger)\, \gamma_t \, dt \\ \nonumber
& \;\; \phantom{=} \; + \tfrac{1}{4} q^\dagger\,(Z_t+\bar{Z}_t^\dagger)\, (\tilde{R}-\tilde{\omega}) \, (Z_t+\bar{Z}_t^\dagger)\, \gamma_t \, dt\\[4mm] \nonumber
q_0^\dagger\, & d\lambda_{t,\text{det}} + q_0^\dagger\, \bar{M}^\dagger_t \, (dZ_t+d\bar{Z}_t^\dagger)\, \gamma_t + q_0^\dagger\, d\bar{M}_t^\dagger \, (Z_t+\bar{Z}_t^\dagger)\, \gamma_t + q_0^\dagger\, \bar{M}^\dagger_t \, (Z_t+\bar{Z}_t^\dagger)\, d\gamma_{t,\text{det}}\\  \label{eq:appl:F8}
+&  q_0^\dagger \left(\bar{M}^\dagger_t \, (Z_t+\bar{Z}_t^\dagger)\, d\gamma_t + d\lambda_t \right)\; \left(\tfrac{d\mu_t}{\mu_t} + \bar{\gamma}^\dagger_t \, (Z_t+\bar{Z}_t^\dagger)\, d\gamma_t \right) \\ \nonumber
& \;\; = q_0^\dagger\, \bar{M}^\dagger_t \, (Z_t+\bar{Z}_t^\dagger)\, \tilde{J}\, u_t \, dt + \tfrac{1}{4}\, q_0^\dagger\,\bar{M}^\dagger_t \, (Z_t+\bar{Z}_t^\dagger)\,(\tilde{R}-\tilde{\omega}) \, (Z_t+\bar{Z}_t^\dagger)\, \gamma_t\\[4mm] \nonumber
&\!\!\!\!\!\! \frac{d\mu_{t,\text{det}}}{\mu_t} + \frac{d(\text{det}(Z_t)^{1/2})}{\text{det}(Z_t)^{1/2}} + \bar{\gamma}^\dagger_t\, dZ_t \, \gamma_t  + \bar{\gamma}^\dagger_t\, (Z_t+\bar{Z}_t^\dagger)\, d\gamma_{t,det} \\ \nonumber
+ & d \bar{\gamma}^\dagger_t\, Z_t \, d\gamma_t + \tfrac{d\mu_t}{\mu_t} \, \bar{\gamma}^\dagger_t\, (Z_t+\bar{Z}_t^\dagger)\, d\gamma_t + \tfrac{1}{2} \left( \bar{\gamma}^\dagger_t\, (Z_t+\bar{Z}_t^\dagger)\, d\gamma_t \right)^2 \\  \label{eq:appl:F9}
& \;\; =  \bar{\gamma}^\dagger_t\, (Z_t+\bar{Z}_t^\dagger)\, \tilde{J}\, u_t\, dt + \tfrac{1}{8} \bar{\gamma}^\dagger_t\, (Z_t+\bar{Z}_t^\dagger)\, (\tilde{R}-\tilde{\omega})\,(Z_t+\bar{Z}_t^\dagger)\, \gamma_t \, dt \\ \nonumber 
& \;\; \phantom{=} \; + \tfrac{1}{2}\, \text{Tr}\left(\, (\mathbf{I}_{2n} + \tfrac{Z_t+\bar{Z}_t^\dagger}{4})\,(\tilde{R}-\tilde{\omega}) \right)\, dt \; .
\end{align}
\item Using \eqref{eq:appl:F4}-\eqref{eq:appl:F6}, we express all the quadratic terms appearing on the left-hand side of the other equations, i.e.~the \Ito~terms:
\begin{align}
\label{eq:appl:S1} & (Z_t+\bar{Z}_t^\dagger)\, d\gamma_t \, d\bar{\gamma}^\dagger_t \, (Z_t+\bar{Z}_t^\dagger) = 4 (\mathbf{I}_{2n} + \tfrac{Z_t+\bar{Z}_t^\dagger}{4})\, B \, (\mathbf{I}_{2n} + \tfrac{Z_t+\bar{Z}_t^\dagger}{4})\, dt \\
\label{eq:appl:S2} & \!\!\left(\, d\lambda_t +\bar{M}^\dagger_t\, (Z_t+\bar{Z}_t^\dagger)\, d\gamma_t\, \right) \!\! \left( d\bar{\lambda}^\dagger_t + d\bar{\gamma}^\dagger_t \, (Z_t+\bar{Z}_t^\dagger)\, M_t \right) = 4 \bar{M}^\dagger_t\,  \tfrac{Z_t+\bar{Z}_t^\dagger}{4} \, B \,  \tfrac{Z_t+\bar{Z}_t^\dagger}{4} \, M_t \, dt \\
\label{eq:appl:S3} & \left(\, d\lambda_t +\bar{M}^\dagger_t\, (Z_t+\bar{Z}_t^\dagger)\, d\gamma_t\, \right) \, d\bar{\gamma}^\dagger_t \, (Z_t+\bar{Z}_t^\dagger) = \bar{M}^\dagger_t\,(Z_t+\bar{Z}_t^\dagger)\, B \, (\mathbf{I}_{2n} + \tfrac{Z_t+\bar{Z}_t^\dagger}{4}) \; dt \\
\label{eq:appl:S4} & (Z_t+\bar{Z}_t^\dagger)\, d\gamma_t \, \left( \tfrac{d\mu_t}{\mu_t} + \bar{\gamma}^\dagger_t (Z_t+\bar{Z}_t^\dagger) d\gamma_t  \right) = 4\, (\mathbf{I}_{2n} + \tfrac{Z_t+\bar{Z}_t^\dagger}{4})\, Z_t\, dt \\ \nonumber
& \phantom{KKKK} \text{ with } Z_t = B \, \tfrac{Z_t+\bar{Z}_t^\dagger}{4}\, \gamma_t  + {\textstyle \sum_l }\; \eta_l s_l \, \beta_{l,A} \\
\label{eq:appl:S5} &  \left(\bar{M}^\dagger_t \, (Z_t+\bar{Z}_t^\dagger)\, d\gamma_t + d\lambda_t \right)\; \left(\tfrac{d\mu_t}{\mu_t} + \bar{\gamma}^\dagger_t \, (Z_t+\bar{Z}_t^\dagger)\, d\gamma_t \right) = 4 \bar{M}^\dagger_t \,  \tfrac{Z_t+\bar{Z}_t^\dagger}{4} \,  Z_t \, dt \\
\label{eq:appl:S6} & d\bar{\gamma}^\dagger_t \, Z_t \, d\gamma_t = 2\, \text{Tr}\left(\,  (Z_t+\bar{Z}_t^\dagger)^{-1} \, (\mathbf{I}_{2n} + \tfrac{Z_t+\bar{Z}_t^\dagger}{4})\, B\, (\mathbf{I}_{2n} + \tfrac{Z_t+\bar{Z}_t^\dagger}{4})\,  \right) \, dt \\
\label{eq:appl:S7} &
 \tfrac{d\mu_t}{\mu_t} \, \bar{\gamma}^\dagger_t\, (Z_t+\bar{Z}_t^\dagger)\, d\gamma_t + \tfrac{1}{2} \left( \bar{\gamma}^\dagger_t\, (Z_t+\bar{Z}_t^\dagger)\, d\gamma_t \right)^2 \\ \nonumber
& \phantom{KKK} = 4 \bar{\gamma}^\dagger_t\,  (\mathbf{I}_{2n} + \tfrac{Z_t+\bar{Z}_t^\dagger}{4})\, Z_t \, dt  - 2 \bar{\gamma}^\dagger_t (\mathbf{I}_{2n} + \tfrac{Z_t+\bar{Z}_t^\dagger}{4})\, B \, (\mathbf{I}_{2n} + \tfrac{Z_t+\bar{Z}_t^\dagger}{4})\, \gamma_t \, dt
\end{align}
Here we have introduced $B = \sum_{l=1}^m \eta_l \beta_{l,A}\beta_{l,A}^\dagger \; \in \mathbb{R}^{2n \times 2n}$. For the single cavity setting of Section \ref{ssec:app:iqm}, we just have $B = \eta\, \mathbf{I}_{2n}$.\\
\end{itemize}

\paragraph*{A set of deterministically evolving variables:} We solve these equations sequentially. 
\begin{itemize}
\item First, from \eqref{eq:appl:F1} and \eqref{eq:appl:S1}, we obtain an autonomous, deterministic equation for $Z_t$, and thus for $S_t$ in the original formulation. Decomposing it in a symmetric part $Z_S=(Z+\bar{Z}^\dagger)/2$ and a skew-symmetric one $Z_A=(Z-\bar{Z}^\dagger)/2$, we get:
\begin{eqnarray} \label{eq:appl:FS}
\tfrac{d}{dt} Z_S &=& [\tilde{\delta},\, Z_S] + \tfrac{1}{2} \{ \tilde{R},\, Z_S\} + 2 \tilde\omega + \tfrac{1}{2} Z_S (\tilde{R}-\tilde{\omega}) Z_S - 2 (\mathbf{I}_{2n}+\tfrac{Z_S}{2}) B (\mathbf{I}_{2n}+\tfrac{Z_S}{2})  \; , \\ \label{eq:appl:FA}
\tfrac{d}{dt} Z_A &=&  [\tilde{\delta},\, Z_A] + \tfrac{1}{2} \{ \tilde{R},\, Z_A\} \; .
\end{eqnarray}
Starting with $Z_{A,0}=0$ at $t=0$, we will have $Z_{A,t}=0$ for all times. We can thus identify $Z_t = Z_{S,t}$. We are not going to explicitly integrate \eqref{eq:appl:FS} here, but one can note that it asymptotically converges (modulo generic appropriate observation conditions) towards the stationary point $Z_t = -2 \, \mathbf{I}_{2n}$. We report below its translation back to $S_t$.

\item Next, replacing the previous point in  \eqref{eq:appl:F3} and using \eqref{eq:appl:S3}, we obtain a deterministic equation for $M_t$, depending on $S_t$:
$$ \tfrac{d}{dt} M_t = \tilde{\delta}\, M_t - 2 Z_t^{-1} \tilde{\omega}\, M_t - \tfrac{1}{2} \tilde{R} \, M_t + 2 Z_t^{-1} (\mathbf{I}_{2n}+\tfrac{Z_t}{2})\, B \, M_t \; .
$$
This is a linear equation, but nontrivial, since $M_{0} \neq 0$ and $Z_t$ is time-varying. Once $Z_t$ has converged to $Z_t = -2 \, \mathbf{I}_{2n}$ (if we can say so, since there is no reason for such timescale separation), it would become a time-independent linear equation, involving rotation of $M_t$ with $\tilde{\delta},\tilde{\omega}$ and exponential decay with rate $\tilde{R}/2$.

\item Next, we can replace the preceding results in \eqref{eq:appl:F2}, and using \eqref{eq:appl:S2}, we obtain $D_t$ as an explicit integral on $M_t$:
$$ \tfrac{d}{dt} D_t = 2 \bar{M}^\dagger_t\, \tilde{\omega}\, M_t - 2 \bar{M}^\dagger_t\, B \, M_t \; .$$
This closes a first set of deterministic variables.

\item After inserting the corresponding \Ito~terms into \eqref{eq:appl:F4},\eqref{eq:appl:F7}, we get the stochastic equation:
\begin{eqnarray}\label{eq:appl:dgamforSing}
d\gamma_t &=& \tilde{J}\, u_t \, dt + \tilde{\delta} \, \gamma_t \, dt - \tfrac{1}{2} \tilde{R} \, \gamma_t \, dt - 2 Z_t^{-1} \tilde{\omega}\, \gamma_t\, dt \\ \nonumber
&& + Z_t^{-1}\, (\mathbf{I}_{2n}+\tfrac{Z_t}{2})\, \left( 2\, B \gamma_t\, dt - {\textstyle \sum_l }\; \sqrt{\eta_l} \, \beta_{l,A} \, dy_{t,l} \right) \; ,
\end{eqnarray}
where thus $dy_{t,l}$ follows \eqref{eq:appl:dy}. These variables are too numerous to be all independent, according to the manifold dimension expected from the algebraic criterion. In fact, each $\gamma_t^{(j \neq j')}$ can be expressed as a deterministic function of $\gamma_t^{(j,j)}$ and $\gamma_t^{(j',j')}$ $\in \mathbb{R}^{2n}$, such that those variables together contribute $2dn$ dimensions to the confining manifold. The deterministic link goes as follows.\vspace{3mm}

First, define $L_t^{(j,j')} = Z_t^{(j,j')}\, \left(\mathbf{I}_{2n}+\tfrac{Z_t^{(j,j')}}{2}\right)^{-1} \, \gamma_t^{(j,j')}$. This results in
\begin{equation}\label{eq:appl:dLjjp}
dL_t^{(j,j')} = Z_t^{(j,j')}\, \left(\mathbf{I}_{2n}+\tfrac{Z_t^{(j,j')}}{2} \right)^{-1}  \tilde{J}\, u_t \, dt -{\textstyle \sum_l}\; \sqrt{\eta_l} \beta_{l,A}\, dy_{t,l} \; + \left( \tilde{\delta}^{(j,j')} - \tilde{\omega}^{(j,j')} + \tfrac{\tilde{R}}{2} \right) \, L_t^{(j,j')} \; .
\end{equation}
Any linear combination  $\sum_{j,j'} \; C^{(j,j')}\; L_t^{(j,j')}$ with coefficients satisfying $\sum_{j,j'} \; C^{(j,j')} = 0$ would cancel the explicit dependence on measurements $dy_{t,l}$.

The remaining issue is to find a linear combination which also provides an autonomous deterministic term, thus avoiding indirect dependence on the measurements.
\begin{itemize}
\item The term in $dy_{t,l}$ will cancel as soon as $\sum_{j,j'} \; C^{(j,j')} = 0$. Note that the property also holds if the $C^{(j,j')} \in \mathbb{C}^{2n \times 2n}$ are time-dependent.
\item The term in $\tilde{J}\, u_t \, dt$ will be an autonomous deterministic contribution, as long as the $C^{(j,j')}_t$ only depend on deterministic variables.
\item The term in $\tilde{R}$ is independent of $(j,j')$. Hence, if we take $X^{(j,j')}$ a scalar linear combination of the $L^{(j,j')}$, then it would lead to a term proportional to $\tilde{R}\; X^{(j,j')}$, which is compatible with an autonomous equation for $X^{(j,j')}$. If the $C^{(j,j')}$ are matrix coefficients, then the result of their commutation with $\tilde{R}$ will have to be investigated.
\item Unfortunately, the term in $\tilde{\delta}^{(j,j')}-\tilde{\omega}^{(j,j')}$ is not of this type, calling for a more complicated solution. However, we observe that (i) these are all rotations and (ii) if we take the triplet $(j\neq j'),\; (j,j),\; (j',j')$, then their corresponding coefficients are linearly related.
\end{itemize}

This last observation is the key towards proposing two things. First, we will search for deterministic variables as a linear combination of said triplet, thus
\begin{equation}\label{eq:appl:Xjcomb}
X^{(j,j')}_t \; = \; C^{(j,j',1)}_t \, L^{(j,j)}_t + C^{(j,j',2)}_t \, L^{(j',j')}_t  + C^{(j,j',3)}_t \, L^{(j,j')}_t \; .
\end{equation}
Second, we will decompose $\tilde{\delta}^{(j,j')}-\tilde{\omega}^{(j,j')}$ into a mean contribution and a deviation:
\begin{eqnarray}\label{eq:appl:fordeltadec}
(\tilde{\delta}-\tilde{\omega})^{(j,j)} &=& \tilde{\delta}^{(j,j')} + \omega^{(j,j')} \otimes J \\ \nonumber
(\tilde{\delta}-\tilde{\omega})^{(j',j')} &=& \tilde{\delta}^{(j,j')} - \omega^{(j,j')} \otimes J \\ \nonumber
(\tilde{\delta}-\tilde{\omega})^{(j,j')} &=& \tilde{\delta}^{(j,j')} + i\omega^{(j,j')} \otimes \mathbf{I}_2 \; . 
\end{eqnarray}
The first term on the right hand side is the same for all three cases, it thus plays a role similar to $\tilde{R}$. The second term depends on the row of \eqref{eq:appl:fordeltadec}, but it generates a rotation; moreover, all these terms commute since $\delta^{(j,j')}$ and $\omega^{(j,j')}$ are diagonal, and we recall $\tilde{\delta}^{(j,j')} =  \delta^{(j,j')} \otimes J$. This suggests to cancel the row-dependent parts of \eqref{eq:appl:fordeltadec} by going to a corresponding rotating frame, which would be transparent for the first, common term. 
The algebraic criterion, fixing the manifold dimension, encourages us the other terms should behave favorably.

We thus write:
\begin{eqnarray}\label{eq:appl:Cvals1}
C^{(j,j',1)}_t &=& C^{(j,j',1)}_0 \, \exp\left(-(\omega^{(j,j')} \otimes J)\, t \right) \;\; , \\ \nonumber
C^{(j,j',2)}_t &=& C^{(j,j',2)}_0 \, \exp\left(+(\omega^{(j,j')} \otimes J)\, t \right) \;\; , \\ \nonumber
C^{(j,j',3)}_t &=& C^{(j,j',3)}_0 \, \exp\left(-i(\omega^{(j,j')} \otimes \mathbf{I}_2)\, t \right) \; . 
\end{eqnarray}
The constraint $\sum_{j,j'} \; C_t^{(j,j')} = 0$ for all $t$ then imposes, up to pre-multiplying by the same constant matrix: 
\begin{equation}\label{eq:appl:Cvals2}
 C^{(j,j',1)}_0 = \mathbf{I}_n \otimes \left(\begin{array}{cc} \tfrac{-1}{2} & \tfrac{i}{2} \\ \tfrac{-i}{2} & \tfrac{-1}{2} \end{array}\right) \; , \quad C^{(j,j',2)}_0 = \mathbf{I}_n \otimes \left(\begin{array}{cc} \tfrac{-1}{2} & \tfrac{-i}{2} \\ \tfrac{i}{2} & \tfrac{-1}{2} \end{array}\right) \; ,  \quad C^{(j,j',3)}_0= \mathbf{I}_{2n} \; . 
 \end{equation}
One checks that these $C^{(j,j',k)}_0$, although not full-rank, commute with $A \otimes J$ for any $A$, and of course with $A \otimes \mathbf{I}_2$. They thus maintain the property 
$$C^{(j,j',1)}_t \, \tilde{\delta}^{(j,j')} \, L^{(j,j)}_t + C^{(j,j',2)}_t \, \tilde{\delta}^{(j,j')} \, L^{(j',j')}_t + C^{(j,j',3)}_t \, \tilde{\delta}^{(j,j')}  \, L^{(j,j')}_t =  \tilde{\delta}^{(j,j')} \, X^{(j,j')}_t \; .$$
Finally, similar commutation properties, as well as the property
\begin{eqnarray} \label{eq:appl:AllScal}
 \exp\left(-(\omega^{(j,j')} \otimes J)\, t \right)\, C^{(j,j',1)}_0 &=& e^{-i (\omega^{(j,j')}\otimes \mathbf{I}_2) t} \, C^{(j,j',1)}_0 \; , \\ \nonumber
 \exp\left((\omega^{(j,j')} \otimes J)\, t \right)\, C^{(j,j',2)}_0 &=& e^{-i (\omega^{(j,j')} \otimes \mathbf{I}_2) t} \, C^{(j,j',2)}_0 \; ,
\end{eqnarray}
lead to the result:
$$C^{(j,j',1)}_t \, \tilde{R} \, L^{(j,j)}_t + C^{(j,j',2)}_t \, \tilde{R} \, L^{(j',j')}_t + C^{(j,j',3)}_t \,\tilde{R} \, L^{(j,j')}_t = e^{\tilde{\omega}^{(j,j')} t} \, \tilde{R}\, e^{-\tilde{\omega}^{(j,j')} t} \, X^{(j,j')}_t \; .$$
In hindsight, \eqref{eq:appl:AllScal} shows that $\tilde{X}_t = e^{-\tilde{\omega}^{(j,j')} t} \, X_t =  \; C^{(j,j',1)}_0\, L^{(j,j)}_t + C^{(j,j',2)}_0 \, L^{(j',j')}_t  + C^{(j,j',3)}_0 \, L^{(j,j')}_t \; $  in fact undergoes deterministic dynamics. Taking all things together, we have
\begin{eqnarray*}
\tfrac{d}{dt}\tilde{X}^{(j,j')}_t & = &  C^{(j,j',1)}_0 \, Z_t^{(j,j)}\, \left(\mathbf{I}_{2n}+\tfrac{Z_t^{(j,j)}}{2} \right)^{-1}  \tilde{J}\, u_t \;+\; C^{(j,j',2)}_0 \, Z_t^{(j',j')}\, \left(\mathbf{I}_{2n}+\tfrac{Z_t^{(j',j')}}{2} \right)^{-1}  \tilde{J}\, u_t \\
&& +  Z_t^{(j,j')}\, \left(\mathbf{I}_{2n}+\tfrac{Z_t^{(j,j')}}{2} \right)^{-1}  \tilde{J}\, u_t
\; +\;\; (\tilde{\delta}-\tilde{\omega}+ \tfrac{\tilde{R}}{2})^{(j,j')} \, \tilde{X}^{(j,j')}_t 
\end{eqnarray*}
as a deterministic variable, which together with the knowledge of $\gamma^{(j,j)}_t$ and $\gamma^{(j',j')}_t$ define:
\begin{eqnarray}\label{eq:appl:gamofX}
\gamma_t^{(j,j')} &=& \frac{\mathbf{I}_{2n}+\tfrac{Z_t^{(j,j')}}{2}}{Z_t^{(j,j')}} \, \Bigg( \tilde{X}^{(j,j')}_t - C^{(j,j',1)}_0\, \frac{Z_t^{(j,j)}}{\mathbf{I}_{2n}+\tfrac{Z_t^{(j,j)}}{2}}\; \gamma_t^{(j,j)} \\ \nonumber
&& \phantom{ \frac{\mathbf{I}_{2n}+\tfrac{Z_t^{(j,j')}}{2}}{Z_t^{(j,j')}} \, \, \Bigg( X^{(j,j')}_t -} - C^{(j,j',2)}_0\, \frac{Z_t^{(j',j')}}{\mathbf{I}_{2n}+\tfrac{Z_t^{(j',j')}}{2}}\; \gamma_t^{(j',j')} \Bigg) \; .
\end{eqnarray}

\item Next, now again separately for each $(j,j')$, consider \eqref{eq:appl:F5},\eqref{eq:appl:F8}, in which $d\gamma_t$ can be inserted from the last point. This readily yields
\begin{equation}\label{eq:appl:dlamforsing}
d\lambda_t = 4 \bar{M}_t^\dagger\, \tilde{\omega} \, \gamma_t\, dt - 2 \bar{M}_t^\dagger\,  \left( 2\, B \gamma_t\, dt - {\textstyle \sum_l }\; \sqrt{\eta_l} \, \beta_{l,A} \, dy_{l,t} \right) \; .
\end{equation}
We recognize that the measurement-dependent term can be eliminated in a linear combination with $d\gamma$: defining
$$h_t = \gamma_t + Z_t^{-1}\, (\mathbf{I}_{2n}+\tfrac{Z_t}{2})\, \tfrac{(\bar{M}_t^\dagger)^{-1}}{2} \, \lambda_t \; ,$$
we obtain the autonomous deterministic evolution:
$$\tfrac{d}{dt} h_t = \tilde{J} \, u_t + \tilde{\delta}\, h_t - \tfrac{1}{2} \tilde{R} \, h_t + \tilde{\omega} \, h_t \; . $$
For $u_t=0$, this would involve a matrix exponential inducing constant rotation and decay, in fact the same as for $M_t$ when $Z_t = -2 \, \mathbf{I}_{2n}$. In any case, it establishes that $\lambda_t^{(j,j')}$ is a purely deterministic function of $\gamma_t^{(j,j')}$, i.e.~once $\gamma_t^{(j,j')}$ is given, the value of $\lambda_t^{(j,j')}$ is independent of the measurement outcomes. 

\item Last but not least, there remains to treat $\mu_t$. Inserting the \Ito~terms and the previous results into \eqref{eq:appl:F6},\eqref{eq:appl:F9} yields
\begin{eqnarray*}
\frac{d\mu_{t}}{\mu_t} + \frac{d(\text{det}(Z_t)^{1/2})}{\text{det}(Z_t)^{1/2}} &=& 2 \sum_l \, \sqrt{\eta_l}\, (\beta_{l,A}^\dagger \gamma_t - s_l) \, dw_{t,l} \\
&& + \tfrac{1}{4} \text{Tr}((\tilde{R}-\tilde{\omega})Z_t)\, dt + 2 \bar{\gamma}_t^\dagger \tilde{\omega}\, \gamma_t\, dt + \tfrac{1}{2} \text{Tr}(\tilde{R}-\tilde{\omega}) \, dt \\
&& - \text{Tr}\left(\, B \, (\mathbf{I}_{2n}+\tfrac{Z_t}{2})\, Z_t^{-1}\, (\mathbf{I}_{2n}+\tfrac{Z_t}{2})\, \right) \, dt \; .
\end{eqnarray*}
Inserting $\frac{d(\text{det}(Z_t)^{1/2})}{\text{det}(Z_t)^{1/2}} = \tfrac{1}{2}\, \text{Tr}(Z_t^{-1}\, dZ_t)$ with $dZ_t$ derived above, reduces this equation to:
$$\frac{d\mu_{t}}{\mu_t}  = 2 \sum_l \, \sqrt{\eta_l}\, (\beta_{l,A}^\dagger \gamma_t - s_l) \, dw_{t,l} - \text{Tr}(\tilde{\omega}\, (\mathbf{I}_{2n}+\tfrac{Z_t}{2})\, Z_t^{-1})\, dt + 2 \bar{\gamma}_t^\dagger \, \tilde{\omega} \, \gamma_t\, dt \; .$$
Taking into account that $\mu^{(j,j')} = \bar{\mu}^{(j',j)}$, these are $d^2$ stochastic equations, whereas there should remain $2d-2$ independent stochastic variables. One way of parameterizing this is to consider the $\mu^{(j=j')}$ as independent stochastic, thus following (recall that $\tilde{\omega}^{(j,j')}=0$ for $j=j'$):
$$\frac{d\mu^{(j,j)}_{t}}{\mu^{(j,j)}_t}  \;=\; 2 \sum_l \, \sqrt{\eta_l}\, (\beta_{l,A}^\dagger \gamma^{(j,j)}_t - s_l) \, dw_{t,l} \; = \; 2 \sum_l \, \sqrt{\eta_l}\, \beta_{l,A}^\dagger (\gamma^{(j,j)}_t - Q_{t}) \, dw_{t,l}\; .$$
These contribute $d-1$ independent variables, as they are subject to one constraint fixing the total normalization. Like for the single-oscillator case, we see that $\mu_t^{(j,j)}$ varies more if its Gaussian position $\gamma^{(j,j)}$ deviates more from the total state mean positions $Q_t$.

Next, we observe that the stochastic term comports one part that is independent of $(j,j')$ --- and could thus cancel when taking linear combinations of $\mu^{(j,j')}$ --- while the other part involves $\gamma^{(j,j)}_t$.  To make such a term appear from another stochastic variable, we consider $\Gamma_t = \bar{\gamma}_t^\dagger \, Z_t\, (\mathbf{I}_{2n}+\tfrac{Z_t}{2})^{-1}\, \gamma_t$. A few computations then indeed yield:
\begin{eqnarray}\label{eq:appl:dGamma}
d(\log(\mu)+\Gamma) &=& - 2 \sum_l\, \sqrt{\eta_l}\, s_l\, dw_{t,l} - 2 \sum_l \, \eta_l \, s_l^2 \, dt \\ \nonumber
&& + \text{Tr}\left( B\, (\mathbf{I}_{2n}+\tfrac{Z_t}{2})\, Z_t^{-1}\, \right) \, dt - \text{Tr}(\tilde{\omega}\, (\mathbf{I}_{2n}+\tfrac{Z_t}{2})\, Z_t^{-1})\, dt \\ \nonumber
&& + 2\, L_t^\dagger \, \tilde{J} \, u_t \, dt \; ,
\end{eqnarray}
with $L_t$ essentially related to $\gamma_t$ and following the stochastic equation \eqref{eq:appl:dLjjp}. The first row of \eqref{eq:appl:dGamma} is now totally independent of $(j,j')$. The second row only contains deterministic variables. In the last row, the presence of $L_t$ still has to be handled. The idea, in order to obtain purely deterministic variables, is to take linear combinations of \eqref{eq:appl:dGamma} for various $(j,j')$: its first row would disappear, while the last row would involve the deterministic variables $\tilde{X}^{(j,j')}$. 

\emph{For $u_t=0$, this is simple:} the last row is absent, so any linear combination of $(\log(\mu)+\Gamma)^{(j,j')}$ for various indices $(j,j')$ would yield a purely deterministic equation, involving only the second row of \eqref{eq:appl:dGamma}, as soon as the coefficients of the linear combination sum to zero. This also shows that the $\mu_t^{j,j}$ are not stochastically independent either in this case, as we have already seen for the single oscillator model.  For $u_t \neq 0$, we have to make this slightly more complicated.

First, consider the real part of \eqref{eq:appl:dGamma}. We would define 
\begin{equation}\label{eq:appl:YR}
Y^{(j,j')}_{R,t} = \tilde{c}^{(j,j',1)} \mathfrak{R}(\log(\mu)+\Gamma)^{(j,j)}+\tilde{c}^{(j,j',2)} \mathfrak{R}(\log(\mu)+\Gamma)^{(j',j')}+\tilde{c}^{(j,j',3)} \mathfrak{R}(\log(\mu)+\Gamma)^{(j,j')}
\end{equation}
with scalars $\tilde{c}^{(j,j',k)} $ satisfying $\sum_k \, \tilde{c}^{(j,j',k)} = 0$, in order for the first row of \eqref{eq:appl:dGamma} to cancel when writing $dY^{(j,j')}_{R,t}$. We further note that all variables associated to $j'=j$ are real, so from \eqref{eq:appl:Cvals2} we can write for the last row of \eqref{eq:appl:dGamma}:  
\begin{eqnarray*}
\tfrac{-1}{2}\, \mathfrak{R}\left( 2\, (L_t^{(j,j)})^\dagger \, \tilde{J} \, u_t \right) &=& \mathfrak{R}\left( 2\, (C_0^{(j,j',1)}\, L_t^{(j,j)})^\dagger \, \tilde{J} \, u_t \right)\; ,\\
\tfrac{-1}{2}\, \mathfrak{R}\left( 2\, (L_t^{(j',j')})^\dagger \, \tilde{J} \, u_t \right) &=& \mathfrak{R}\left( 2\, (C_0^{(j,j',2)}\, L_t^{(j',j')})^\dagger \, \tilde{J} \, u_t \right) \; ,
\end{eqnarray*}
while $C_0^{(j,j',3)}= \mathbf{I}_{2n}$ trivially implies:
$$\mathfrak{R}\left( 2\, (L_t^{(j,j')})^\dagger \, \tilde{J} \, u_t \right) = \mathfrak{R}\left( 2\, (C_0^{(j,j',3)}\, L_t^{(j,j')})^\dagger \, \tilde{J} \, u_t \right)
 \; .$$
Therefore, taking $\tilde{c}^{(j,j',1)}= \tilde{c}^{(j,j',2)}=-1/2$ and  $\tilde{c}^{(j,j',3)}=1$ in \eqref{eq:appl:YR} is compatible with a fully deterministic evolution:
\begin{eqnarray*}
\tfrac{d}{dt} Y^{(j,j')}_{R,t} &=& \frac{-1}{2}\, \text{Tr}\left( B\, (\mathbf{I}_{2n}+\tfrac{Z^{(j,j)}_t}{2})\, (Z_t^{(j,j)})^{-1}\, \right) - \frac{1}{2}\,  \text{Tr}\left( B\, (\mathbf{I}_{2n}+\tfrac{Z^{(j',j')}_t}{2})\, (Z_t^{(j',j')})^{-1}\, \right) \\
&& + \mathfrak{R}\left(\,  \text{Tr}\left( \,(B-\tilde{\omega}^{(j,j')})\, (\mathbf{I}_{2n}+\tfrac{Z^{(j,j')}_t}{2})\, (Z_t^{(j,j')})^{-1}\, \right) \,\right)\\
&& + 2\, \mathfrak{R}(\tilde{X}_t^{(j,j')})^\dagger \, \tilde{J} \, u_t \, .
\end{eqnarray*}
Since $\mathfrak{R}\log(\mu) = \log(|\mu|)$, this shows how the weights $|\mu_t^{(j\neq j')}|$ are deterministically linked to the $\mu_t^{(j,j)}$ and $\gamma_t^{(j,j)}$.

There remains to find deterministic variables among the $\mathfrak{I}\log(\mu^{(j,j')}) = \text{phase}(\mu^{(j,j')})$. Taking the imaginary part of \eqref{eq:appl:dGamma}, the first row drops rightaway. Furthermore, from \eqref{eq:appl:Cvals2}, we note that for any triplet $j,j',j''$ one has:
$$\mathfrak{I}(L^{(j,j')}_t + L^{(j',j'')}_t + L^{(j'',j)}_t ) = \mathfrak{I}(\tilde{X}^{(j,j')}_t + \tilde{X}^{(j',j'')}_t + \tilde{X}^{(j'',j)}_t )\; . $$
Therefore, we can define
\begin{equation}\label{eq:appl:YI}
Y^{(j,j',j'')}_{I,t} = \mathfrak{I}(\log(\mu)+\Gamma)^{(j,j')}+ \mathfrak{I}(\log(\mu)+\Gamma)^{(j',j'')}+ \mathfrak{I}(\log(\mu)+\Gamma)^{(j'',j)}
\end{equation}
and obtain its deterministic evolution:
\begin{eqnarray*}
\tfrac{d}{dt}\, Y^{(j,j',j'')}_{I,t} &=& \mathfrak{I}\left(\,  \text{Tr}\left( \,(B-\tilde{\omega}^{(j,j')})\, (\mathbf{I}_{2n}+\tfrac{Z^{(j,j')}_t}{2})\, (Z_t^{(j,j')})^{-1}\, \right) \,\right)\\
&& + \mathfrak{I}\left(\,  \text{Tr}\left( \,(B-\tilde{\omega}^{(j',j'')})\, (\mathbf{I}_{2n}+\tfrac{Z^{(j',j'')}_t}{2})\, (Z_t^{(j',j'')})^{-1}\, \right) \,\right)\\
&& + \mathfrak{I}\left(\,  \text{Tr}\left( \,(B-\tilde{\omega}^{(j'',j)})\, (\mathbf{I}_{2n}+\tfrac{Z^{(j'',j)}_t}{2})\, (Z_t^{(j'',j)})^{-1}\, \right) \,\right)\\
&& + 2 \, \mathfrak{I}(\tilde{X}^{(j,j')}_t + \tilde{X}^{(j',j'')}_t + \tilde{X}^{(j'',j)}_t )^\dagger \, \tilde{J} \, u_t \, .
\end{eqnarray*}
This expresses how the phases of the $\mu^{(j,j')}_t$ are deterministically linked. One can check that for instance the phases of $\mu^{(1,2)}_t,\,\mu^{(1,3)}_t,...,\mu^{(1,d)}_t$ can be chosen independently, then constraining all the other phases through the $Y^{(j,j',j'')}_{I,t}$, in agreement with the fact that there remained to find $(d-1)$ stochastic variables and all other deterministic.
\end{itemize}
This concludes the computations. There remains to define the initial conditions, by imposing the Gaussian kernel to coincide with the Dirac-peak limit at $t=0$. We have thus established the following result.\\


\begin{prop}\label{prop:LLast}
The quantum SDE \eqref{eq:PRmodel0},\eqref{eq:PRmodel1}, for a general initial condition, admits the explicit solution:
$$\rho_{t} = \sum_{j,j'=1}^d\; \q{j}\qd{j'} \otimes \rho_{t,(j,j')} \; ,$$
where each $\rho_{t,(j,j')}$ is an operator on the Hilbert space of the $n$ cavities. The evolution of the latter is described as follows in its Wigner function representation on $q=(x_1;p_1;x_2;p_2;...;x_n;p_n) \in \mathbb{R}^{2n}$:
\begin{eqnarray}\label{eq:applS:WtoG}
\mathcal{W}^{\{\rho_{t,(j,j')}\}}(q) &=& \frac{\mu_t^{(j,j')}}{\pi^{n}\mathrm{det}(S_t^{(j,j')})^{1/2}}\int \mathcal{W}^{\{\rho_0^{(j,j')}/\mu_0^{(j,j')}\}}(q_0) \, K^{(j,j')}_{t}(q,q_0)\, dq_0 \; , \\ \nonumber
K^{(j,j')}_{t}(q,q_0) &=& \exp\Bigg[-\left(q-\bar{\gamma}^{(j,j')}_t-\bar{M}^{(j,j')}_t q_0\right)^\dagger\; (S_t^{(j,j')})^{-1}\; \left(q-\gamma^{(j,j')}_t-M^{(j,j')}_t q_0\right) \\ \nonumber
&& \phantom{KKKKKKKKK} \; + q_0^\dagger D_t^{(j,j')} q_0 + (\bar{\lambda}^{(j,j')})^\dagger q_0 \; \Bigg] \; .
\end{eqnarray}
Defining the constants:
\begin{align*}
& \tilde{R} = \sum_{l=1}^m  \; R^{(re)}_l \otimes I_2 + R^{(im)}_l \otimes J\quad \in \mathbb{R}^{2n \times 2n}\\
& \qquad \qquad \text{with} \; R^{(re)}_{l,(k,k')} = \mathfrak{R}(\beta_{l,k}\bar{\beta}_{l,k'})\;  \text{ and }  \; R^{(im)}_{l,(k,k')} = \mathfrak{I}(\beta_{l,k}\bar{\beta}_{l,k'}) \; ; \\
& \tilde{J} = \mathbf{I}_n \otimes [0 \; ,\; 1 \; ;\; -1 \; ,\; 0 ] \quad \in \mathbb{R}^{2n \times 2n} \; ; \\
& \tilde\delta^{(j,j')} = \mathrm{diag}_k\left(\Delta_k + \frac{\chi_{j,k}+\chi_{j',k}}{2}\right) \otimes [0 \; ,\; 1 \; ;\; -1 \; ,\; 0 ] \quad \in \mathbb{R}^{2n \times 2n} \; ; \\
& \tilde\omega^{(j,j')} = -i\, \mathrm{diag}_k \left( \frac{\chi_{j,k}-\chi_{j',k}}{2} \right) \otimes \mathbf{I}_2 \quad \in \mathbb{C}^{2n \times 2n}\; ; \\
& \beta_{l,A} = [\mathfrak{R}(\beta_{l,1});\; -\mathfrak{I}(\beta_{l,1});\; \mathfrak{R}(\beta_{l,2});\; -\mathfrak{I}(\beta_{l,2});\; ...  ] \quad \in \mathbb{R}^{2n} \; ; \\
& B = \sum_{l=1}^m \; \eta_l \beta_{l,A} \, \beta_{l,A}^\dagger \quad \in \mathbb{R}^{2n \times 2n} \; ,
\end{align*}
the Gaussian kernels $K^{(j,j')}_{t}(q,q_0)$ contain the following parameters with purely deterministic evolution:
\begin{eqnarray}\label{eq:applS:dS}
\tfrac{d}{dt} S_t^{(j,j')} &=& [\tilde{\delta}^{(j,j')},\, S^{(j,j')}_t] - \tfrac{1}{2} \{ \tilde{R},\, S^{(j,j')}_t\} + \tfrac{1}{2} (\tilde{R}-\tilde{\omega}^{(j,j')})  \\ \nonumber
&& + 2 S^{(j,j')}_t\, \tilde\omega^{(j,j')}\, S^{(j,j')}_t - 2 (S^{(j,j')}_t - \tfrac{\mathbf{I}_{2n}}{2}) \, B\, (S^{(j,j')}_t - \tfrac{\mathbf{I}_{2n}}{2})  \; ,\\[3mm]
 \tfrac{d}{dt} M^{(j,j')}_t &=& \left(\, \tilde{\delta}^{(j,j')} + 2 S^{(j,j')}_t \, \tilde{\omega}^{(j,j')} - \tfrac{1}{2} \tilde{R}  - 2 (S^{(j,j')}_t - \tfrac{\mathbf{I}_{2n}}{2})\, B \right) \, M^{(j,j')}_t \; ,\\[3mm]
\tfrac{d}{dt} D_t &=& 2 \bar{M}^\dagger_t\, \tilde{\omega}\, M_t - 2 \bar{M}^\dagger_t\, B \, M_t \; ,
\end{eqnarray}
initialized with $S^{(j,j')}_0=D^{(j,j')}_0=\mathbf{0}_{2n}$ and $M^{(j,j')}_0 = \mathbf{I}_{2n}$.
\newline It features the following independent stochastic variables:
\begin{eqnarray}
d\gamma^{(j=j')}_t &=& \tilde{J}\, u_t \, dt + \tilde{\delta}^{(j=j')} \, \gamma^{(j=j')}_t \, dt - \tfrac{1}{2} \tilde{R} \, \gamma^{(j=j')}_t \, dt \\ \nonumber
&& - \, (S_t^{(j=j')}-\tfrac{\mathbf{I}_{2n}}{2})\, \left( 2\, B \gamma^{(j=j')}_t\, dt - {\textstyle \sum_{l=1}^m }\; \sqrt{\eta_l} \, \beta_{l,A} \, dy_{t,l} \right) \; ,\\[3mm]
d\mu^{(j=j')}_{t} &=& 2 \, \mu^{(j=j')}_{t} \; \sum_{l=1}^m \, \sqrt{\eta_l}\, \beta_{l,A}^\dagger \left(\gamma^{(j=j')}_t - Q_{t}\right) \, \left( dy_{t,l} - 2 \sqrt{\eta_l}\, \beta_{l,A}^\dagger \, Q_{t} \; dt \right)\; , \\[3mm]
 \nonumber
d(\mathrm{phase}(\mu_t^{(1,j')})) &=& 2 \sum_{l=1}^m \; \sqrt{\eta_l}\, \beta_{l,A}^\dagger\, \mathfrak{I}(\gamma^{(1,j')}_t) \; dy_{t,l} - 2 \mathfrak{I}\left((\bar{\gamma}^{(1,j')}_t)^\dagger \,(B-\tilde{\omega}^{(1,j')})\, \gamma^{(1,j')}_t \right) \, dt \\
&& + \mathfrak{I}\left(\mathrm{Tr}\left( \tilde{\omega}^{(1,j')} \, (S^{(1,j')}_t - \tfrac{\mathbf{I}_{2n}}{2}) \right) \right) \, dt \; ,
\end{eqnarray}
initialized with $\gamma^{(j,j')}_0=0$ and $\mu_0^{(j,j')} = \text{Tr}(\rho_{0,(j,j')})$,  where the measurements give $dy_{t,l} = 2 \sqrt{\eta_l}\, \beta_{l,A}^\dagger \, Q_{t} \; dt + dw_{t,l}$  and  $Q_{t} = [\,\text{Tr}(\rho_t\mathbf{X}_1);\;\text{Tr}(\rho_t\mathbf{P}_1);\;\text{Tr}(\rho_t\mathbf{X}_2);\;\text{Tr}(\rho_t\mathbf{P}_2);\;... \,] \; \in \mathbb{R}^{2n}$.
\newline
The remaining parameters are deterministically linked to those stochastic variables as follows:
\begin{eqnarray}
\lambda_t^{(j,j')} &=& 2 \bar{M}_t^{(j,j')\dagger}\,(\tfrac{\mathbf{I}_{2n}}{2}-S_t^{(j,j')})^{-1}\, (h_t^{(j,j')}- \gamma_t^{(j,j')}) \\ \nonumber
&& \text{where}\\ \nonumber
&& \!\!\!\!\! \tfrac{d}{dt} h_t^{(j,j')} = \tilde{J} \, u_t + (\tilde{\delta}+\tilde{\omega}- \tfrac{\tilde{R}}{2})^{(j,j')} \, h_t^{(j,j')} \; , \\[3mm]
\gamma_t^{(j,j')} &=& (\tfrac{\mathbf{I}_{2n}}{2}-S_t^{(j,j')}) \Bigg( \tilde{X}^{(j,j')}_t - C_1\, (\tfrac{\mathbf{I}_{2n}}{2}-S_t^{(j,j)})^{-1}\; \gamma_t^{(j,j)} - C_2\, (\tfrac{\mathbf{I}_{2n}}{2}-S_t^{(j',j')})^{-1}\, \gamma_t^{(j',j')} \Bigg) \; \nonumber \\ \nonumber
&& \text{where}\\
&& \!\!\!\!\! \tfrac{d}{dt}\tilde{X}^{(j,j')}_t  =   C^{(j,j',1)}_0 \left(\tfrac{\mathbf{I}_{2n}}{2}-S_t^{(j,j)} \right)^{-1}  \tilde{J}\, u_t \;+\; C^{(j,j',2)}_0 \, \left(\tfrac{\mathbf{I}_{2n}}{2}-S_t^{(j',j')} \right)^{-1}  \tilde{J}\, u_t \\ \nonumber
&&\phantom{\tfrac{d}{dt}\tilde{X}^{(j,j')}_t  =    k} +  \left(\tfrac{\mathbf{I}_{2n}}{2}-S_t^{(j,j')} \right)^{-1}  \tilde{J}\, u_t
\; +\;\; (\tilde{\delta}-\tilde{\omega}+ \tfrac{\tilde{R}}{2})^{(j,j')} \, \tilde{X}^{(j,j')}_t  \; , \\[3mm]
\log(|\mu_t^{(j,j')}|) &=& Y^{(j,j')}_{R,t} + \tfrac{1}{2}(\log(\mu)+\Gamma)_t^{(j,j)} +\tfrac{1}{2}(\log(\mu)+\Gamma)_t^{(j',j')} - \mathfrak{R}(\Gamma_t^{(j,j')}) \\ \nonumber
&& \text{where } \Gamma_t = \bar{\gamma}_t^\dagger \, (\tfrac{\mathbf{I}_{2n}}{2}-S_t^{(j,j)})^{-1}\, \gamma_t \; \text{ and }\\ \nonumber
&& \!\!\!\!\! \tfrac{d}{dt} Y^{(j,j')}_{R,t} = \frac{1}{2}\, \text{Tr}\left( B\, (S^{(j,j)}_t-\tfrac{\mathbf{I}_{2n}}{2} + S^{(j',j')}_t - \tfrac{\mathbf{I}_{2n}}{2})\, \right)  \\
&& \phantom{\tfrac{d}{dt} Y^{(j,j')}_{R,t} =  k} + \mathfrak{R}\left(\, 2(\tilde{X}_t^{(j,j')})^\dagger \, \tilde{J} \, u_t - \text{Tr}\left( \,(B-\tilde{\omega}^{(j,j')})\,(S^{(j,j')}_t-\tfrac{\mathbf{I}_{2n}}{2})\, \right) \,\right) \; , \\[3mm]
\mathrm{phase}(\mu_t^{(j,j')}) &=& Y^{(j,j',1)}_{I,t}  + \mathrm{phase}(\mu_t^{(1,j')})- \mathrm{phase}(\mu_t^{(1,j)}) -\mathfrak{I}(\Gamma_t^{(j',1)}+\Gamma_t^{(1,j)}+\Gamma_t^{(j,j')}) \\ \nonumber
&& \text{where}\\ \nonumber
&& \!\!\!\!\! \nonumber
\tfrac{d}{dt}\, Y^{(j,j',1)}_{I,t} = \mathfrak{I}\left(\,  \text{Tr}\left( \,(B-\tilde{\omega}^{(j,j')})\, (\tfrac{\mathbf{I}_{2n}}{2}-S^{(j,j')}_t)\, \right) \,\right)\\ \nonumber
&& \phantom{\tfrac{d}{dt}\, Y^{}_{I,t}} + \mathfrak{I}\left(\,  \text{Tr}\left( \,(B-\tilde{\omega}^{(j',1)})\, (\tfrac{\mathbf{I}_{2n}}{2}-S^{(j',1)}_t) + (B-\tilde{\omega}^{(1,j)})\, (\tfrac{\mathbf{I}_{2n}}{2}-S^{(1,j)}_t)\, \right) \,\right)\\ \nonumber
&& \phantom{\tfrac{d}{dt}\, Y^{}_{I,t}} + 2 \, \mathfrak{I}(\tilde{X}^{(j,j')}_t + \tilde{X}^{(j',1)}_t + \tilde{X}^{(1,j)}_t )^\dagger \, \tilde{J} \, u_t \, ,
\end{eqnarray}
with $C_1 = -\mathbf{I}_{2n}/2 + i\,\tilde{J}/2$ and  $C_2 = -\mathbf{I}_{2n}/2 - i\,\tilde{J}/2$, and with the initialization:
\begin{eqnarray*}
h_0^{(j,j')} = 0 \; , \quad \tilde{X}^{(j,j')}_0 = 0 \; , \quad  Y^{(j,j')}_{R,0} = \log\left(\tfrac{|\mu_0^{(j,j')}|}{\sqrt{\mu_0^{(j,j)}\mu_0^{(j',j')} }}\right) \; , \quad Y^{(j,j',1)}_{I,0} = \mathrm{phase}\left(\tfrac{\mu_0^{(j,j')}\mu_0^{(1,j,)}}{\mu_0^{(1,j')}} \right) \; .
\end{eqnarray*}
\end{prop}
\vspace{5mm}

Some of the above expressions are undefined when $S_t - \mathbf{I}_{2n}\, /2$ becomes singular; the interested reader is invited to go back through the derivations above in order to smoothen out these points.\\

Note that each $R_l^{(re)}$ above is positive semidefinite, while $q^\dagger (R_l^{(im)}\otimes J) \, q = \overline{q^\dagger (R_l^{(im)\dagger}\otimes J^\dagger) \, q } = \overline{q^\dagger (R_l^{(im)}\otimes (-J)) q}= -q^\dagger (R_l^{(im)}\otimes J) \, q = 0$, proving that $\tilde{R}$ is positive semi-definite. It governs the rate at which $S_t$ converges towards $\mathbf{I}_{2n}\, /2$ and $M_t$ converges towards $\mathbf{0}_{2n}$. 

Assuming that this has happened (e.g.~$\tilde{R}$ positive definite), the Gaussian kernel factorizes into a function of $q_0$ and a Gaussian in $q$. The former integrates out with the initial state as a constant, and the latter expresses the product of coherent states towards which the cavity has converged, indexed by $j,j'$. In this limit, from \eqref{eq:appl:dgamforSing}, the $\gamma^{(j',j)}$ undergo no stochastic motion anymore: they follow deterministic dynamics which depend on $(j,j')$ but not on the measurement record (nor, consistently, on the $\eta_l$). These correspond to the coherent state amplitudes. The $D^{(j,j')}$ and the $\lambda^{(j,j')}$ (see \eqref{eq:appl:dlamforsing} to resolve the undefiniteness) don't vary anymore either and the relative weight $\text{Tr}(\rho_{t,(j,j)})$ of qudit level $j$ is governed by $\mu_{t}^{(j,j)}$. Like for the single oscillator case, we can write $\tilde{\mu}_{t}^{(j,j)} = \text{Tr}(\rho_{t,(j,j)})$ with $\tfrac{d\tilde{\mu}}{\tilde{\mu}} = \tfrac{d\mu}{\mu}$, such that $Q_t = \sum_{j'} \tilde{\mu}_t^{(j',j')} \, \gamma_t^{(j',j')}$ while $\gamma_t^{(j,j)} = \gamma_t^{(j,j)} \, \sum_{j'} \tilde{\mu}_t^{(j',j')}$. From these observations, we obtain:
$$
d\tilde{\mu}_t^{(j,j)} = 2 \, \tilde{\mu}_t^{(j,j)} \, \sum_{l=1}^m \, \sqrt{\eta_l} \, \beta_{l,A}^\dagger \, \left(\sum_{j'=1}^d \, ( \gamma_t^{(j,j)}- \gamma_t^{(j',j')})\, \tilde{\mu}_t^{(j',j')}  \right) \, dw_{t,l} \; .
$$
This generalization of \eqref{eq:appfl:mugetsQND} again matches the form \eqref{eq:VinceRes}. It expresses how the setting has become equivalent, in the limit, to a \emph{direct} QND measurement of the $\tilde{\mu}_t^{(j,j)}$, with measurement operators $L_l(t) = \sum_{j=1}^d \, \q{j}\qd{j}\,(\beta_{l,A}^\dagger \gamma_t^{(j,j)})$.




\end{document}